\def\mean#1{\left< #1 \right>}

\documentclass[twocolumn]{aastex631}

\usepackage{multirow}
\usepackage{amsmath}
\usepackage{mathrsfs}
\usepackage{graphicx}
\usepackage{color,subfigure}
\usepackage{makecell}
\usepackage{hyperref}

\begin{document}

\title{The Formation of Direct Collapse Black Holes  at Cosmic Dawn and 21 cm Global Spectrum
}

\correspondingauthor{Bin Yue}
\email{yuebin@nao.cas.cn}

\author{Meng Zhang}
\affiliation{National Astronomical Observatories, Chinese Academy of Sciences, 20A Datun Road, Chaoyang District, Beijing 100101, China}
\affiliation{School of Astronomy and Space Science, University of Chinese Academy of Sciences, No.1 Yanqihu East Rd, Huairou District, Beijing 101408, China}

\author{Bin Yue}
\affiliation{National Astronomical Observatories, Chinese Academy of Sciences, 20A Datun Road, Chaoyang District, Beijing 100101, China}
\affiliation{Key Laboratory of Radio Astronomy and Technology, Chinese Academy of Sciences, 20A Datun
Road, Chaoyang District, Beijing 100101, China}

\author{Yidong Xu}
\affiliation{National Astronomical Observatories, Chinese Academy of Sciences, 20A Datun Road, Chaoyang District, Beijing 100101, China}
\affiliation{Key Laboratory of Radio Astronomy and Technology, Chinese Academy of Sciences, 20A Datun
Road, Chaoyang District, Beijing 100101, China}

\author{Andrea Ferrara}
\affiliation{Scuola Normale Superiore, Piazza dei Cavalieri 7, 56126 Pisa, Italy}




\begin{abstract}
JWST reveals numerous high-z galaxies and Supermassive Black Holes (SMBHs), suggesting that stars and SMBH seeds formation at $z \gtrsim 10$ may be more efficient than previously derived. One popular SMBH seed scenario is the Direct Collapse Black Holes (DCBHs) formed in pristine atomic-cooling halos irradiated by nearby galaxies. Therefore, the efficient star formation likely facilitates the formation of DCBH. We calculate the first critical $k_{\rm H_2}-k_{\rm H^-}$ curves for DCBH formation under the influence of X-ray radiation using the one-zone model. We then build the UV luminosity function consistent with JWST observations and incorporate it into the model that calculates the DCBH-triggering probability. We confirm that enhanced star formation promotes the DCBH formation. However, the DCBH abundance $n_{\rm DCBH}$ is significantly influenced by the X-ray radiation that is also related to star formation. Since the 21 cm global spectrum is also X-ray dependent, the 21 cm absorption depth $\delta T_b^{\rm trough}$ at Cosmic Dawn encodes the DCBH abundance information. We provide a tentative trend in the $n_{\rm DCBH}$ - $\delta T^{\rm trough}_{\rm b}$ relation, which could be a useful guide. In our fiducial model, if $\delta T_b^{\rm trough}\gtrsim -100$ mK, then the DCBH is rather rare; if $-150~{\rm mK} \lesssim \delta T_b^{\rm trough}\lesssim -100$ mK, $ n_{\rm DCBH} \sim \mathcal{O}(10^{-2}-10^{-3})$ cMpc$^{-3}$ (comoving $\rm Mpc^{-3}$), consistent with the HST/JWST observed SMBHs abundance at $z\gtrsim 6$; if $\delta T_b^{\rm trough}\lesssim -150$ mK, $n_{\rm DCBH}$ can largely exceed $\mathcal{O}(10^{-2})$ cMpc$^{-3}$.  The 21 cm global spectrum observations will help to constrain the DCBH abundance.
\end{abstract}



\section{Introduction}\label{sec:intro}
The First black holes are seeds of the Supermassive Black Holes (SMBHs) with mass $\gtrsim 10^8~M_{\odot}$ at $z\gtrsim 6$
(e.g., \citealt{Banados+2024,
Tripodi+2024, Fan+2023}).
They are believed to form in the Cosmic Dawn and exhibit rapid growth (see e.g., \citealt{Woods+2019, Inayoshi+2020} and references therein for a review).
One promising mechanism to form the first black holes is the direct collapse process in metal-free atomic-cooling halos (ACHs, virial temperature $\gtrsim 10^{4}\,\rm K$)\citep{Omukai+2001,Oh2002ApJ,Begelman2006MNRAS,Shang+2010,Latif2013MNRAS}.

The formation of Direct Collapse Black Hole (DCBH) requires suppressing the H$_2$ cooling to maintain the gas cloud in an atomic-cooling state, preventing fragmentation. This demands an external radiation field strong enough to dissociate H$_2$ and detach H$^{-}$, a key intermediary for H$_2$ formation. This is described by a critical Lyman-Werner (LW) radiation intensity $J_{\rm crit}$, defined as the mean intensity between $11.2-13.6$ eV \footnote{In some literature the critical intensity is defined at 13.6 eV.}, and a DCBH can form in the metal-free gas cloud irradiated by supercritical radiation field.
 
$J_{\rm crit}$ depends on the spectrum of the radiation field since both H$_2$ dissociation and H$^-$ detachment are essential for suppressing the H$_2$ cooling. One-zone simulations show that for blackbody spectrum with effective temperature of $10^4$ ($10^5$) K , $J_{\rm crit}\sim 30$ ($\sim10^2-10^4$), in units of $10^{-21}\,\rm erg\ s^{-1}cm^{-2}Hz^{-1}sr^{-1}$. For a realistic first galaxy spectrum consisting of stars with various temperatures, $J_{\rm crit} \sim 100-1000$ \citep{Sugimura+2014, Latif+2015, Yue+2017}. 

However, the 3D full hydrodynamic simulations give $\gtrsim 10-100$ times higher $J_{\rm crit}$, probably because shocks and hydrodynamic effects promote H$_2$ and H$^-$ formation via additional ionization \citep{Shang+2010, Regan+2014, Latif+2014, Latif+2015}. Furthermore, X-ray radiation ionizes and heats the gas cloud, generating more electrons, thus facilitating the formation of H$_2$ and enhancing $J_{\rm crit}$ \citep{ Inayoshi+2015, Latif+2015, Glover+2016, Regan+2016, Yue+2017}.  
\citet{Inayoshi+2015} gave the dependence of $J_{\rm crit}$ on the X-ray intensity, and found that when the X-ray intensity $J_X \gtrsim 10^{-2}$ (in units of $10^{-21}\, \rm erg\ s^{-1}\ cm^{-2}\ sr^{-1}\ Hz^{-1}$) at 1 $\rm keV$, the $J_{\rm crit}$ is significantly boosted.

Sometimes it is not feasible to solely use $J_{\rm crit}$ as the DCBH formation criterion, as one must specify the spectral energy distribution (SED). Alternatively, a critical curve in the $k_{\rm H_2}-k_{\rm H^-}$ plane has been proposed and confirmed by both one-zone simulations \citep{Agarwal+2016, Wolcott-Green+2017} and 3D hydrodynamic simulations \citep{Luo+2020}, where $k_{\rm H_2}$ is the photodissociation rate of H$_2$, $k_{\rm H^-}$ is the photodetachment rate of H$^-$.
DCBH can form as long as the external radiation field has $k_{\rm H_2}$ and $k_{\rm H^-}$ above the critical curve. Fitting formulae are provided in these works, but there are still some discrepancies in the detailed form.

The critical curve should also depend on the X-ray radiation. Unfortunately, until now we did not find such kind of investigation. However, such critical curves are quite necessary for estimating the DCBH formation rate more reasonably. 

The number density of DCBHs has long been debated, primarily due to uncertainties in $J_{\rm crit}$. Some studies suggest DCBHs are rare, with densities as low as $\lesssim 10^{-6}\, \rm cMpc^{-3}$(comoving $\rm Mpc^{-3}$) \citep{Dijkstra+2008, Dijkstra+2014, Habouzit+2016}, while others estimate much higher densities, up to $\sim 10^{-3}-10^{-1}\, \rm cMpc^{-3}$ \citep{Agarwal+2012, Agarwal+2013, Yue+2014, Pacucci+2015_spectral, Yue+2017}. \citet{Yue+2014} proposed that the radiation from initial DCBHs could trigger the subsequent formation of many more DCBHs \citep{Yue+2017}.

Observations of high-$z$ AGNs set up lower limits on the DCBH abundance if each SMBH has at least a DCBH progenitor at the earlier redshift. Previously, the observed high-$z$ AGNs were rare, with $\sim 10^{-9}-10^{-6}$ cMpc$^{-3}$ (e.g. \citealt{Jiang+2016, Jiang+2022, Matsuoka+2023}), which seems favor a low DCBH density scenario. However, recent JWST observations reveal a large number of faint AGN at $z \sim 4-8$, with the measured number density $\approx 10^{-5}-10^{-3}\,\rm cMpc^{-3}$ \citep{Harikane+2023_AGN, Kocevski+2023, Onoue+2023, Scholtz+2023, Maiolino+2023, Furtak+2024, Greene+2024, Kocevski+2024, Matthee+2024}. 
Particularly, a recent search for faint AGNs via photometric variability in the Hubble Ultra-Deep Field has established a larger lower limit on the number density of SMBHs at $z \sim 6$, with $n_{\rm SMBH} \gtrsim 8\times 10^{-3}\,\rm cMpc^{-3}$ 
\citep[][hereafter H24]{Hayes+2024}. Expanding on this, a more extensive statistical analysis of variability-selected AGN candidates in the HUDF has been conducted by \citet{Cammelli+2025}, the estimated number density reaches  $\gtrsim 10^{-2}\,\rm cMpc^{-3}$ at $z \gtrsim 6$. This sheds light on the formation of DCBHs and their impact on the IGM in the early Universe.

A search for DCBH candidates in the JWST cycle-1 PEARLS survey set a conservative upper limit of $5\times 10^{-4}\,\rm cMpc^{-3}$ on the number density of DCBHs \citep{Nabizadeh+2024}. This is lower than the above lower limit, probably because a SMBH has a low duty-cycle during the pure DCBH phase.

Recent JWST observations have also revealed numerous galaxy candidates with photometric redshifts $z\gtrsim10$ (e.g. \citealt{Donnan+2023_candidates, Donnan+2023_UVLF, McLeod+2024, Adams+2023, Naidu+2022, Castellano+2022, Harikane+2023_LF, Finkelstein+2022, Austin+2023, Hainline+2024}), some of which have already been spectroscopically confirmed \citep{Curtis-Lake+2023, Haro+2023, Harikane+2024, Wang+2023, Castellano+2024}. This suggests that high-redshift galaxies may be more abundant and star formation more efficient than previously predicted \citep{Boylan-Kolchin+2023, Dekel+2023, Andalman+2024}, potentially enhancing DCBH formation. Investigating DCBH formation under UV luminosity function (LF) models constrained by recent JWST observations is therefore essential.

The previous works that gave the promising results on DCBH abundance did not involve the X-ray feedback. X-rays from the nearby X-ray sources (including galaxies and DCBHs) and a background may reduce the number density of DCBHs.

Direct observing the X-ray properties of the first luminous objects at Cosmic Dawn remains challenging. The 21 cm signal, however, may provide valuable information of X-ray emissions and serve as a prospective tool for studying 
the thermal and ionization evolution of the Universe and the first luminous objects \citep[see][for reviews]{Furlanetto+2006, Morales+2010, Pritchard+2012, McQuinn+2016}.

Several dedicated single-antenna instruments have been constructed or proposed to detect this signal, such as EDGES \citep{Monsalve+2017},  SARAS \citep{Singh+2018}, REACH \citep{Eloy+2019}, and so on. Notably, EDGES has reported the detection of an absorption feature centered at $z\sim 17$ \citep{Bowman+2018} with depth of $\sim - 500$ mK. The depth and shape of the absorption trough are unexpected in standard models \citep{Cohen+2017, Reis+2021}. The absorption can be interpreted by radio background excess \citep{Feng+2018, Fialkov+2019}, potentially sourced from the first luminous objects \citep{Ewall-Wice+2018,Mirocha+2019,Ziparo+2022, Zhang+2023}.
However, SARAS3 claimed non-detection of such a feature \citep{Singh+2022}.

In this paper, we first investigate the critical curves for DCBH formation in the presence of X-ray radiation using one-zone model. We then  estimate the DCBH formation rate and the 21 cm global spectrum at Cosmic Dawn, and try to find the connections between them.  
The layout of this paper is as follows: In Sec.~\ref{sec:one-zone}, we present the results of one-zone simulations. In Sec. \ref{sec:LF_and_n_DCBH}, we show the UV LF models and the DCBH formation rate. In Sec. \ref{sec:X-ray}, we investigate how the X-ray influences the DCBH formation rate. In Sec. \ref{sec:21-cm},  we investigate the connections between  21 cm absorption and the DCBH population abundance.  We finally give the discussion and summary in Sec. \ref{sec:discussion} and \ref{sec:summary}. Throughout this paper, we adopt the cosmological parameters from \citet{Planck2016}: $\Omega_{\rm m}=0.308, \Omega_{\rm \Lambda} = 0.692, \Omega_{\rm b}=0.048, h=0.678, \sigma_{8}=0.815$, and $n_{\rm s}=0.968$.

\section{The One-zone Model and critical curves for DCBH formation
}\label{sec:one-zone}

The one-zone model is commonly used to study the evolution of a system by considering it as a single, well-mixed zone, ignoring spatial variations within that zone. These models are particularly useful when trying to capture the overall behavior of a system. Many studies have employed this approach to tackle a series of problems, especially for the formation of DCBHs (e.g., \citealt{Omukai+2008, Wolcott-Green+2017, Yue+2017}) and Pop III stars (e.g., \citealt{Omukai+2000, Visbal+2014_star_formation}).

In this section, we use the one-zone model to derive the critical curve for DCBH formation and explore its dependence on X-ray radiation. The one-zone model adopted here is a modified version of \citet{Yue+2017}. We update some reaction rates according to \citet{Wolcott-Green+2017, Wolcott-Green+2021}, as listed in Appendix \ref{sec:appendix}. We also update the self-shielding factor according to \citet{Wolcott-Green+2017}.

As mentioned in Sec.~\ref{sec:intro}, X-ray radiation ionizes and heats the gas cloud, generating electrons, which facilitate the formation of H$_{2}$. This process may shift the location of the critical curve in $k_{\rm H_2}-k_{\rm H^-}$ plane. The X-ray radiation generally has a  
power-law form above a frequency $\nu_{\rm min}$,
\begin{equation}
J_{X}(\nu)=J_0\left(  \frac{\nu}{\nu_0}  \right)^{-\alpha_X},
\end{equation}
where $J_0$ is the specific intensity at the normalization frequency $\nu_0$, in units of $10^{-21} \,\rm erg\ s^{-1}\ cm^{-2}\ Hz^{-1}\ sr^{-1}$. We set $\nu_0=\nu_{\rm min}$ as the frequency of a photon with energy 1 keV. 
$\alpha_X$ is the spectral index and we adopt $\alpha_X \sim 1.5$, same to the fiducial model in \citet{Mesinger+2011}.

\begin{figure}
\centering{
\subfigure{\includegraphics[width=0.5\textwidth]{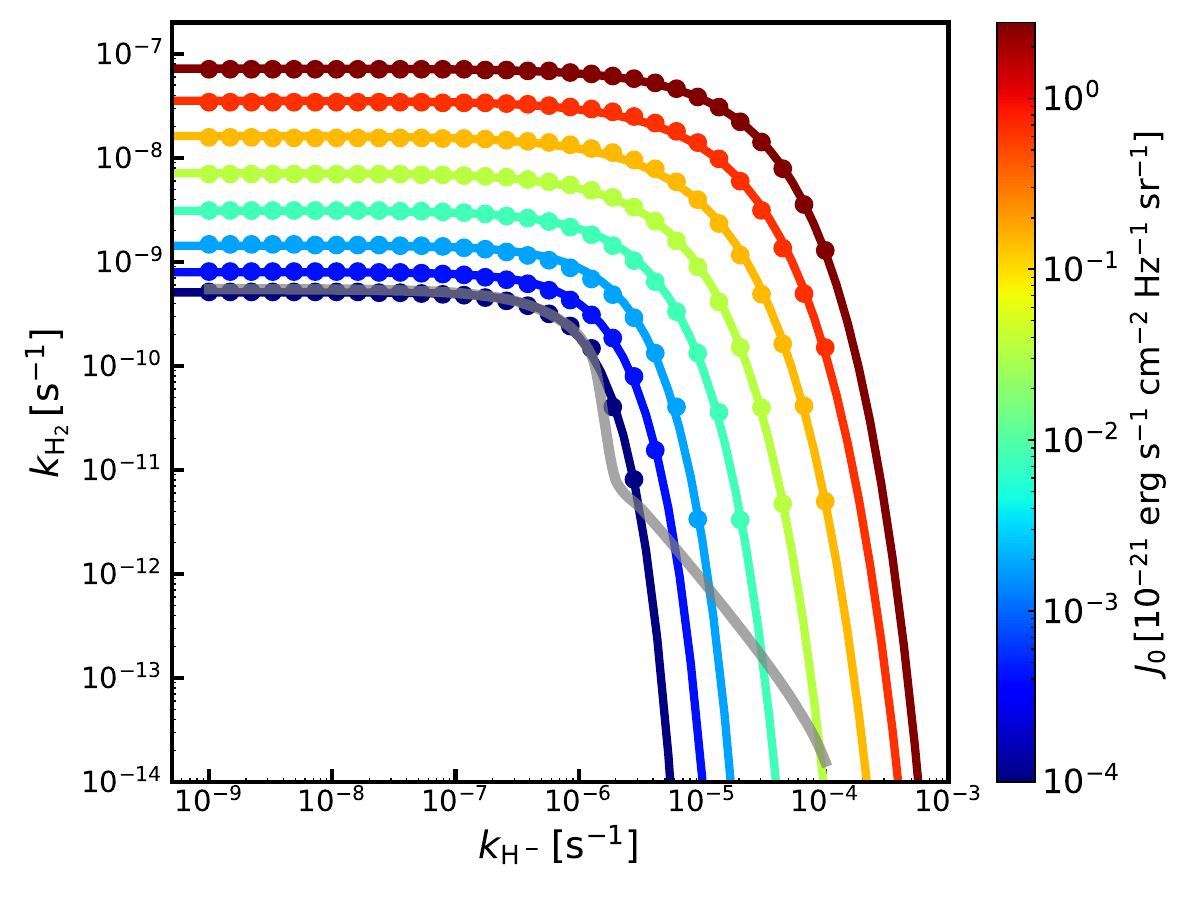}}
\caption{The critical curve as a function of X-ray intensity, with data points from our one-zone simulations and lines representing the fitting curves. The grey solid line represents the critical curve in the absence of X-ray radiation.
}
\label{fig:critical curve}
}
\end{figure}

In Fig. \ref{fig:critical curve} we show the critical curve in $k_{\rm H_2}-k_{\rm H^-}$ plane for an X-ray background with varying amplitude $J_0$. 
When X-ray radiation is negligible, the curve is nearly flat for $k_{\rm H^-}\lesssim 10^{-6}$ s$^{-1}$, which means that the H$_2$ dissociation is the dominant process for suppressing the H$_2$ formation, and the ability to suppress H$_2$ formation is basically proportional to the LW flux. For $k_{\rm H^-}\gtrsim 10^{-6}$ s$^{-1}$, the critical $k_{\rm H_2}$ drops dramatically, suggesting that the H$^-$ detachment plays the key role and lowers the requirement on $k_{\rm H_2}$.

As the X-ray intensity increases, the critical curve shifts roughly along the $y$-axis. It means that the X-ray radiation mainly boosts the H$_2$ formation.
Given $\alpha_X$, the critical curve for varying $J_0$ can be very well-fitted \footnote{{When X-ray radiation is very weak, i.e, $J_{0}\lesssim 2\times 10^{-4}$, the critical curves exhibit a turnover for $k_{\rm H^{-}}\gtrsim 2\times 10^{-6}\,\rm s^{-1}$ and $k_{\rm H_2}\lesssim 10^{-11}\,\rm s^{-1}$. In practice, this phenomenon can be negligible, as the SED shape is unlikely to fall within this region.}} by a function in log-log space 
\begin{equation}
\label{eq:critical_curve}
    \log (k_{\rm H_2}) = \frac{A}{1+e^{-B\log(k_{\rm H^-})}} + C
\end{equation}
both A, B, and C, are functions of $J_{0}$, 
\begin{equation}
\label{eq:parameter_func_A_B}
    Y = f_1(x) \exp (-k J_0) + [1-\exp(-k J_0)] f_2(x),
\end{equation}
where $Y\in[\log(-A), B]$, $x = \log(J_0)$, $f_1 (x) = a_0 x + a_1$ and $f_2(x) = a_2 x+a_3$. For $C$, the relationship is given by 
\begin{equation}
\label{eq:parameter_func_C}
    C = a_{0} + a_{1} \log(J_{0}) + a_{2}\log^{2} (J_{0}).
\end{equation}
Note that Eqs.(\ref{eq:parameter_func_A_B}, \ref{eq:parameter_func_C}) are only valid for $J_0 \in [10^{-4}, 3]$. We list the fitting parameters in Table~\ref{tab:fitting-parameters}.

\begin{table}
\caption{Fitting Parameters in Eqs. (\ref{eq:parameter_func_A_B}, \ref{eq:parameter_func_C})}
\label{tab:fitting-parameters}
\hspace{-2.5cm}
\renewcommand{\arraystretch}{1.5}
\begin{tabular}{llllllllllllllll}
\hline\hline
 & $a_0$ \qquad \qquad   & $a_1$ \qquad \qquad  & $a_2$ \qquad \qquad & $a_3$ \qquad \qquad & $k$ \\
\hline
A & -1.11  & 3.60 & -0.11 & 3.37 & 20.75 \\
B & -0.37 & 1.75 &  0.12 & 1.73 & 15.25 \\
C & -7.40 & 0.73 & 0.06 & - & - \\
\hline
\end{tabular}
\end{table}

\section{The UV luminosity function of the first galaxies and DCBH formation rate}\label{sec:LF_and_n_DCBH}

We derive the UV LF of the first galaxies from dark matter halo mass function $dn/dM_h$ \citep{Sheth+2002} via the $L_{\rm UV}-{\rm SFR}$ and ${\rm SFR}-M_h$ relations, where $L_{\rm UV}$ is the UV luminosity defined at 1600\AA, SFR is the star formation rate and $M_h$ is the dark matter halo mass.
Throughout this paper, we use the SED (luminosity per SFR)  
derived from a stellar population synthesis model in STARBURST99 \citep{Leitherer+1999,Vazquez+2005,Leitherer+2010} for continuous star formation mode, with a Salpeter initial stellar mass function in the range of $0.1-100\, M_{\odot}$, metallicity of $Z = 0.004$, and age of 100 $\rm Myr$, as the fiducial SED of a typical first galaxy.

The accretion rate of dark matter halos in the growth history is well-fitted by \citep{McBride2009MNRAS}\footnote{\citet{McBride2009MNRAS} provided the fittings of both the mean accretion rate and the median of accretion rate distribution.  Here we adopted the median, as we will add scatters on the UV and LW luminosities derived from the halo accretion rate.}
\begin{align}
\dot{M}_h(M_h,z)&=24.1\left(\frac{M_h}{10^{12}~M_\odot}\right)^{1.094} \times \nonumber \\
&(1+1.75z)\sqrt{\Omega_m(1+z)^3+\Omega_\Lambda}~~[M_\odot {\rm yr}^{-1}].   
\end{align}
For the first galaxies (galaxies with Population II stars), the star formation efficiency (SFE) is generally described by a double-power law formula \citep{Yang+2003}
\begin{equation}
f_{*,\rm II}(M_h)=\frac{2f_0}{\left( \frac{M_h}{M_p}  \right)^{-\gamma_{\rm lo}} +\left( \frac{M_h}{M_p} 
 \right)^{\gamma_{\rm hi}}  },
\end{equation}
where $f_0$, $\gamma_{\rm lo}$, $\gamma_{\rm hi}$ and $M_p$ are free parameters. Specifically, $f_0$ is the SFE for halos with mass $M_p$.

Generally, this SFE is considered to be independent of redshift and the observed LFs at $5\lesssim z \lesssim 10$ are well predicted. However, at $z\gtrsim 10$, JWST observed more galaxies than the expectation under the assumption of redshift-independent SFE. Therefore, we also consider the possibility that the star formation in halos $\lesssim M_p$ may have higher SFE at $z\gtrsim 10$. In both cases $\gamma_{\rm lo}$ is formally written as
\begin{equation}
\gamma_{\rm lo}(z)=\frac{\gamma_{\rm lo}^0}{a\left( \frac{1+z}{11}\right)^\eta+1}.
\end{equation}

Then the SFR of the first galaxies in halos with mass $M_h$ is
\begin{equation}
{\rm SFR}_{\rm II}(M_h,z)=f_{*,\rm II}(M_h,z)\frac{\Omega_b}{\Omega_m} \dot{M}_h.
\end{equation}
The center of their UV luminosity distribution  
\begin{equation}
\hat{L}_{\rm UV, II}={\rm SFR}_{\rm II}/\mathcal{K}_{\rm UV},
\label{eq:L_UV}
\end{equation}
where $\mathcal{K}_{\rm UV} = 1.17 \times 10^{-28}\,\rm  M_{\odot}\ yr^{-1}/(erg\ s^{-1}\ Hz^{-1})$ is derived from  our fiducial SED for the first galaxies.

We consider a Gaussian scatter in the actual UV luminosities around  $\log \hat{L}_{\rm UV,II}$, that is 
\begin{equation}
P_{\log 
 L_{\rm UV}}=\frac{1}{\sqrt{2\pi}\sigma_{\rm UV}}\exp\left(-\frac{(\log L_{\rm UV,II}-\log \hat{L}_{\rm UV,II})^2 }{2\sigma_{\rm UV}^2}    \right).
\end{equation}
Then the UV LF is  
\begin{align}
\frac{dn}{d\log L_{\rm UV,II}}=\int^{ \infty}_{\rm M_{4}}  dM_h f_{\rm dy,II}(M_h,z) \frac{dn}{dM_h} P_{\log 
 L_{\rm UV}},  
 \label{eq:UVLF}
\end{align}
where $f_{\rm d, II}$ is the duty cycle. According to \citet{O'Shea+2015}, 
\begin{equation}
f_{\rm d,II}=\left[1+(2^{\alpha/3}-1)\left(\frac{M_h}{M_c}\right)^{-\alpha}  \right]^{-3/\alpha},
\label{eq:f_duty}
\end{equation}
where $M_{\rm c} = 6\times 10^7~M_\odot$, $\alpha=1.5$.

The parameters $f_0$, $\gamma^0_{\rm lo}$, $\gamma_{\rm hi}$, $M_p$ and $\sigma_{\rm UV}$ are calibrated to match the observed UV LFs.
In this paper, we investigate three models: the SFE is independent of mass for $M_h \lesssim M_p$ ($\gamma_{\rm lo}^0=0$, model A); the SFE depends on halo mass but is independent of redshift ($a=0$, model B); the SFE depends on both halo mass and redshift ($a=0.5$ and $\eta=3$, model C). Their parameters are listed in Tab. \ref{tab:params}. Basically, when $z\gtrsim 10$, the SFE of model C is close to model A, otherwise it is close to model B.

\begin{figure}
\centering{
\subfigure{\includegraphics[width=0.5\textwidth]{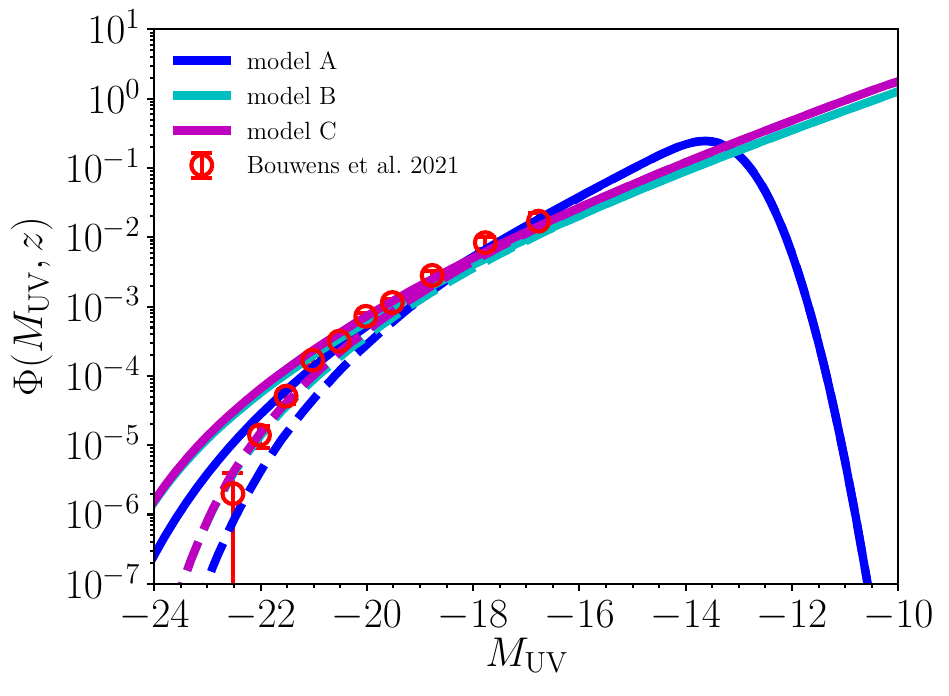}}
\subfigure{\includegraphics[width=0.5\textwidth]{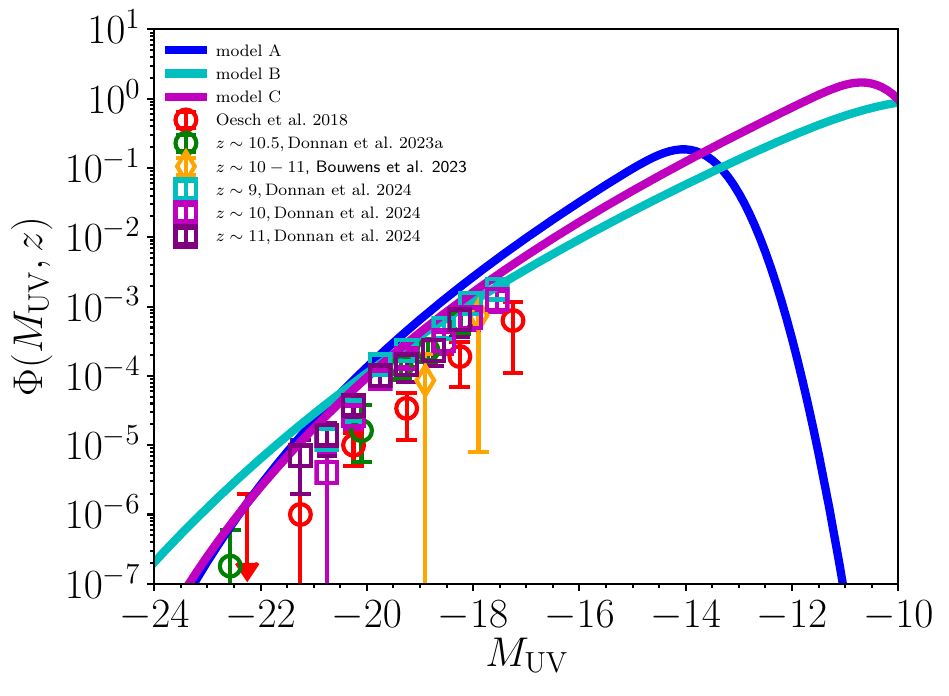}}
\subfigure{\includegraphics[width=0.5\textwidth]{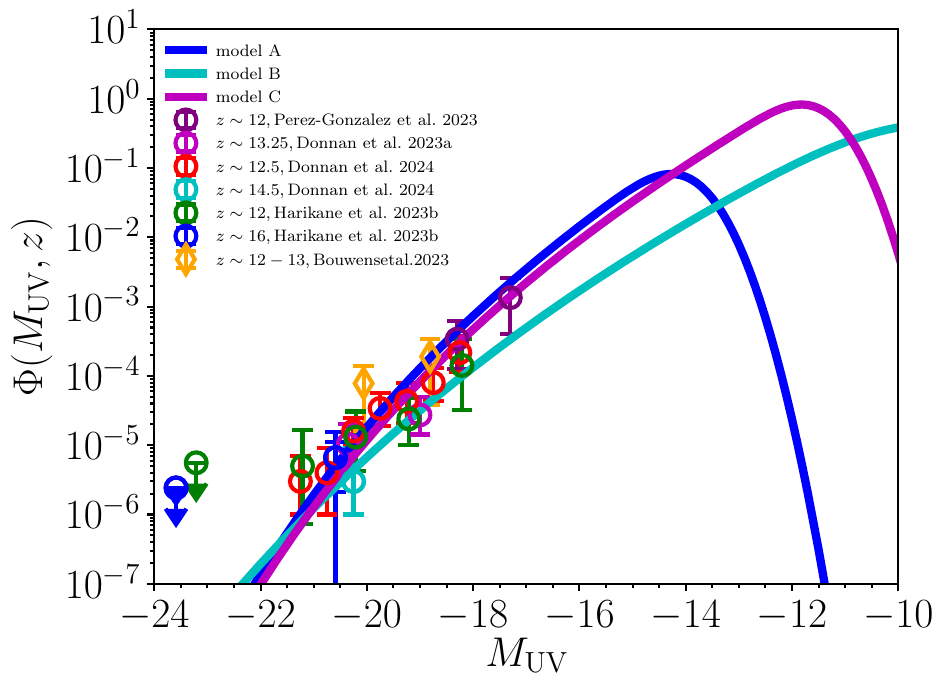}}
\caption{The UV LFs of our model A, B, and C at $z\sim6$ (top), $z\sim10$ (middle) and $z\sim13.5$ (bottom). For comparison, 
we also include observational data from HST and JWST at similar redshifts.
In the $z\sim6$ panel we also plot the UV LFs with dust attenuation by dashed lines.
At $z\gtrsim 10$, however, dust attenuation is negligible. Data points are from \citet{Bouwens2021AJ,Oesch+2018,Bouwens+2023b,Donnan+2023_UVLF,Harikane+2023_LF,Perez-Gonzalez2023ApJ,Donnan+2024}
}
\label{fig:UVLF}
}
\end{figure}

The derived UV LFs for models A, B and C are given in Fig. \ref{fig:UVLF}, compared with the latest observations of HST and JWST.
For comparison, we also present the UV LFs with dust attenuation in the $z\sim6$ panel. The dust attenuation model follows \citet{Williams2018} and \citet{Vogelsberger2019}. At $z\sim6$, model B and C are almost identical. However, model C starts to deviate from model B since $z\sim 10$.

The derivation of LW luminosity follows a similar approach to that of UV luminosity. The Gaussian distribution has the center 
\begin{align}
\hat{L}_{\rm LW,II}(M_h,z)&=s_{\rm LW,II}\times {\rm SFR}_{\rm II}(M_h,z) \nonumber \\
&=s_{\rm LW,II}\times \mathcal{K}_{\rm UV}\hat{L}_{\rm UV,II}(M_h,z) 
\end{align}
where $s_{\rm LW, II} = 7.54 \times 10^{27}\,\rm (erg~s^{-1} Hz)/(M_{\odot} yr^{-1})$ is the specific LW luminosity per unit SFR for the first galaxies, which we derive from the fiducial first galaxies SED used in this work. The scatter is assumed the same as for UV luminosity,  $\sigma_{\rm  LW}=\sigma_{\rm UV}$.

Therefore the emissivity of LW radiation from the first galaxies is
\begin{align}
\epsilon_{\rm LW,II}(z)&=  \int
d\log L_{\rm UV,II}\frac{dn}{d\log L_{\rm UV,II}}  L_{\rm LW,II}.
\end{align}

For Pop III stars, we simply assume that the SFE $f_{*,\rm III}$ and duty cycle $f_{\rm dy,III}$ are constant,
\begin{align}
L_{\rm LW,III}&=s_{\rm LW,III}\times {\rm SFR}_{\rm III} \nonumber \\
&=s_{\rm LW,III}\times f_{*,\rm III}\frac{\Omega_b}{\Omega_m}\dot{M}_h,
\end{align}
and
\begin{equation}
\epsilon_{\rm LW,\rm III}(z)=
 f_{\rm d,III}\int_{M_{\rm crit}(z)}^{M_4} 
dM_h \frac{dn}{dM_h} L_{\rm LW,III}(M_h,z), 
\end{equation}
where $s_{\rm LW, III} = 1.2 \times 10^{28}\,\rm (erg~ s^{-1} Hz)/(M_{\odot} yr^{-1}) $ is the specific LW luminosity per unit SFR for Pop III stars with a Salpeter IMF in the range of $1 - 500\,\rm M_{\odot}$ and age of $100\,\rm Myr$\citep{Schaerer+2003}\footnote{\url{https://cdsarc.cds.unistra.fr/ftp/VI/109/}}. 
$M_4$ is the halo mass corresponding to a virial temperature of $10^4$ K. $M_{\rm crit}(z)$ is the minimum mass of mini-halos that allows Pop III stars to form. We use the fitting formula from \citet{Kulkarni2021ApJ}  and 
self-consistently model its evolution in the presence of LW radiation from Pop III stars, the first galaxies and DCBHs.
All the parameters of the first galaxies and Pop III stars in our DCBH formation model are listed in Tab. \ref{tab:params}.

The SED of a DCBH depends on the column density of the surrounding medium. We adopt the SEDs produced by 1D radiative transfer simulations for $N_{\rm H}^{\rm BH}=1.3\times 10^{25}$ cm$^{-2}$ and $N_{\rm H}^{\rm BH}=5\times 10^{24}$ cm$^{-2}$ \citep{Pacucci+2015_spectral}. We assume the mass of DCBHs follows a log-normal distribution in the range $M_{\rm BH, min}=10^4 \, M_\odot$ and $M_{\rm BH, max}=10^7 \, M_\odot$, with a central value of $\log \hat{M}_{\rm BH}=10^6\, M_\odot$ and a scatter $\sigma_{\rm BH}=1$.
We normalize the DCBH luminosity to be the Eddington luminosity. However, in simulations, the DCBH can always sustain super-Eddington accretion \citep{Pacucci+2015_simulation}, so the real mass can be smaller.

The DCBH formation rate is given by
\begin{align}
    \frac{dn_{\rm DCBH}}{dz} &= p_{\rm DCBH}(z)\frac{dn_{\rm h}}{dz} \nonumber \\
    & \approx [p_{1}(z) + p_{2}(n_{\rm DCBH}, z)] \frac{dn_{\rm h}}{dz},
    \label{eq:dn_DCBH_dz}
\end{align}
where $n_{\rm h}(z)$ is the number density of the ACHs at redshift $z$.
The probability of DCBH formation, $p_{\rm DCBH}(z)$, consists of two parts.  The first part, 
$p_1$, represents the probability that an ACH receives LW flux above the critical curve when the radiation originates from the first galaxies. The second part, $p_2$,
represents the probability that an ACH receives LW flux
above the critical curve when the radiation comes from DCBHs.  

We ignore the effect of metal pollution in ACHs. According to the fiducial simulation in \citet{Hicks2024}, $\sim 16\%$ of halos with $10^6~M_\odot < M_h < 10^9~M_\odot$ will be chemically enriched by $z\sim12$. The DCBH formation rate should be reduced by a similar fraction. Nevertheless, we believe that incorporating metal pollution will not change our conclusion. For simplicity, we ignore this effect.
 
We then calculate the $p_1$ and $p_2$ using the Monte Carlo simulations in \citet{Dijkstra+2008,Dijkstra+2014}, incorporating the LW luminosities and critical curves provided above. We combine the one-zone simulations, Monte Carlo simulations, and Eq. (\ref{eq:dn_DCBH_dz}) to find the full evolution of DCBH abundance.
 
\begin{figure*}
\centering{
\subfigure{\includegraphics[width=0.45\textwidth]{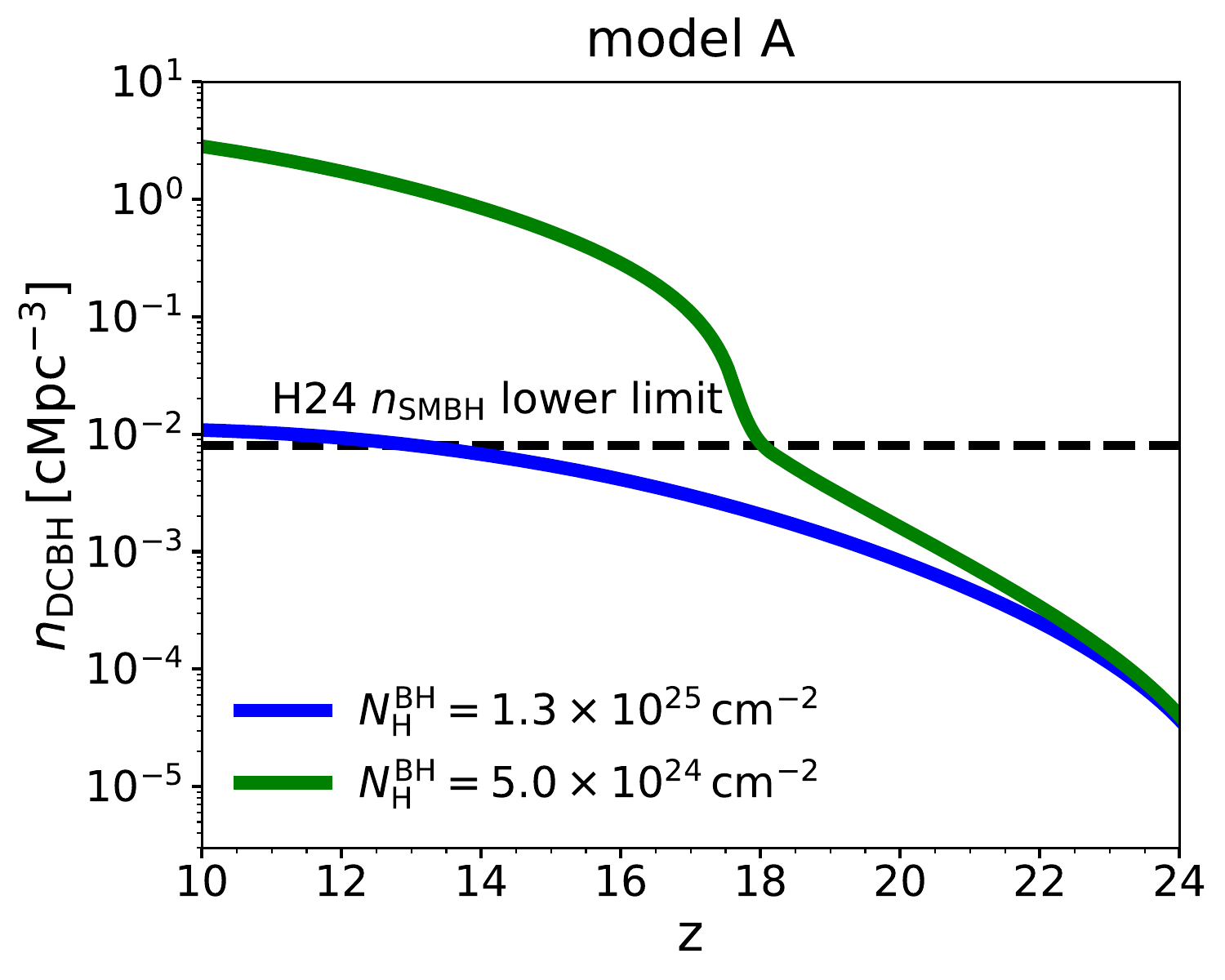}}
\subfigure{\includegraphics[width=0.45\textwidth]{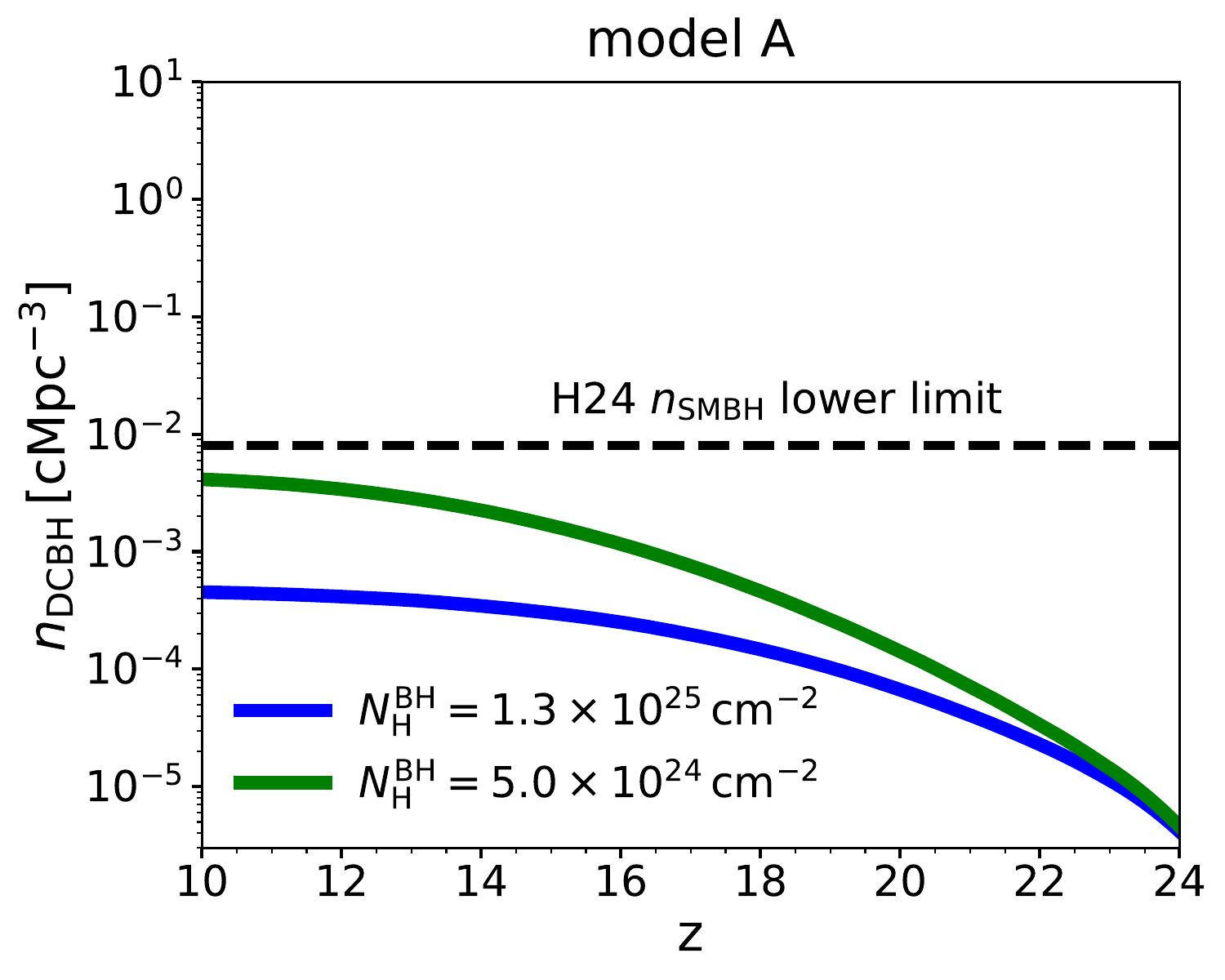}}

\subfigure{\includegraphics[width=0.45\textwidth]{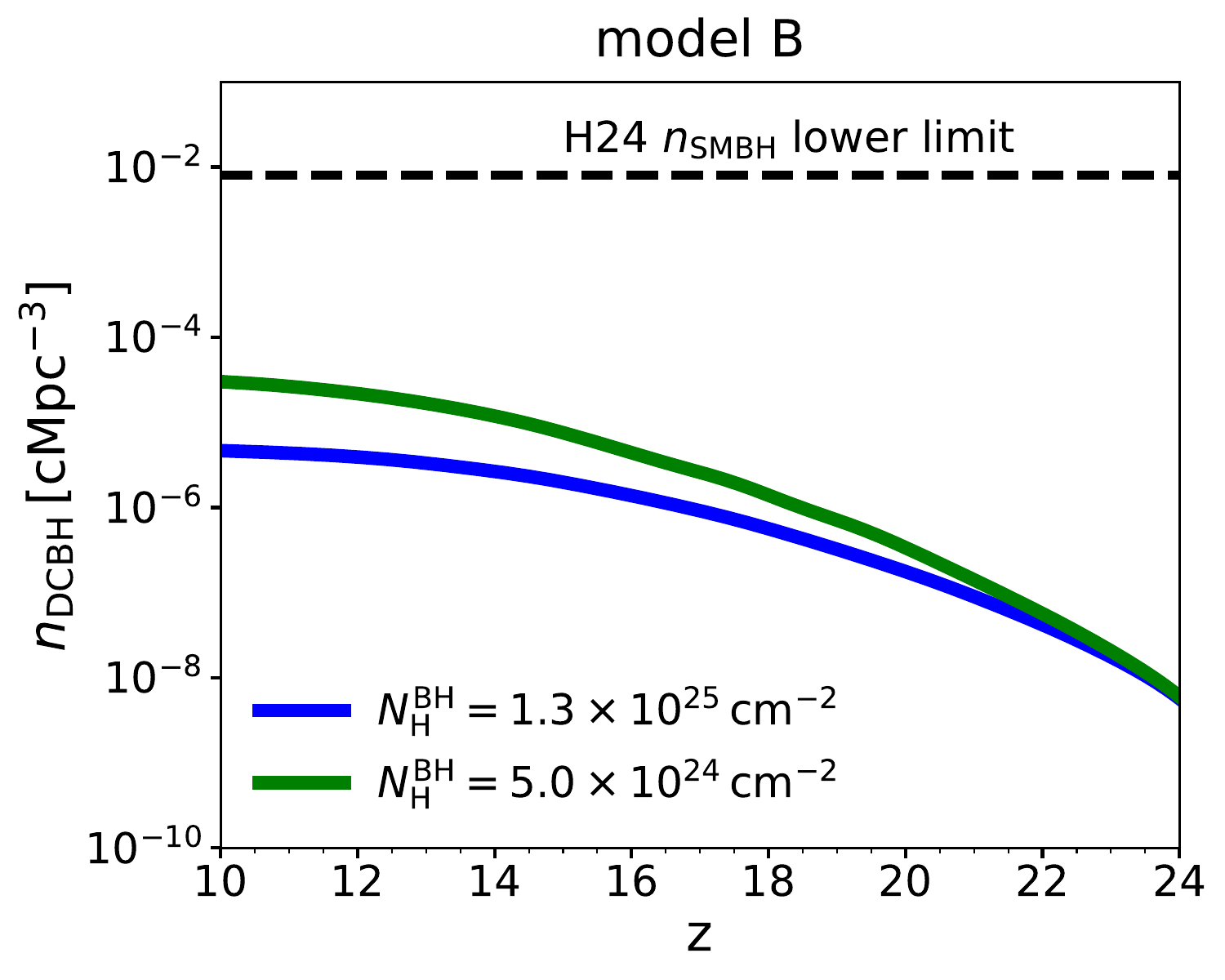}}
\subfigure{\includegraphics[width=0.45\textwidth]{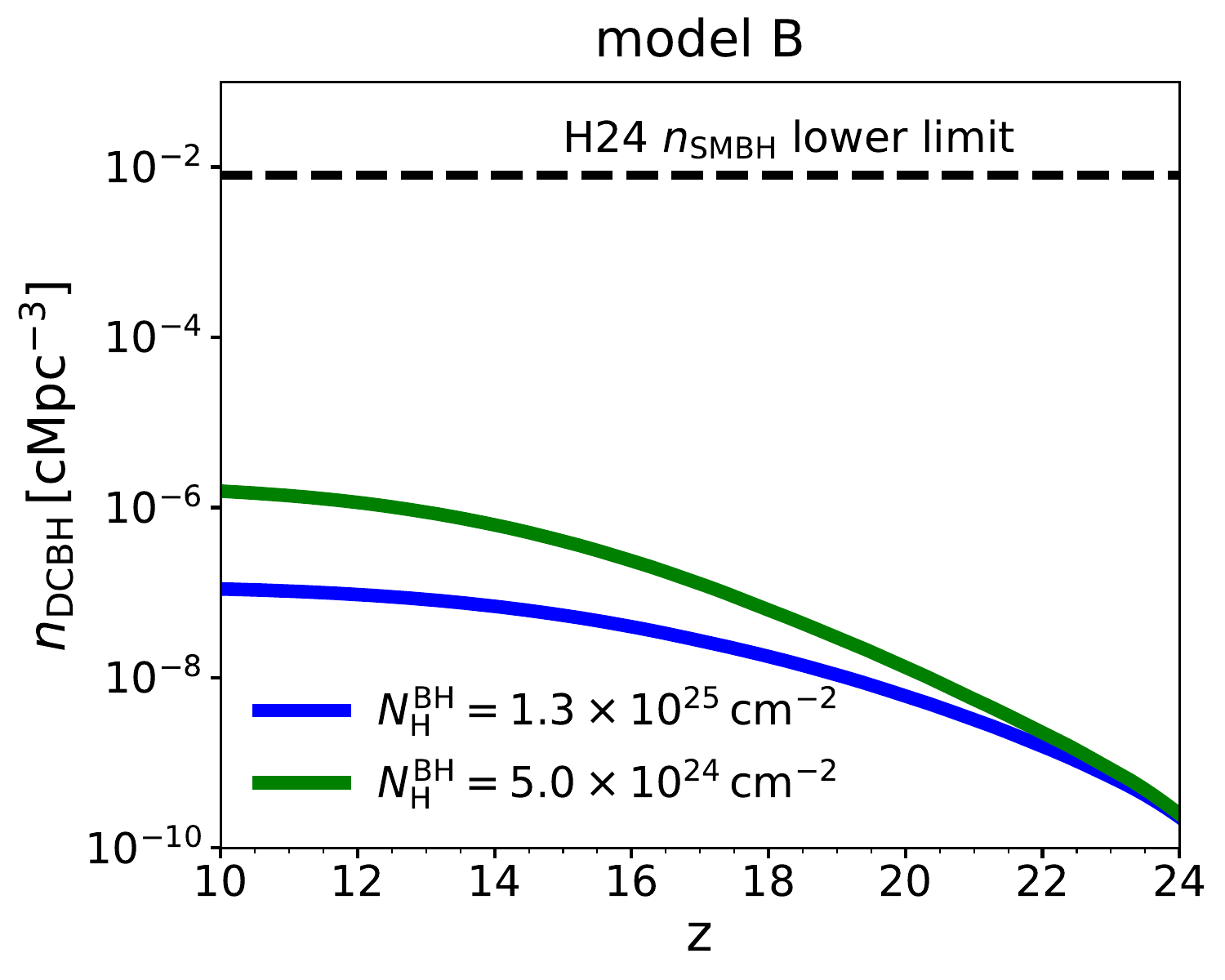}}

\subfigure{\includegraphics[width=0.45\textwidth]{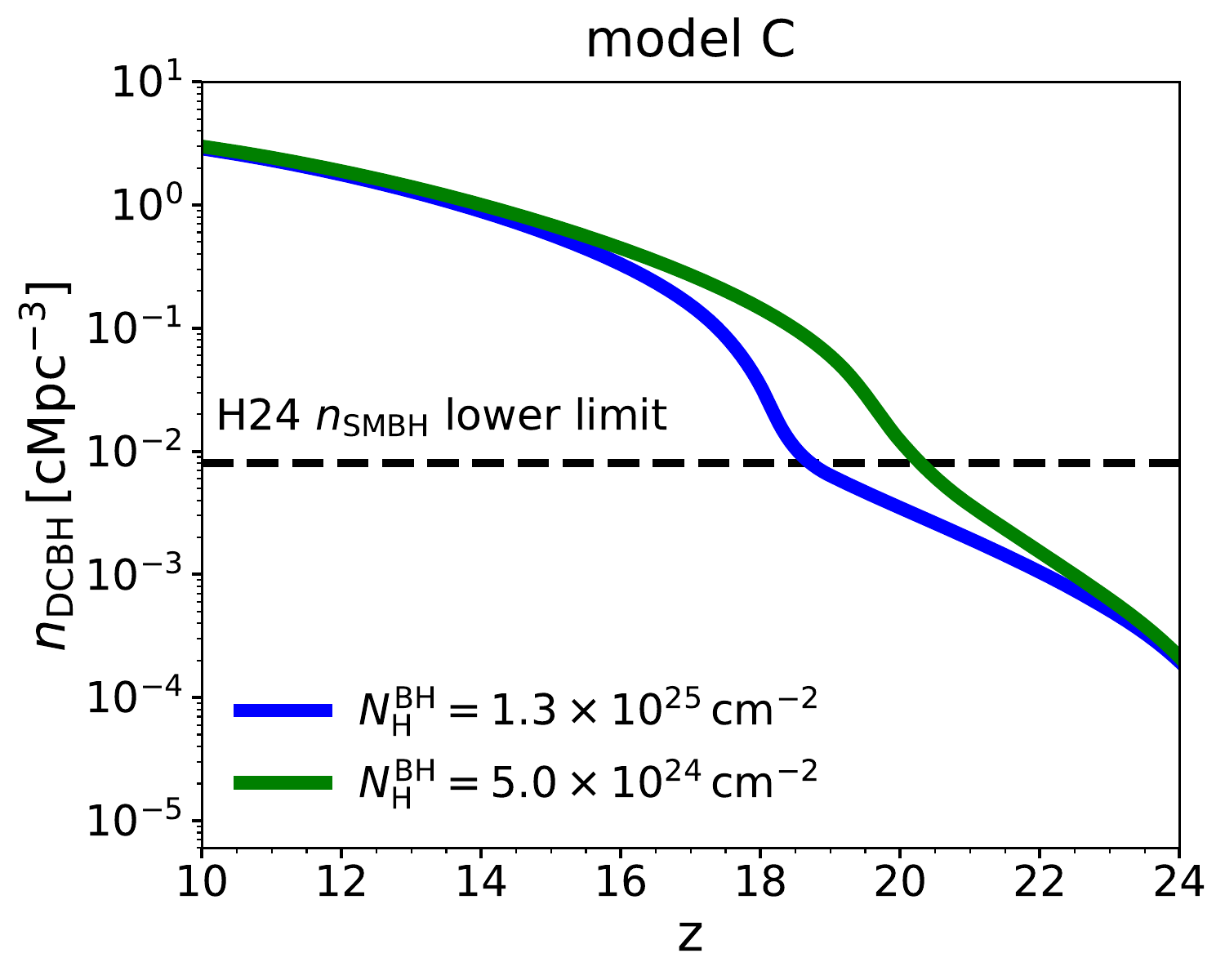}}
\subfigure{\includegraphics[width=0.45\textwidth]{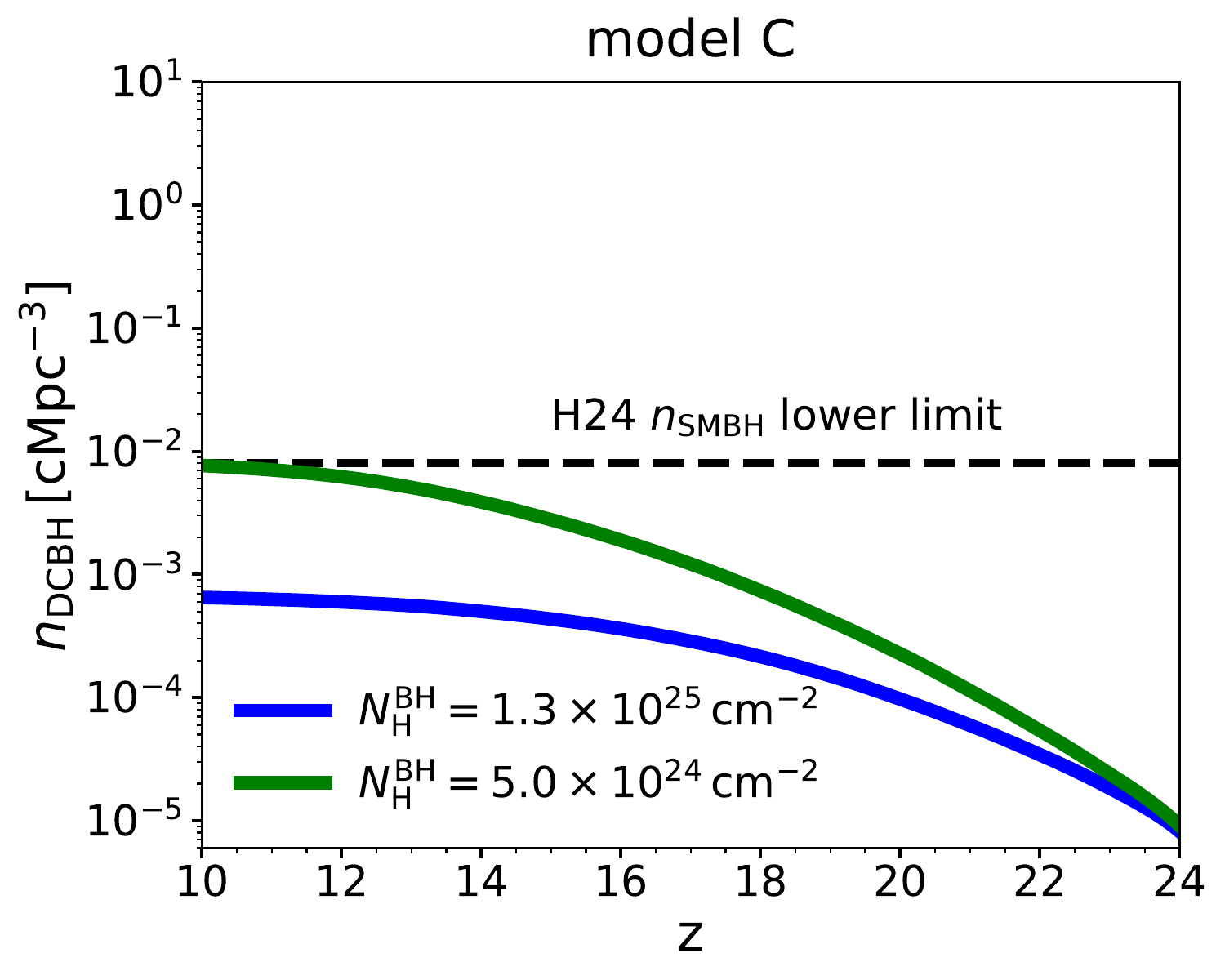}}
\caption{The $n_{\rm DCBH}$ as a function of redshift $z$ without X-ray (left column) and with X-ray (right column) for model A, model B and model C, respectively. The black dashed lines represent the observational constraints on $n_{\rm SMBH}$ in H24.
}
\label{fig:n_DCBH}
}
\end{figure*}

We show the evolution of the $n_{\rm DCBH}$ in the absence of X-rays for three different first galaxies models in the first row of Fig.~\ref{fig:n_DCBH}. The SFE of small halos ($T_{\rm vir}\sim 10^4$ K) is most important, because they are far more numerous than massive halos and play a major role in providing the H$_2$-dissociation/H$^-$-detachment radiation, even though they are less clustered.

In model A, $f_{*,\rm II}$ is almost constant for halos with $M_h \lesssim M_p$, small halos have star formation activity as efficient as halos with peak efficiency ($M_h\sim M_p$), allowing DCBHs to form rapidly. In model B however, $f_{\rm *,II}\propto M_h^{0.46}$ when $M_h\lesssim M_p$, small halos have SFE much smaller than the massive ones, therefore the DCBH formation rate is much smaller. Model C has the same SFE as model B at $z\lesssim 10$, but it increases dramatically at $z\gtrsim 10$, as inspired by JWST observations. The DCBH formation rate is largely boosted compared with model B.

In each first galaxies model, we present the results for DCBH SEDs with $N_{\rm H}^{\rm BH} = 1.3 \times 10^{25},  5.0\times 10^{24} \,\rm cm^{-2}$.  
The results are quite different, implying that the DCBH also plays a significant role in triggering the formation of new DCBH, and different DCBH SEDs have different abilities.

The radiation from DCBHs can promote the rapid growth of  $n_{\rm DCBH}$. Since this is a positive feedback process, once $n_{\rm DCBH}$ reach a certain threshold $\mathcal{O}(10^{-3}-10^{-2})$\,\rm  cMpc$^{-3}$, the DCBH formation is a runaway process and finally $n_{\rm DCBH} $ reaches an upper limit that all newly-formed pristine ACHs are occupied by DCBHs \citep{Yue+2014,Yue+2017}. It sounds unrealistic, however, we note that the observed lower limit from SMBHs is $\mathcal{O}(10^{-2})$\,\rm  cMpc$^{-2}$ in H24. 
So we can not rule out the possibility that such a mechanism have a chance to work in the Universe.

However, in Sec. \ref{sec:one-zone} we have learned that X-ray can efficiently suppress the formation of DCBH. In addition to Pop III stars and the first galaxies, DCBHs with lower $N_{\rm H}^{\rm BH}$ can produce X-ray radiation and suppress the formation of new DCBHs as well.  We will investigate the evolution of $n_{\rm DCBH}(z)$ when X-rays from Pop III stars, the first galaxies and DCBHs are taken into account in the next section.  

\begin{table}
\caption{Parameters of model A, B and C. 
Units for $M_p$: $1M_\odot$; for $\mathcal{K}_{\rm UV}$: $10^{-28}(M_\odot {\rm yr}^{-1})/({\rm erg~s^{-1} Hz^{-1} })$; for $s_{\rm LW,II}$ and $s_{\rm LW, III}$: 
$10^{27}({\rm erg~s^{-1} Hz^{-1}
})/(M_\odot {\rm yr}^{-1})
$.
}
\hspace{-3cm}
\begin{tabular}
{l|llllllllllllllllll}

 & model A & model B & model  C \\
\hline
$f_0$ & 0.14 & 0.14 & 0.14 \\
$\gamma_{\rm lo}$ & 0 & 0.46 &  $\frac{0.46}{0.5[(1+z)/11]^3+1}$ \\
$\log M_p$ & 12.3 & 12.3 & 12.3 \\
$\gamma_{\rm hi}$ & 0.82 & 0.82 & 0.82 \\
$f_{\rm d, II}$ & 0.1 & Eq. (\ref{eq:f_duty}) & Eq. (\ref{eq:f_duty}) \\
$\sigma_{\rm UV}$ & 0.2 & 0.5  &0.2  \\
$f_{*,\rm  III}$ & 0.005  &  0.005 &  0.005 \\
$f_{\rm d, III}$ &  0.1 &  0.1 & 0.1 \\
$f_{X,\rm II}$ & 1 & 1 & 1 \\
$f_{X,\rm III}$ & 1 & 1& 1 \\
$\mathcal{K}_{\rm UV}$ & 1.17 & 1.17 & 1.17 \\
$s_{\rm LW,  II}$ & 7.54 & 7.54 & 7.54 \\
$s_{\rm LW, III}$ & 12 & 12 & 12 \\
\hline
\end{tabular}
\label{tab:params}
\end{table}

\section{The X-rays}\label{sec:X-ray}

The X-rays influencing DCBH formation primarily originate from the cosmic X-ray background (CXB) and nearby X-ray sources. The CXB provides a diffuse and isotropic X-ray radiation that permeates the universe, while localized sources can produce intense X-ray emissions. In this section, we will describe how to incorporate the effects of X-rays from the CXB and nearby X-ray sources into our model for DCBH formation.

The realistic X-ray background comprises contributions from Pop III stars, the first galaxies, and previously formed DCBHs,
\begin{equation}
    J_{\rm X}(\nu, z) = \frac{(1+z)^{3}}{4\pi} \int_{z}^{\infty} \epsilon_{\rm X}(\nu^{\prime}, z^{\prime}) \frac{cdz^{\prime}}{H(z^{\prime})(1+z^{'})},
\end{equation}
where $\nu^{\prime} = \nu (1+z^{\prime})/(1+z)$, $H(z)$ is the Hubble parameter. The emissivity $\epsilon_{\rm X}(\nu, z)$ is 
\begin{equation}
\epsilon_X(\nu,z)=\epsilon_{X,\rm III}(\nu,z)+\epsilon_{X,\rm II}(\nu,z)+\epsilon_{X,\rm BH}(\nu,z).
\end{equation}

For both Pop III stars and the first galaxies, the X-ray emission is actually from the BH remnants of the short-lived massive stars. So the X-ray luminosity is tightly related to the SFR \citep{Grimm+2003, Ranalli+2003, Gilfanov+2004, Mineo+2012}, we adopt the $L_{\rm X}-\rm SFR$ relation from \citet{Grimm+2003},
\begin{equation}
   L_{\rm 2-10keV} = 6.9 \times 10^{40} f_{\rm X} \left(\frac{\rm SFR}{\rm M_{\odot}\ yr^{-1} }\right)\,\rm erg\ s^{-1},
\end{equation}

Where $f_{X}$ is the normalization factor. This factor can be larger than 1, as \citet{Abdurashidova+2022} suggest that galaxies discovered at higher redshifts ($z = 8$ and $10$) exhibit a more efficient X-ray production compared to local ones. We further assume that the X-ray spectrum follows a power-law form $\propto \nu ^{-\alpha_X}$ with $\alpha_{\rm X} = 1.5$, same as the fiducial model in \citet{Mesinger+2011}.

The contribution of Pop III stars to the emissivity is  
\begin{equation}
\epsilon_{X,\rm III}(\nu,z)=
 f_{\rm d,III}\int_{M_{\rm crit}(z)}^{M_4} 
dM_h \frac{dn}{dM_h} L_{X,\rm III}(\nu,M_h,z), 
\label{eq:epsilon_X_III}
\end{equation}

For the first galaxies,
\begin{align}
&\epsilon_{X,\rm II}(\nu)=f_{\rm esc, X}(\nu) \int d\log L_{\rm UV,II}
\frac{dn}{d\log L_{\rm UV,II}}
L_{X,\rm II}(\nu, {\rm SFR}_{\rm II} ),
\label{eq:epsilon_X_II}
\end{align}
where $dn/d\log L_{\rm UV,II}$ is the UV LF of the first galaxies calculated from Eq. (\ref{eq:UVLF}), and 
\begin{equation}
{\rm SFR}_{\rm II}=\mathcal{K}_{\rm UV}L_{\rm UV,II}.
\end{equation}

The X-ray escape fraction is given by
\begin{equation}
    f_{\rm esc, X}(\nu) = \exp \left[-(\sigma_{\rm X, 1}(\nu)N_{\rm H}^{\rm gal} + \sigma_{\rm X, 2}(\nu)ZN_{\rm H}^{\rm gal})\right], 
\end{equation}
where $N_{\rm H}^{\rm gal} = 10^{21}\,\rm cm^{-2}$ is the typical hydrogen column density obtained in numerical simulations of the first galaxies \citep{Das+2017}. $\sigma_{\rm X, 1}$  is the synthesis photoelectric cross-section of neutral hydrogen and helium, while $\sigma_{\rm X,2}$ accounts for 17 metal elements with solar abundance \citep{Balucinska-Church+1992}.

For the DCBHs, we use SED templates for various column densities $N_{\rm H}^{\rm BH}$ \citep{Pacucci+2014}. Generally, when $N_{\rm H}^{\rm BH}\gtrsim 10^{24}\,\rm cm^{-2}$, the escaped X-ray photons from DCBHs are too hard to heat the IGM gas and to influence subsequent DCBH formation. The total DCBH luminosity is normalized to be the Eddington luminosity.

The X-ray emissivity from DCBHs is given by

\begin{equation}
\epsilon_{X,\rm BH}(\nu,z)=n_{\rm BH}(z)
\int d\log M_{\rm BH}  P_{\log M_{\rm BH}} L_{X,\rm BH}(M_{\rm BH}),
\label{eq:epsilon_X_BH}
\end{equation}
where
\begin{equation}
P_{\log M_{\rm BH}}=\frac{1}{\sqrt{2\pi}\sigma_{\rm BH}}\exp\left(-\frac{(\log M_{\rm BH}-\log \hat{M}_{\rm BH})^2}{2\sigma^2_{\rm BH}}\right).
\end{equation}

The formation of DCBH is triggered by galaxy radiation from very nearby sources, the X-rays of these sources are generally much stronger than the background
and works efficiently in promoting H$_2$ cooling \citep{Inayoshi+2015}.
Therefore, we need to find the DCBH formation criterion for the full galaxy SED, including the LW, H$^-$-detachment, and X-ray radiation. Unlike the X-ray background, such nearby X-rays are proportional to the LW or H$^-$-detachment radiation from the same sources. When the X-rays of the nearby sources are involved, we obtain the new critical galaxy radiation by the following steps: We first find the critical galaxy radiation in the absence of the nearby X-rays,  then take into account of the nearby X-rays corresponding to this critical galaxy radiation. By increasing the strength of the galaxy radiation gradually and running the one-zone simulation for each strength, we find the new critical galaxy radiation when the X-rays from the same sources are involved. If the X-rays are too strong, the new critical galaxy radiation may do not exist, meaning that such galaxy SED cannot trigger the formation of DCBH. For the DCBH, we use the same method to obtain the critical radiation.

In the second row of Fig.~\ref{fig:n_DCBH}, we show the evolution of DCBH number density in the presence of X-ray radiation for $f_{X,\rm II}=f_{X,\rm III}=
f_{\rm X} \sim 1.3$. The DCBH abundance is significantly reduced and the number density is limited to  $\mathcal{O}(10^{-3}-10^{-2})$ cMpc$^{-3}$ in model A and model C, and matches the observational level by \citet{Nabizadeh+2024} and H24. We also investigate other $f_X$ values, and find that when $f_X\lesssim 0.1$, the influence on DCBH formation is negligible; when $f_X \gtrsim 5$, the DCBH formation is fully suppressed.

The two DCBH SEDs adopted in this paper are all Compton-thick, the X-rays from nearby DCBHs have just negligible influence on the DCBH formation even when $N_{\rm H}^{\rm BH}=5\times 10^{24}~$cm$^{-2}$. Therefore the DCBH number density is still higher in this case. However, if further decrease the $N_{\rm H}^{\rm BH}$ and finally the SED becomes Compton-thin, then X-rays from nearby DCBHs would fully suppress the DCBH formation.

We note that in both model A and model C, the $n_{\rm DCBH}$ exceeds the observed upper limit set by \citet{Nabizadeh+2024}. However, this should be the upper limit of directly observable DCBHs, rather than the cumulative number of DCBHs ever-formed, as they can evolve into SMBHs shortly after formation. Models with a number density exceeding this upper limit remain feasible.

In addition to reducing the formation rate of DCBHs, X-ray radiation can also heat the IGM and reduce the absorption of the 21 cm global signal. So there may be some correlations between the DCBH formation rate and the 21 cm global signal. We will investigate this in the next section.

\section{
The 21 cm global spectrum
}\label{sec:21-cm}

At Cosmic Dawn, X-ray heating is crucial for the 21 cm signal as it determines the IGM kinetic temperature $T_k$ and then the spin temperature $T_s$. X-ray also ionizes the Hydrogen and Helium but the ionizing effect on 21 cm signal is minor at this stage \citep[see][and reference therein]{Furlanetto+2006, Pritchard+2012, Mesinger+2019}. The evolution of $T_k$ and ionization fractions follow (we ignore the HeIII):  
\begin{align}
    \frac{dT_{\rm k}}{dt}  = & -2 H(z) T_{\rm k} -  \frac{1}{n} \left[n_{\rm H} \frac{dx_{\rm HII}}{dt} + n_{\rm He} \frac{dx_{\rm HeII}}{dt}\right] T_{\rm k} \nonumber \\
    &- \frac{2}{3k_{\rm B}n} \Lambda(T_{\rm k}) + \frac{2}{3k_{\rm B}n} f_{\rm heat} \rho_{\rm X} \nonumber \\
  \frac{dx_{\rm HII}}{dt} = &\left[\gamma_{\rm H}(T_{\rm k})(1-x_{\rm HII})-\alpha_{\rm H}(T_k)x_{\rm HII} \right] n_{\rm e}\nonumber \\ 
    &+ f_{\rm ion, H} \frac{\rho_{\rm X}}{n_{\rm H}E_{\rm H}} \nonumber \\
\frac{dx_{\rm HeII}}{dt} = &\left[\gamma_{\rm He}(T_{\rm k})(1-x_{\rm HeII})-\alpha_{\rm He}(T_k)x_{\rm HeII} \right] n_{\rm e} \nonumber \\
    &+ f_{\rm ion, He} \frac{\rho_{\rm X}}{n_{\rm He}E_{\rm He}},
    \label{eq:T_K}
\end{align}
where $n_{\rm H}$, $n_{\rm He}$  are physical cosmic densities of Hydrogen and Helium and respectively. $n_e=n_{\rm H}x_{\rm  HII}+n_{\rm He}x_{\rm HeII}$ is density of electrons, and $n=n_e+n_{\rm H}+n_{\rm He}$ is the total density of all particles. $k_{\rm B}$ is the Boltzmann constant. $\Lambda(T)$ is the total cooling function \citep{Maselli+2003} including the collisional ionization cooling, collisional excitation, Bremsstrahlung cooling, and Compton heating/cooling. $f_{\rm heat}$, $f_{\rm ion, H}$ and $f_{\rm ion, He}$ are fractions of X-ray energy deposited into heating, hydrogen ionization and helium ionization respectively \citep{Valdes+2008}.  $\gamma_{\rm H, He}(T_k)$ and $\alpha_{\rm H, He}(T_k)$ are rates of collision ionization and recombination for hydrogen and helium, respectively \citep{Maselli+2003}. 
$E_{\rm H} = 13.6\,\rm eV$, $E_{\rm He} = 24.6\,\rm eV$ are ionization energy for hydrogen and helium.  

$\rho_{\rm X}$  is the physical energy density of X-ray photons deposited into the IGM, given by
\begin{equation}
    \rho_{\rm X} = (1+z)^{3} \int_{\nu_{\rm min}}^{\infty} \epsilon_{\rm X}(\nu) [1-e^{-\tau_{\rm IGM}(\nu)}]d\nu,
\label{eq:rho_X}    
\end{equation}
here $\tau_{\rm IGM}(\nu)$ is the IGM optical depth for X-ray emission, 
\begin{align}
    \tau_{\rm IGM}(\nu) = &\left[\sigma_{\rm H}(E)n_{\rm H}(1-x_{\rm HII}) + \sigma_{\rm He}(E) n_{\rm He} (1-x_{\rm HeII})  \right] \nonumber \\
    & \times (c/H(z))
\end{align}
$\sigma_{\rm H}(E)$ and $\sigma_{\rm He}(E)$ are the photon-ionization cross sections for hydrogen \citep{Verner+1996} and helium \citep{Yan+1998}.

The differential brightness temperature of the 21 cm global signal is given by
\begin{align}
    \delta T_{\rm b} & \approx 27 x_{\rm HI} \left(\frac{\Omega_{\rm b}h^{2}}{0.023}\right) \left(\frac{0.15}{\Omega_{\rm m}h^{2}} \frac{1+z}{10}\right)^{1/2} \nonumber \\
    & \times \left(\frac{T_{\rm S}-T_{\rm R}}{T_{\rm S}}\right)\,\rm mk, 
    \label{eq:delta T_b}
\end{align}
where the temperature of radio radiation $T_{\rm R} = T_{\rm CMB}$. The spin temperature is 
\begin{equation}
    T_{\rm S}^{-1} = \frac{T_{\rm R}^{-1}+x_{\rm \alpha}T_{\rm K}^{-1}+x_{\rm c}T_{\rm K}^{-1}}{1+x_{\rm \alpha} + x_{\rm c}},
\end{equation}
where $x_{\rm c}$ is the collisional coupling coefficient \citep{Pritchard+2012} and $x_\alpha$ is the Ly$\alpha$ coupling coefficient that is proportional to the specific number intensity of Ly$\alpha$ background \citet{Hirata+2006}.

The Ly$\alpha$ background is directly from the redshifted Lyman band (between Ly$\alpha$ frequency $\nu_\alpha$ and Lyman limit frequency $\nu_{\rm LL}$) emission of Pop III stars, the first galaxies, and DCBHs, or produced indirectly by X-ray photons, say
\begin{equation}
    J_{\rm \alpha} (z) = J_{\alpha, \rm Ly}(z) + J_{\alpha, \rm X}(z),
\end{equation}
where   
\begin{equation}
    J_{\alpha, \rm Ly}(z) =\frac{1}{h_p \nu_\alpha}    
    \frac{c(1+z)^{3}}{4\pi} \int_{z}^{z_{\rm max}}  \frac{\epsilon_{\alpha}(\nu^{\prime}, z^{\prime})}{H(z^{\prime})(1+z^{\prime})}dz^{\prime},
\end{equation}
here $\nu^{\prime} = \nu_{\rm \alpha} (1+z^{\prime})/(1+z)$,  $z_{\rm max} = (\nu_{\rm LL}/\nu_{\alpha})(1+z)$, $h_p$ is the Planck constant. 

The comoving emissivity is  
\begin{equation}
    \epsilon_{\alpha}(\nu^{\prime}, z^{\prime}) = \epsilon_{\alpha, \rm III}(\nu^{\prime}, z^{\prime}) + \epsilon_{\alpha, \rm II}(\nu^{\prime}, z^{\prime}) + \epsilon_{\alpha, \rm BH}(\nu^{\prime}, z^{\prime})
\end{equation}
where $\epsilon_{\alpha, \rm III}(\nu^{\prime}, z^{\prime})$, $\epsilon_{\alpha, \rm II}(\nu^{\prime}, z^{\prime})$, $\epsilon_{\alpha, \rm BH}(\nu^{\prime}, z^{\prime})$ are calculated analogous to Eqs. (\ref{eq:epsilon_X_III}, \ref{eq:epsilon_X_II}, \ref{eq:epsilon_X_BH}), with X-ray luminosity replaced by luminosity $\ge \nu_\alpha$.

In addition, a fraction $f_{\rm \alpha}$ of the X-ray energy is deposited into Ly$\alpha$ photons \citep{Valdes+2008}, thereby contributing to the Ly$\alpha$ background,
\begin{equation}
    J_{\rm \alpha, X}(z) = \frac{c}{4\pi} \frac{f_{\rm \alpha} \rho_{\rm X}}{h_{\rm P}\nu_{\rm \alpha}} \frac{1}{H(z)\nu_{\rm \alpha}},
\end{equation}
where $\rho_X$ is given by Eq. (\ref{eq:rho_X}).

\begin{figure*}
\centering{
\subfigure{\includegraphics[width=0.45\textwidth]{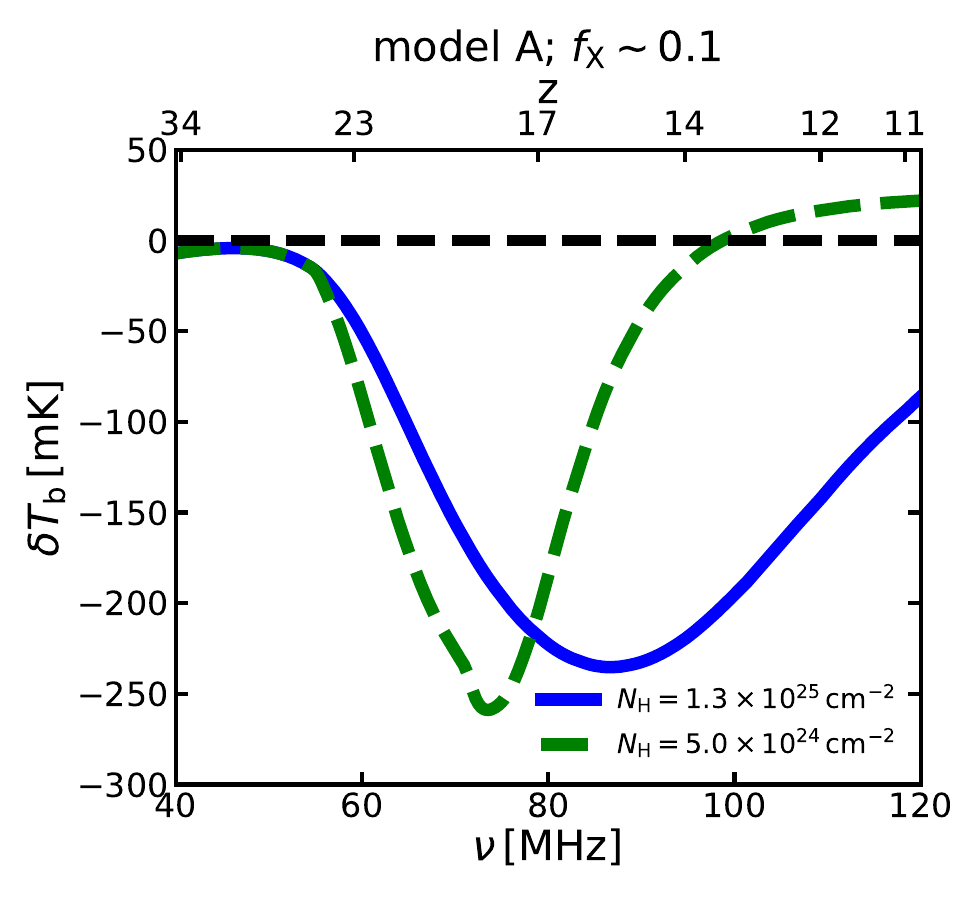}}
\subfigure{\includegraphics[width=0.45\textwidth]{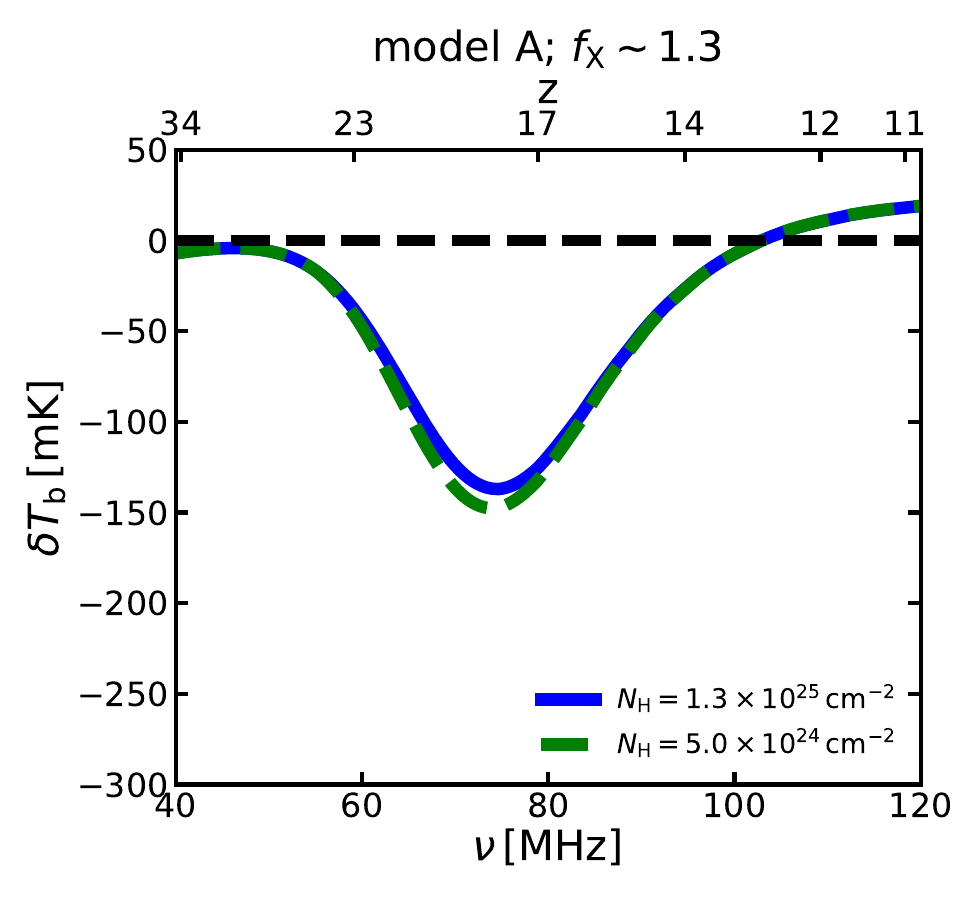}}

\subfigure{\includegraphics[width=0.45\textwidth]{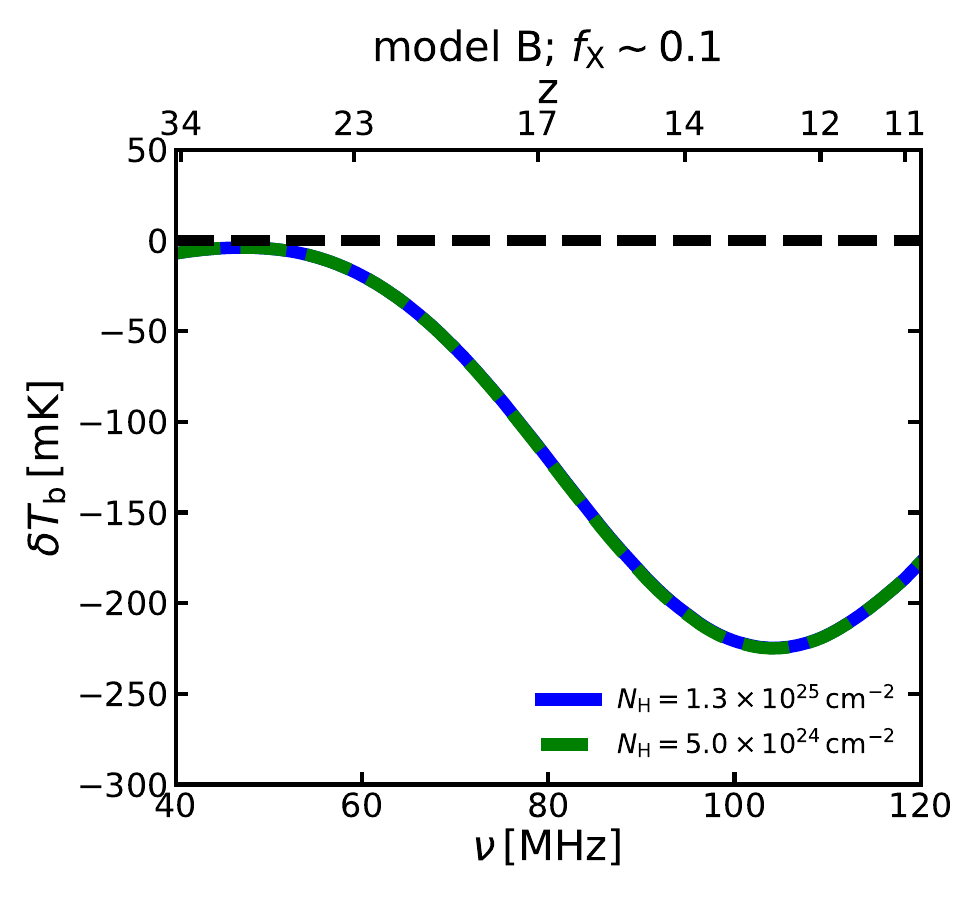}}
\subfigure{\includegraphics[width=0.45\textwidth]{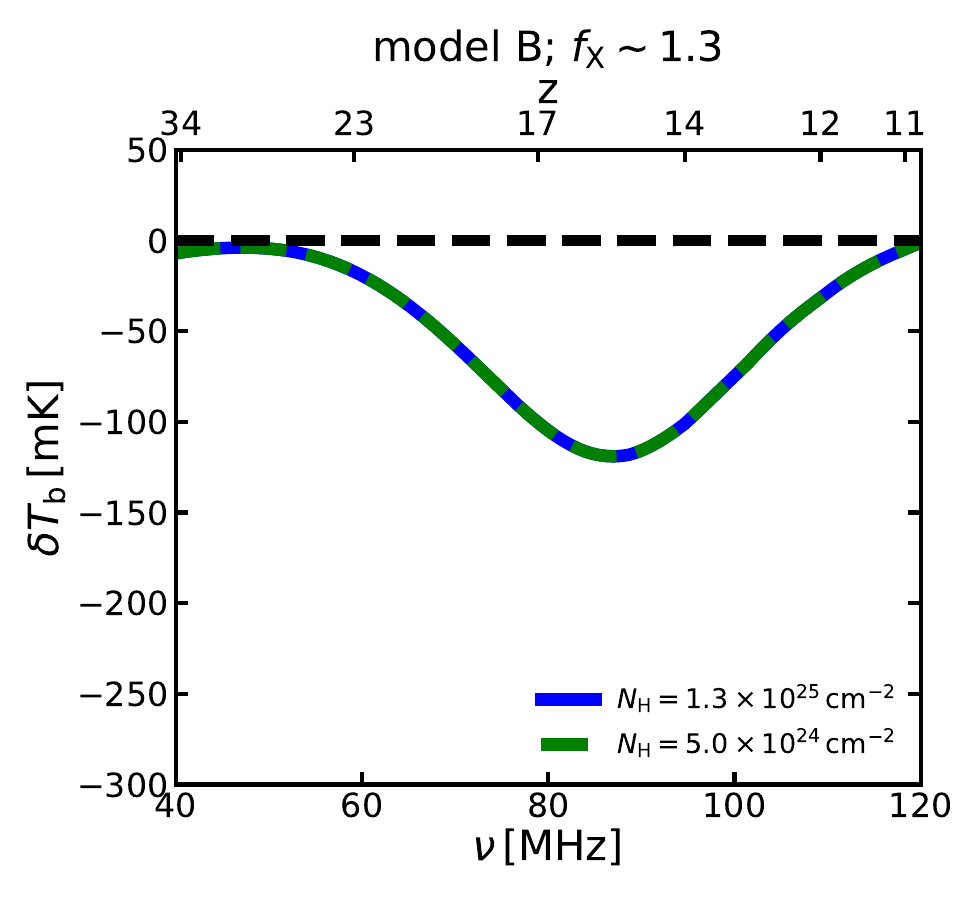}}

\subfigure{\includegraphics[width=0.45\textwidth]{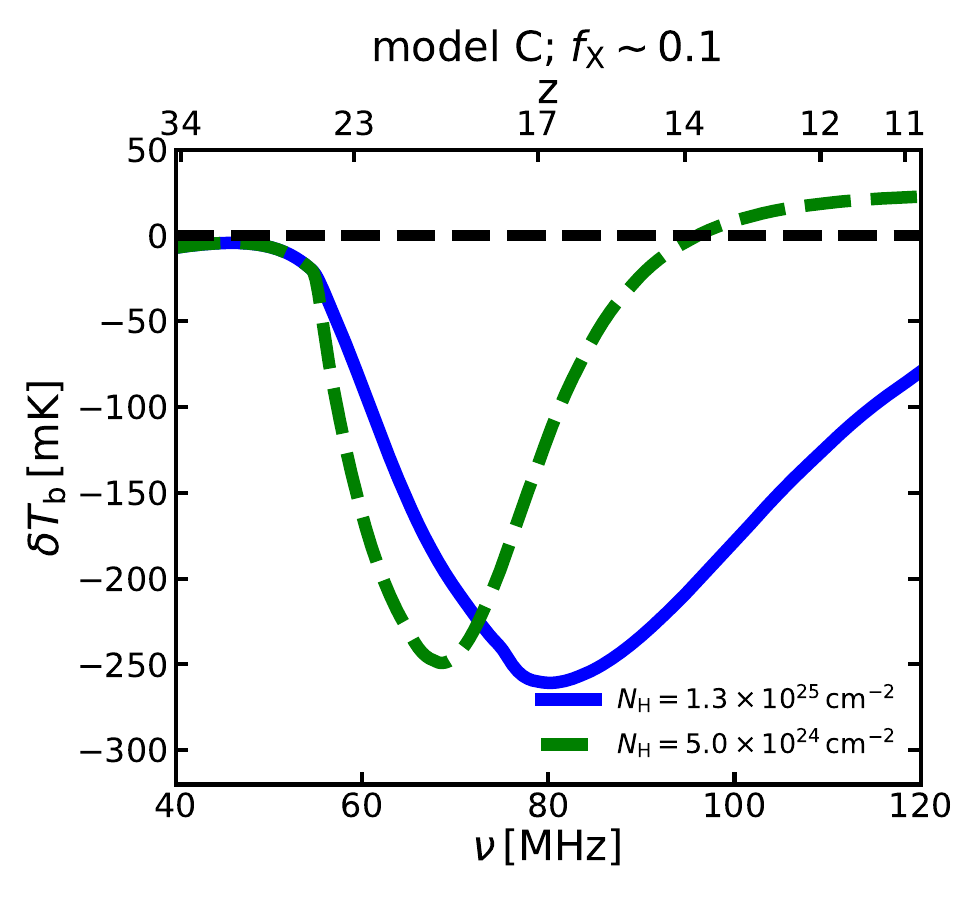}}
\subfigure{\includegraphics[width=0.45\textwidth]{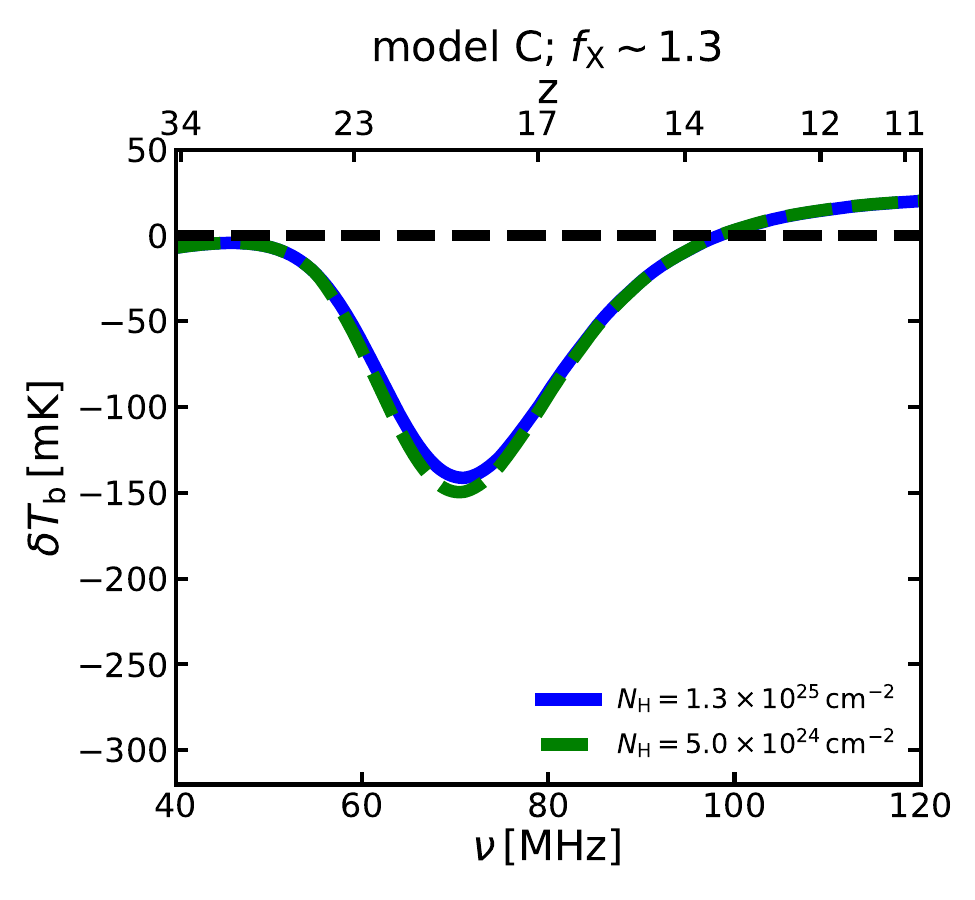}}

\caption{The 21 cm global spectrum for model A, model B and model C. We plot the case for $f_{\rm X} \sim$ 0.1 and 1.3.  In each panel, we plot the results for two DCBH SEDs. 
\label{fig:21cm_signal_new}
}
}
\end{figure*}

We plot the 21 cm global signal results in Fig. \ref{fig:21cm_signal_new}. When $f_{X,\rm II}=f_{X,\rm III}=f_X \sim 0.1$, the X-ray radiation from Pop III stars and the first galaxies just moderately heats the IGM, the 21 cm signal exhibits a strong absorption feature with $\delta T_{\rm b} \sim -230\,\rm mK$. Such weak X-ray radiation also 
has negligible influence on DCBH formation and allows the DCBH number density to exceed $\mathcal{O}( 10^{-3}-10^{-2})$ cMpc$^{-3}$.  So the DCBH also has the chance to influence the 21 cm global spectrum as long as it is not extremely Compton-thick.
For example, in model A \& C, when $N_{\rm H}^{\rm BH} = 5.0\times 10^{24}\,\rm cm^{-2}$, the 21 cm signal absorption depth starts to decrease rapidly after $z\sim 20$, due to the heating by X-ray radiation from DCBHs. However, if   $N_{\rm H}^{\rm BH} = 1.3\times 10^{25}\,\rm cm^{-2}$, even their number density is larger than $\sim10^{-3}$ cMpc$^{-3}$, the influence on the 21 cm signal is still negligible.

When $f_X \sim 1.3$, the X-ray radiation reduces the 21 cm absorption feature  to $\delta T_{\rm b} \sim -150 \,\rm mK$ and lowers the DCBH number density to below  $\mathcal{O}(10^{-3}-10^{-2})$ cMpc$^{-3}$. In this case, the Compton-thick DCBH has a negligible influence on the 21 cm global signal.

DCBHs are hard to form due to the intense X-ray radiation from the nearby first galaxies when $f_X\gtrsim 5$. Such intense X-ray radiation also heats the IGM heavily and reduced the 21 cm absorption to $\sim 100$ mK. So, as long as the 21 cm signal shows an absorption feature weaker than $\sim 100$ mK, the DCBH population would be rare, no matter whether they are Compton-thick or Compton-thin.

In short summary, the 21 cm global signal may reflect the information of the DCBH population abundance.
A signal with a strong absorption feature implies weak X-ray radiation from Pop III stars and the first galaxies, coupled with either a high number density of Compton-thick DCBHs with $n_{\rm DCBH} \gtrsim 10^{-3}$ cMpc$^{-3}$ or a low number density of Compton-thin DCBHs. If the signal exhibits weak absorption, this implies either strong X-ray radiation from Pop III stars and the first galaxies with a low number density of DCBHs, or weak X-ray radiation from Pop III stars and the first galaxies with a high number density of Compton-thin DCBHs.

\section{discussion}\label{sec:discussion}

\subsection{Alternative to increasing star formation efficiency}

We note that alternative models have been proposed to explain the JWST observed stellar mass functions and UV LFs without increasing the SFE at $z\gtrsim 10$ \citep[e.g,][]{Ferrara+2024a, Ferrara+2024b, Harikane+2023_LF}. For example, \citet{Donnan+2025} developed a model that can explain the UV LFs from $z\sim 6$ to $z\sim 13$ without evolving the SFE, but assuming stars are younger at higher redshifts, therefore are more luminous.  

Our model is based on the fact that not only abundant high-$z$ galaxies, but also abundant $z\gtrsim 6$ SMBHs (up to $\gtrsim 10^{-4}-8\times 10^{-3}$ cMpc$^{-3}$, were detected \citep{Harikane+2023_AGN,Hayes+2024}), and the assumption that the seeds of these SMBHs are triggered by progenitors of the observed high-$z$ galaxies. In this case, increasing the SFE, instead of the IMF (like the top-heavy Pop III stars), is straightforward. Because the critical LW intensity for top-heavy Pop III stars is much higher than regular galaxies, therefore not possible to trigger many DCBHs to form. 
In our calculations, the model (model B) uses the same SFE from $z\sim 6$ to $z\gtrsim 10$ cannot produce a DCBH number density of $\gtrsim 10^{-4}$ cMpc$^{-3}$.

Increasing the SFE may result in more metals production. This may pollute the potential pristine ACHs. However, the SED for galaxies with higher metallicity is softer, therefore can suppress the H$_2$ formation more efficiently, because the H$^-$ detachment becomes more efficient. The net effect is not necessarily negative.   

Nevertheless, since currently the model increasing SFE at $z\gtrsim 10$ is still not ruled out, and it is a straightforward method to trigger more SMBH seeds to form, this work is still a reasonable investigation.

\subsection{The Cosmic Radio Background from DCBHs}

It is still not yet known whether the DCBHs are like the radio-loud AGNs or not. 
However, according to previous investigations, to produce sufficient radio background that boosts the 21 cm global signal absorption \citep{Ewall-Wice+2018, Ewall-Wice2020, Mirabel+2022}, or even be detected directly by next-generation radio telescopes \citep{Yue+2021,Whalen2021ApJ}, they (or at least a fraction of them) must produce radio emission as efficiently as radio-loud AGNs. Generally, in observations, radio-loud AGNs have radio loudness $\gtrsim 10^2$ larger than radio-quiet ones (e.g. \citealt{Ivezic+2002,Wang+2006,Bariuan+2022}).

We use the following formula for the radio-loud sample in \citet{Bariuan+2022} to estimate the radio luminosity of the DCBHs, 
\begin{equation}
\log L_R=1.12\log L_X-0.20\log M_{\rm BH}-5.64 \pm \Delta\log R,
\label{eq:LR}
\end{equation}
where $L_R=\nu_5 L_R(\nu_5)$ is the cumulative radio luminosity at 5 GHz, in units of erg s$^{-1}$. $L_X$ is the X-ray luminosity in $2-10$ keV, in units of erg s$^{-1}$ as well. We assume $L_{\rm X} = K_{\rm x}^{-1} L_{\rm edd}$, where the bolometric correction factor is $K_{\rm X} \approx 11$ \citep{Duras+2020}. $M_{\rm BH }$ is the black hole mass in units of solar mass. 
 
$\Delta \log R$ is a free parameter that describes the deviation from the mean log radio loudness ($\log R$) of the radio-loud AGNs in \citet{Bariuan+2022}. They have  $\mean{\log R} \approx 3$ and a standard deviation $\sigma_{\log R} \approx 0.6$. Here, we set $\Delta \log R$ as a free parameter in our model.

We assume the radio emission follows a power-law spectrum
\begin{equation}
    L_{\rm R}(\nu) = \frac{L_{\rm R}}{\nu_{5}} \left(\frac{\nu}{\nu_{5}}\right)^{-0.7}
\end{equation}

The radio emissivity is 
\begin{equation}
    \epsilon_{\rm R}(\nu, z) = n_{\rm BH}(z) \int d\log M_{\rm BH}L_R(\nu,M_{\rm BH}) P_{\log M_{\rm BH}} 
\end{equation}

Then the radio radiation intensity is 
\begin{equation}
    J_{\rm R}(\nu, z) = \frac{(1+z)^{3}}{4\pi} \int_{z}^{\infty} \epsilon_{\rm R}(\nu^{\prime}, z^{\prime}) \frac{cdz^{\prime}}{H(z^{\prime})(1+z^{\prime})}
\end{equation}

\begin{figure}
\centering{
\subfigure{\includegraphics[width=0.5\textwidth]{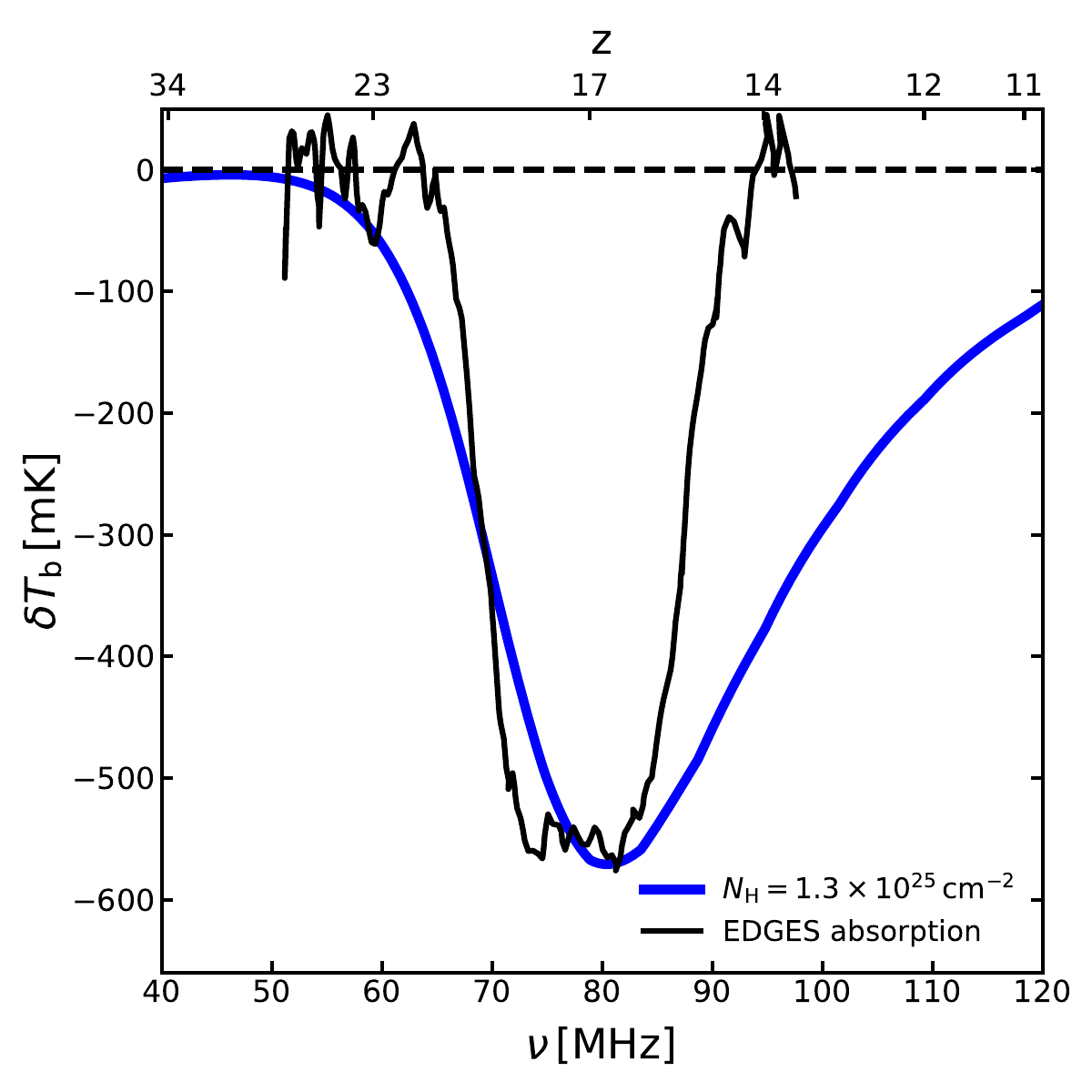}}

\caption{The 21 cm signal as a function of frequency for model C with radio-loud-like DCBHs. We plot the case for 
$f_{X,\rm II}=f_{X,\rm III}=f_{\rm X} \sim 1.3, \Delta \log R = 0.7, N_{\rm H}^{\rm BH} = 1.3\times 10^{25}\,\rm cm^{-2}$.
}
\label{fig:21cm_signal_model_C_with_DCBH}
}
\end{figure}

In Fig.~\ref{fig:21cm_signal_model_C_with_DCBH}, we show the 21 cm absorption measured by EDGES, and the 21 cm global signal for model C with 
$f_{X,\rm II}=f_{X,\rm III}=f_{\rm X} \sim 1.3 $ and $N_{\rm H}^{\rm BH} = 1.3\times 10^{25}\,\rm cm^{-2}$. We add a term $\Delta \log R = 0.7$ to  Eq. (\ref{eq:LR}), i.e., boosting the standard $L_R-L_X$ relation by a factor $\sim 5$.   
We find that if DCBHs are strong radio sources, they can produce 21 cm absorption at Cosmic Dawn as deep as that detected by EDGES. However, unless the DCBH formation rate at $z\lesssim 17$ is largely reduced by mechanisms not involved in this paper,  it is hard to explain the rapid decrease of absorption after $z\sim17$. 
One possible mechanism could be due to the decreasing of $N_{\rm H}^{\rm BH}$, this can suppress the following DCBH formation and heat the IGM, resulting in reduced 21 cm absorption.

\subsection{The influence of Pop III stars}

Although Pop III stars negligibly trigger the DCBH formation, they produce an X-ray background that heats the IGM and reduces 21 cm absorption, therefore loosening the correlation between the 21 cm absorption depth and the DCBH abundance. Here we investigate this effect by varying the X-ray background from Pop III stars. To avoid complicating our discussions, we fix the $f_{*,\rm III}f_{d,\rm III}=5\times 10^{-4}$ and only vary the $f_{X,\rm III}$ to produce {different levels of} X-ray background \footnote{Because $L_X\propto {\rm SFR}$, $f_{\rm *, III}$ and $f_{\rm d, III}$ are completely degenerate in our model,  $f_{*, \rm III} f_{\rm d, III}$ is treated as a joint parameter. This parameter, together with $f_{\rm X, III}$, governs the X-ray background produced by Pop III stars.}.  
 
\begin{figure}
\centering{
\subfigure{\includegraphics[width=0.5\textwidth]{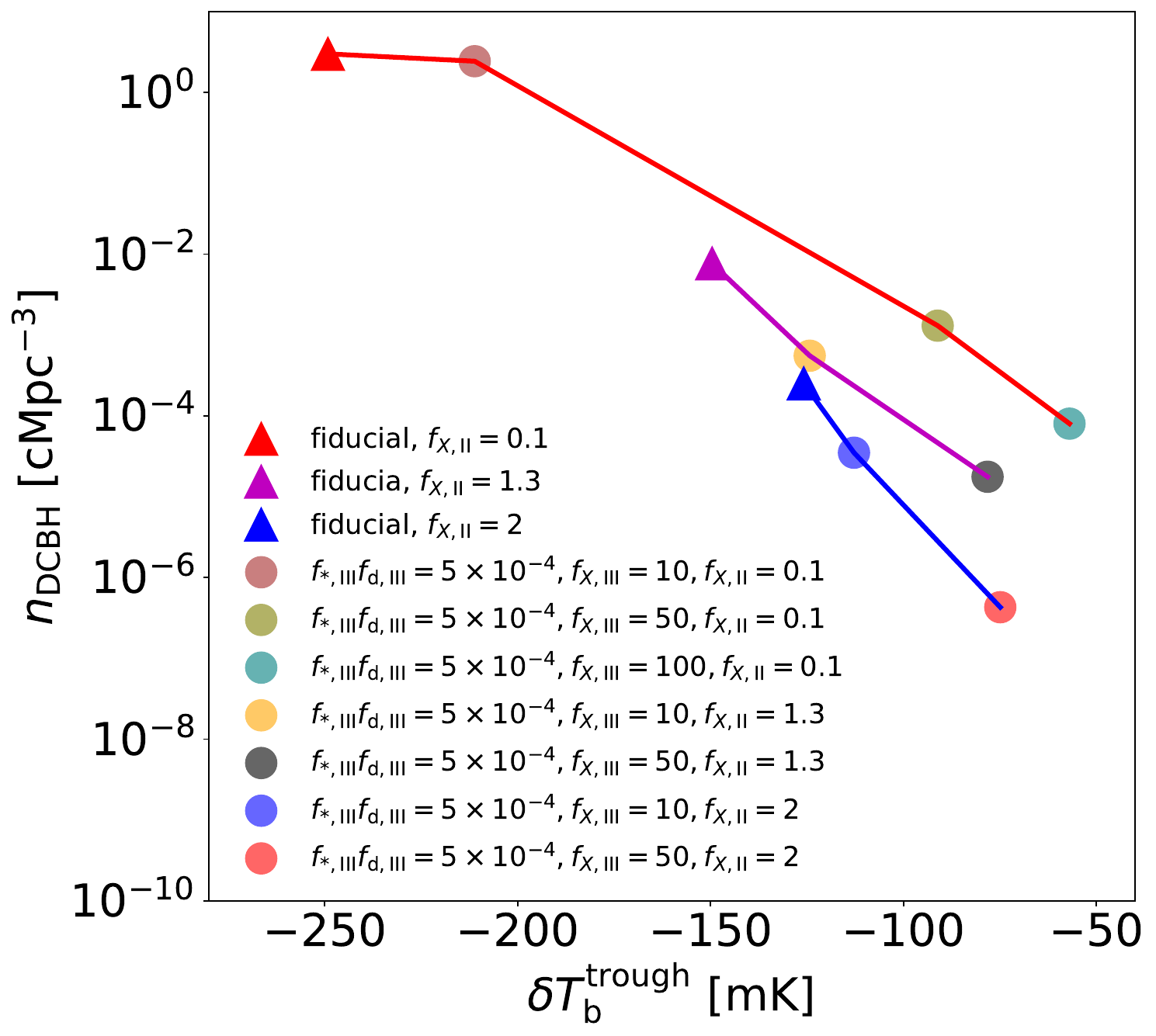}}
\caption{
The $n_{\rm DCBH} $ vs. $ \delta T_b^{\rm trough}$ for varying Pop III stars properties. Here $N_H^{\rm BH}=5\times 10^{24}$ cm$^{-2}$.
}
\label{fig:n_DCBH_vs_delta_Tb}
}
\end{figure}

In Fig.~\ref{fig:n_DCBH_vs_delta_Tb} we plot the $n_{\rm DCBH}$ vs. $\delta T^{\rm trough}_{\rm b}$ for different X-ray backgrounds from Pop III stars. We find that indeed Pop III stars would interfere with the correlation between $n_{\rm DCBH}$ and $\delta T_b^{\rm trough}$. When $\delta T_b^{\rm trough}\sim -150$ mK, the scatter on $n_{\rm DCBH}$ is about one order of magnitude; when $\delta T_b^{\rm trough} \sim -75$ mK, the scatter reaches about two orders of magnitude. However, the general correlation that $n_{\rm DCBH}$ decreases with decreasing 21 cm absorption depth always holds.

$f_{X, \rm III}$  and $f_{*, \rm III} f_{\rm d, III}$ are not degenerate. For the same  $f_{*, \rm III} f_{\rm d, III}\times f_{X, \rm III}$ value, a larger $f_{X, \rm III}$ and a smaller $f_{*, \rm III} f_{\rm d, III}$ may produce stronger X-ray background than smaller $f_{X, \rm III}$ and larger $f_{*, \rm III} f_{\rm d, III}$. This is because the formation of Pop III stars is in self-limited mode, models with larger $f_{*, \rm III} f_{\rm d, III}$ have strong LW feedback and significantly suppress the Pop III stars formation and their X-ray background. More precise determination on $n_{\rm DCBH}$ relies on the knowledge of Pop III stars properties.

In addition to Pop III stars, other factors such as the shock heating (e.g. \citealt{Xu2018ApJ,Xu2021ApJ}) and variations in the Ly$\alpha$ emission compared to X-ray emission (e.g., due to metallicity evolution), may further introduce scatters on DCBH abundance when the 21 cm absorption is weak. However, this is beyond the scope of this paper and we leave this open for future investigation.

\subsection{Discuss about the one-zone model}\label{sec:one-zone-discussion}

Our work uses the one-zone model to find the critical radiation for DCBH formation. In such a model, the whole system is treated as a single region with uniform physical quantities, without accounting for the complex spatial structures. 
It has been pointed out that the one-zone model may underestimate the critical radiation strength \citep{Shang+2010, Latif+2014, Latif+2015,Luo+2020}. The critical intensities found by 3D hydrodynamic simulations are generally $\gtrsim 10 -100$ times larger than the one-zone model. Using hydrodynamic simulations, \citet{Luo+2020} found that the  critical curve also shifts to higher $k_{\rm H_2}$, especially when $k_{\rm H^-}\lesssim 10^{-7}$ s$^{-1}$. Specifically, their critical curve corresponds to $J_{\rm crit} \approx 2500$ for the first galaxies with our fiducial SED and $J_{\rm crit} \approx 240$ for DCBHs with $N_{\rm H}^{\rm BH} = 5 \times 10^{24}$ cm$^{-2}$, in the absence of X-rays.
And the critical radiation for galaxy SEDs found by \citet{Latif+2015} is an order of magnitude higher than that of \citet{Luo+2020}.

The one-zone model underestimates the DCBH criteria probably because the shock heating missed in it enhances the ionization in hydrodynamical simulations \citep{Latif+2014}. As a tentative attempt toward improving the one-zone model, we enhance the ionization rate of the collisional ionization reactions in our one-zone model. We find that the critical curve is boosted by about two orders of magnitude, roughly consistent with the hydrodynamic results \citep{Luo+2020}. However, the overall shape of the critical curve remains distinct, suggesting that other hydrodynamic effects or directional dependence of self-shielding \citep{Shang+2010} may also contribute.

The effects of local X-ray radiation are more complex when detailed dependence on distance is considered, as found by \citet{Regan+2016} in 3D hydrodynamical simulations. They pointed out that, distant X-ray sources ($\gtrsim 1 \,\rm kpc$) heat the gas and reduce the mass inflow rate, significantly suppressing the DCBH formation; whereas nearby sources ($ \lesssim 1\,\rm  kpc$) have weaker impacts because the LW radiation is already very strong and the gas is well heated. We did not find this phenomenon in our one-zone model, probably because one-zone misses some key processes. 

Additionally, spatial variations within the ACH are also important, as found in \citet{Aykutalp+2020}. If there is a massive black hole at the center of an ACH, its X-ray radiation generates partially ionized regions around it and promotes the H$_2$ formation therein. Star formation happens at the edge of ionized regions, which expand with time, and is finally quenched, leaving the central black hole dominating the overall radiation. This is very interesting and highlights the importance of using 3D hydrodynamical simulations. The one-zone method is not able to model this effect directly as it treats the full gas cloud as a single point and ignore the detailed spatial effects of X-ray propagation.

The physical conditions for DCBH formation are more complex. In addition to the supercritical external radiation, there are many alternative mechanisms that can promote/facilitate the DCBH formation as well, some of these mechanisms are explicitly introduced in the reviews by \citet{Woods+2019} and \citet{Inayoshi+2020}. \citet{Inayoshi_2012MNRAS.422.2539I} and \citet{Inayoshi_2015MNRAS.453.1692I} found that the collision between dense accretion inflows or high velocity protogalaxies results in a hot and dense core, where the H$_2$ formation is suppressed. \citet{Visbal+2014_DCBH} proposed the synchronized close pair of ACHs scenario. If the formation of stars in one halo is immediately followed by the partner reaching cooling threshold, then the latter becomes a DCBH-forming halo since it is irradiated by strong radiation from the former. 
This mechanism not only suppresses the H$_2$ cooling, but also avoids the photoevaporation and metal pollution. \citet{Schauer+2017} point out that in regions with $\gtrsim 3 \sigma$ baryon-dark matter streaming velocities, H$_2$ cooling and star formation in mini-halos are delayed, further facilitating the DCBH formation in synchronized halo pairs. 
\citet{Johnson+2017} find that if the Ly$\alpha$ photons produced during the atomic-cooling process are trapped, they can further detach H$^-$.  It lowers the requirement on LW radiation by a factor of a few, and enhances the predictions on DCBH abundance. \citet{Wise+2019} investigated the formation of supermassive stars (finally collapse directly into DCBHs) in halos exposed to strong LW radiation and experiencing rapid growth in early evolution. They found that the dynamical heating amplifies the H$_2$ suppression, reducing the requirement for LW radiation in DCBH formation. They conclude that, it is mainly the dynamical effect, rather than the LW radiation, that drives the formation of DCBHs, thereby the number density of DCBHs can be much higher than predictions purely rely on LW radiation.  
Using hydrodynamical simulations, \citet{Latif+2022} show that the violent supersonic turbulence driven by strong cold accretion inflow suppresses star formation and finally triggers catastrophic baryon collapse and the formation of a massive black hole seed.
The magnetic field also has positive effects on DCBH formation.
\citet{Hirano+2021,Hirano+2023} find that early accretion amplifies the magnetic field that in turn promotes the coalescence of protostars and enhances mass accretion, in both metal-free and slightly metal-enriched ACHs.
\citet{Latif2023ApJ...945..137L} and \citet{Latif2023ApJ...952L...9L} find that strong accretion shocks amplify the magnetic field that is helpful to reduce fragmentation and to transport angular momentum. And \citet{Begelman+2023} find that the magnetic field can channel gas toward the center to maintain a stable accretion flow (even enable super-Eddington accretion), which leads to DCBH formation.
In a series of works \citep{Mayer+2010,Mayer+2015,Zwick_2023MNRAS.518.2076Z,Mayer_2024ApJ...961...76M}, they point out that even the gas is metal enriched,  massive black hole seeds can form directly via merger-driven inflows. 
\citet{Chiaki+2023} found that if the gas is enriched to about 0.1\% of Solar metallicity, the internal radiation from stars in the same DCBH-forming halo can enhance  the DCBH number density, and they claimed that the DCBH scenario can explain the observed SMBH abundance.

All these mechanisms can enhance the number density of DCBHs, although determining the explicit magnitude requires more detailed investigations in the future. Anyway, they make the DCBH scenario more optimistic when the number of observed high-$z$ SMBHs continues to increase. Most of them reply on hydrodynamical effects to work, it is a challenge to investigate them in the current one-zone model.

Nevertheless, the one-zone model is still a quite economical tool for exploring large parameter spaces in the framework of semi-analytical galaxy formation model. For the purposes of this paper, the one-zone model, despite its simplifications, provides a computationally efficient tool to systematically explore the impact of radiation on H$_2$ suppression under different assumptions. It remains useful for understanding general trends and parameter dependencies in DCBH formation, although further validation with 3D hydrodynamical simulations is necessary.

\section{Summary}\label{sec:summary}

We calculated the critical $k_{\rm H_2}-k_{\rm H^-}$ curves for the DCBH formation as functions of X-ray intensity and provided their analytical fitting formulae. To our knowledge, such curves were not provided in previous works that either presented the $J_{\rm crit}$ as a function of X-ray intensity (e.g. \citealt{Latif+2015,Glover+2016,Yue+2017}) or the X-ray/LW intensity ratio \citep{Inayoshi+2015}, or the $k_{\rm H_2}-k_{\rm H^-}$ curves in the absence of X-ray radiation (e.g. \citealt{Agarwal+2016,Wolcott-Green+2017,Luo+2020}). It is feasible to use our results in semi-analytical models that investigates the DCBH formation in the presence of X-ray radiation. 

We investigated the DCBH formation at $z\gtrsim 10$ in models with various SFEs for the first galaxies and with different DCBH SEDs, and examined their influence on the evolution of the 21 cm global signal. We combined the one-zone calculation for DCBH formation with the Monte Carlo simulation for computing the probability that ACHs receive super-critical H$_2$-dissociation/H$^-$-detachment radiation. We provided quantitative results of DCBH abundance under various conditions. We found:

\begin{itemize}

\item If the halos with $T_{\rm vir}\sim 10^4$ K have the same SFE as those with peak SFE (model A), DCBHs can form efficiently in the absence of X-ray. In some cases, they may even undergo the runaway formation process if previously-formed DCBHs also provide H$_2$-dissociation/H$^-$-detachment radiation. 

However, if the SFE decreases as $\propto M_h^{0.46}$ (model B), as motivated by the slope of observed UV LFs, DCBH formation becomes less efficient. The number density remains $\lesssim \mathcal{O}(10^{-4})$ cMpc$^{-3}$ even in the absence of X-ray and even DCBHs can also produce H$_2$-dissociation/H$^-$-detachment radiation.

If the star formation is more efficient in the first galaxies at $z\gtrsim 10$, as inspired by JWST observations but still consistent with HST observations at $z\lesssim 10$ (model C), DCBH formation can also be efficient. In this scenario, 
the number density can be $\gtrsim \mathcal{O}(10^{-2})$ cMpc$^{-3}$ in the absence of X-ray. So this may help interpret the observed excess of both galaxies at $z\gtrsim 10$ and SMBHs at $z\gtrsim6$.  

\item {

If the X-ray radiation from Pop III stars and  the first galaxies is small, 
$f_{X,\rm II}=f_{X,\rm III}=f_X\lesssim 0.1$, then for Compton-thick DCBHs, their number density is close to the case without X-ray radiation. 
However, for Compton-thin DCBHs, self-regulation of number density may occur due to the intense X-ray radiation they emit.

If the X-ray radiation from Pop III stars and the first galaxies is moderate, $f_X\sim 1.3$, then the DCBH number density can be limited to $\mathcal{O}(10^{-3}-10^{-2})$ cMpc$^{-3}$ for Compton-thick populations, where X-rays from DCBHs have a negligible effect. While this study only utilizes the SED of Compton-thick DCBHs, and our results specifically reflect this scenario, a reasonable inference for Compton-thin DCBHs is that their more intense X-ray radiation would further inhibit DCBH formation.

If the X-ray radiation from Pop III stars and the first galaxies is strong with $f_X \gtrsim 5$, DCBH formation is prevented, since the X-rays from galaxies are too strong to trigger the formation of DCBH. 

No matter whether the X-ray radiation is strong or weak, in models with higher SFE in the first galaxies, the DCBH number density is higher.

}

\item {In addition to reducing the DCBH formation rate, X-ray radiation also heats the IGM and reduces the 21 cm absorption signal. If the 21 cm signal shows a strong absorption feature, then it implies both the Pop III stars and the first galaxies have weak X-ray radiation, coupled with either a high number density of Compton-thick DCBHs with $n_{\rm DCBH} \gtrsim 10^{-3}$ cMpc$^{-3}$ or a low number density of Compton-thin DCBHs. However, if the 21 cm signal has weak absorption, then either Pop III stars and the first galaxies have strong X-ray radiation and a low number density of DCBH, or the DCBHs are Compton-thin. Although our results about the DCBH number density vs. 21 cm absorption depth are just tentative and not yet very conclusive, mainly due to uncertainties on Pop III stars, they are still useful. At the very least, they could help to constrain or rule out some extreme DCBH formation models.

Since the DCBH number density may related to the 21 cm signal, future 21 cm observational objects, not just the global spectrum experiment but also the interferometer array like SKA, may provide useful information about the DCBH formation.  

}

\end{itemize}

\section*{Acknowledgments}

We thank the anonymous referee for very helpful comments/suggestions for improving this manuscript. 
This work is supported by the National SKA Program of China Nos. 2020SKA0110402 and 2020SKA0110401, the National Natural Science Foundation of China Grant No. 11973047, and NSFC International (Regional) Cooperation and Exchange Project No. 12361141814.

%






\appendix

\section{updates in the one-zone chemistry network}\label{sec:appendix}

\subsection{update reaction rates}
The reaction rates updated compared with \citet{Yue+2017} are listed here

\begin{enumerate}
    \item The radiative recombination of H$^+$: $\rm H^{+} + e^{-} \xrightarrow{k_3} H + \gamma$. Here we use the Case B rate from \citet{Hui+1997}. 
    \begin{equation}
        k_{\rm 3,case B} = 2.753\times 10^{-14} {\rm cm^{3}\ s^{-1}} \frac{\lambda_{\rm HI}^{1.5}}{[1+(\lambda_{\rm HI}/2.74)^{0.407}]^{2.242}}
    \end{equation}
    where $\lambda_{\rm HI} = 315614/T$, T is in units of $\rm K$.
    \item The radiative association of H and $e^{-}$: $\rm H+e^{-} \xrightarrow{k_{7}} H^- + \gamma$.  We adopt the reaction rates from  \citet{Abel+1997}. 
    \begin{align}
        k_{7} & = 1.429 \times 10^{-18} T^{0.7620} T^{0.1523 \log(T)} T^{-3.274\times 10^{-2} (\log T)^{2}} \quad  T \leq 6000\,\rm K \nonumber \\
        & = 3.802 \times 10^{-17} T^{0.1998 \log T} {\rm dex} (4.0415\times 10^{-5} (\log T)^{6} - 5.447 \times 10^{-3} (\log T)^{4}) \quad T > 6000\,\rm K
    \end{align}
    \item The associative detachment of H$^-$ with H: $\rm H + H^- \xrightarrow{k_{8}} H_2 + e^-$. The reaction rate data comes from \citet{Kreckel+2010} and the fitting formulae come from \citet{Suazo+2019} 
    \begin{equation}
        k_{8} = a_1 (T^{a_2} + a_3 T^{a_4} + a_5 T^{a_6})/(1+a_7 T^{a_8} + a_9 T^{a_{10}} + a_{11} T^{a_{12}})
    \end{equation}
    where $a_1 = 1.35\times 10^{-9}, a_2 = 9.8493 \times 10^{-2}, a_3 = 3.2852 \times 10^{-1}, a_4 = 5.5610 \times 10^{-1}, a_5 = 2.7710\times 10^{-7}, a_6=2.1826, a_7 = 6.1910 \times 10^{-3}, a_8 = 1.0461, a_9 = 8.9712 \times 10^{-11},  a_{10}=3.0424, a_{11}= 3.2576 \times 10^{-14}, a_{12} = 3.7741$.
    \item The mutual neutralization rate: $\rm H^- + H^+ \xrightarrow{k_{11}} 2H$. We adopt the reaction rates \citep{Stenrup+2009}.
    \begin{equation}
        k_{11} = A \left(\frac{2\mu}{\pi kT}\right)^{1/2} + B + 2C \left(\frac{2kT}{\pi \mu}\right)^{1/2} + \frac{3DkT}{\mu}
    \end{equation}
    where $A = 4.77\times 10^{-2}\, \rm cm^{4}\ s^{-2}; B = -1.73 \times 10^{-9}\,\rm cm^{3}\ s^{-1}; C=1.22\times10^{-14}\,\rm cm^{2}; D = -1.57 \times 10^{-21}\,\rm cm\ s$.  $\mu$ is the reduced mass of two H nuclei. $T$ is in units of $K$. 
\end{enumerate}

\subsection{additional reactions} 
We add additional two reactions, the ionization of atomic hydrogen by H–H collisions
\begin{equation}
    \rm H + H \xrightarrow{k_{25}} H^{+} + e^{-} + H, 
\end{equation}
with reaction rates $k_{25}$ \citep{Lenzuni+1991}
\begin{equation}
    k_{25} = 1.75\times 10^{-17} T^{1.3} \exp(\frac{-157800}{T})\,\rm cm^{3}\ s^{-1},
    \label{eq: k_25}
\end{equation}
and H–He collisions,
\begin{equation}
    \rm H + He \xrightarrow{k_{26}} H^{+} + e^{-} + He,
\end{equation}
with reaction rates $k_{26}$ \citep{Lenzuni+1991}
\begin{equation}
    k_{26} = 1.75\times 10^{-17} T^{1.3} \exp(\frac{-157800}{T})\,\rm cm^{3}\ s^{-1},
    \label{eq: k_26}
\end{equation}
These reactions are shown to be important by \citet{Glover+2015_chemical_model, Glover+2015_rate_coefficient}, and adopted in \citet{Wolcott-Green+2017}.






\bibliography{reference}{}

\begin{thebibliography}{}
\expandafter\ifx\csname natexlab\endcsname\relax\def\natexlab#1{#1}\fi
\providecommand{\url}[1]{\href{#1}{#1}}
\providecommand{\dodoi}[1]{doi:~\href{http://doi.org/#1}{\nolinkurl{#1}}}
\providecommand{\doeprint}[1]{\href{http://ascl.net/#1}{\nolinkurl{http://ascl.net/#1}}}
\providecommand{\doarXiv}[1]{\href{https://arxiv.org/abs/#1}{\nolinkurl{https://arxiv.org/abs/#1}}}

\bibitem[{{Abdurashidova} {et~al.}(2022){Abdurashidova}, {Aguirre}, {Alexander}, {Ali}, {Balfour}, {Barkana}, {Beardsley}, {Bernardi}, {Billings}, {Bowman}, {Bradley}, {Bull}, {Burba}, {Carey}, {Carilli}, {Cheng}, {DeBoer}, {Dexter}, {de Lera Acedo}, {Dillon}, {Ely}, {Ewall-Wice}, {Fagnoni}, {Fialkov}, {Fritz}, {Furlanetto}, {Gale-Sides}, {Glendenning}, {Gorthi}, {Greig}, {Grobbelaar}, {Halday}, {Hazelton}, {Heimersheim}, {Hewitt}, {Hickish}, {Jacobs}, {Julius}, {Kern}, {Kerrigan}, {Kittiwisit}, {Kohn}, {Kolopanis}, {Lanman}, {La Plante}, {Lekalake}, {Lewis}, {Liu}, {Ma}, {MacMahon}, {Malan}, {Malgas}, {Maree}, {Martinot}, {Matsetela}, {Mesinger}, {Mirocha}, {Molewa}, {Morales}, {Mosiane}, {Mu{\~n}oz}, {Murray}, {Neben}, {Nikolic}, {Nunhokee}, {Parsons}, {Patra}, {Pieterse}, {Pober}, {Qin}, {Razavi-Ghods}, {Reis}, {Ringuette}, {Robnett}, {Rosie}, {Santos}, {Sikder}, {Sims}, {Smith}, {Syce}, {Thyagarajan}, {Williams}, \& {Zheng}}]{Abdurashidova+2022}
{Abdurashidova}, Z., {Aguirre}, J.~E., {Alexander}, P., {et~al.} 2022, \apj, 924, 51, \dodoi{10.3847/1538-4357/ac2ffc}

\bibitem[{{Abel} {et~al.}(1997){Abel}, {Anninos}, {Zhang}, \& {Norman}}]{Abel+1997}
{Abel}, T., {Anninos}, P., {Zhang}, Y., \& {Norman}, M.~L. 1997, \na, 2, 181, \dodoi{10.1016/S1384-1076(97)00010-9}

\bibitem[{{Adams} {et~al.}(2023){Adams}, {Conselice}, {Ferreira}, {Austin}, {Trussler}, {Juod{\v{z}}balis}, {Wilkins}, {Caruana}, {Dayal}, {Verma}, \& {Vijayan}}]{Adams+2023}
{Adams}, N.~J., {Conselice}, C.~J., {Ferreira}, L., {et~al.} 2023, \mnras, 518, 4755, \dodoi{10.1093/mnras/stac3347}

\bibitem[{{Agarwal} {et~al.}(2013){Agarwal}, {Davis}, {Khochfar}, {Natarajan}, \& {Dunlop}}]{Agarwal+2013}
{Agarwal}, B., {Davis}, A.~J., {Khochfar}, S., {Natarajan}, P., \& {Dunlop}, J.~S. 2013, \mnras, 432, 3438, \dodoi{10.1093/mnras/stt696}

\bibitem[{{Agarwal} {et~al.}(2012){Agarwal}, {Khochfar}, {Johnson}, {Neistein}, {Dalla Vecchia}, \& {Livio}}]{Agarwal+2012}
{Agarwal}, B., {Khochfar}, S., {Johnson}, J.~L., {et~al.} 2012, \mnras, 425, 2854, \dodoi{10.1111/j.1365-2966.2012.21651.x}

\bibitem[{{Agarwal} {et~al.}(2016){Agarwal}, {Smith}, {Glover}, {Natarajan}, \& {Khochfar}}]{Agarwal+2016}
{Agarwal}, B., {Smith}, B., {Glover}, S., {Natarajan}, P., \& {Khochfar}, S. 2016, \mnras, 459, 4209, \dodoi{10.1093/mnras/stw929}

\bibitem[{{Andalman} {et~al.}(2024){Andalman}, {Teyssier}, \& {Dekel}}]{Andalman+2024}
{Andalman}, Z.~L., {Teyssier}, R., \& {Dekel}, A. 2024, arXiv e-prints, arXiv:2410.20530, \dodoi{10.48550/arXiv.2410.20530}

\bibitem[{{Arrabal Haro} {et~al.}(2023){Arrabal Haro}, {Dickinson}, {Finkelstein}, {Kartaltepe}, {Donnan}, {Burgarella}, {Carnall}, {Cullen}, {Dunlop}, {Fern{\'a}ndez}, {Fujimoto}, {Jung}, {Krips}, {Larson}, {Papovich}, {P{\'e}rez-Gonz{\'a}lez}, {Amor{\'\i}n}, {Bagley}, {Buat}, {Casey}, {Chworowsky}, {Cohen}, {Ferguson}, {Giavalisco}, {Huertas-Company}, {Hutchison}, {Kocevski}, {Koekemoer}, {Lucas}, {McLeod}, {McLure}, {Pirzkal}, {Seill{\'e}}, {Trump}, {Weiner}, {Wilkins}, \& {Zavala}}]{Haro+2023}
{Arrabal Haro}, P., {Dickinson}, M., {Finkelstein}, S.~L., {et~al.} 2023, \nat, 622, 707, \dodoi{10.1038/s41586-023-06521-7}

\bibitem[{{Austin} {et~al.}(2023){Austin}, {Adams}, {Conselice}, {Harvey}, {Ormerod}, {Trussler}, {Li}, {Ferreira}, {Dayal}, \& {Juod{\v{z}}balis}}]{Austin+2023}
{Austin}, D., {Adams}, N., {Conselice}, C.~J., {et~al.} 2023, \apjl, 952, L7, \dodoi{10.3847/2041-8213/ace18d}

\bibitem[{{Aykutalp} {et~al.}(2020){Aykutalp}, {Barrow}, {Wise}, \& {Johnson}}]{Aykutalp+2020}
{Aykutalp}, A., {Barrow}, K. S.~S., {Wise}, J.~H., \& {Johnson}, J.~L. 2020, \apjl, 898, L53, \dodoi{10.3847/2041-8213/aba62f}

\bibitem[{{Balucinska-Church} \& {McCammon}(1992)}]{Balucinska-Church+1992}
{Balucinska-Church}, M., \& {McCammon}, D. 1992, \apj, 400, 699, \dodoi{10.1086/172032}

\bibitem[{{Banados} {et~al.}(2024){Banados}, {Momjian}, {Connor}, {Belladitta}, {Decarli}, {Mazzucchelli}, {Venemans}, {Walter}, {Wang}, {Xie}, {Barth}, {Eilers}, {Fan}, {Khusanova}, {Schindler}, {Stern}, {Yang}, {Taufik Andika}, {Carilli}, {Farina}, {Fabian}, {Hennawi}, {Pensabene}, \& {Rojas-Ruiz}}]{Banados+2024}
{Banados}, E., {Momjian}, E., {Connor}, T., {et~al.} 2024, arXiv e-prints, arXiv:2407.07236, \dodoi{10.48550/arXiv.2407.07236}

\bibitem[{{Bariuan} {et~al.}(2022){Bariuan}, {Snios}, {Sobolewska}, {Siemiginowska}, \& {Schwartz}}]{Bariuan+2022}
{Bariuan}, L. G.~C., {Snios}, B., {Sobolewska}, M., {Siemiginowska}, A., \& {Schwartz}, D.~A. 2022, \mnras, 513, 4673, \dodoi{10.1093/mnras/stac1153}

\bibitem[{{Begelman} \& {Silk}(2023)}]{Begelman+2023}
{Begelman}, M.~C., \& {Silk}, J. 2023, \mnras, 526, L94, \dodoi{10.1093/mnrasl/slad124}

\bibitem[{{Begelman} {et~al.}(2006){Begelman}, {Volonteri}, \& {Rees}}]{Begelman2006MNRAS}
{Begelman}, M.~C., {Volonteri}, M., \& {Rees}, M.~J. 2006, \mnras, 370, 289, \dodoi{10.1111/j.1365-2966.2006.10467.x}

\bibitem[{{Bouwens} {et~al.}(2021){Bouwens}, {Oesch}, {Stefanon}, {Illingworth}, {Labb{\'e}}, {Reddy}, {Atek}, {Montes}, {Naidu}, {Nanayakkara}, {Nelson}, \& {Wilkins}}]{Bouwens2021AJ}
{Bouwens}, R.~J., {Oesch}, P.~A., {Stefanon}, M., {et~al.} 2021, \aj, 162, 47, \dodoi{10.3847/1538-3881/abf83e}

\bibitem[{{Bouwens} {et~al.}(2023){Bouwens}, {Stefanon}, {Brammer}, {Oesch}, {Herard-Demanche}, {Illingworth}, {Matthee}, {Naidu}, {van Dokkum}, \& {van Leeuwen}}]{Bouwens+2023b}
{Bouwens}, R.~J., {Stefanon}, M., {Brammer}, G., {et~al.} 2023, \mnras, 523, 1036, \dodoi{10.1093/mnras/stad1145}

\bibitem[{{Bowman} {et~al.}(2018){Bowman}, {Rogers}, {Monsalve}, {Mozdzen}, \& {Mahesh}}]{Bowman+2018}
{Bowman}, J.~D., {Rogers}, A. E.~E., {Monsalve}, R.~A., {Mozdzen}, T.~J., \& {Mahesh}, N. 2018, \nat, 555, 67, \dodoi{10.1038/nature25792}

\bibitem[{{Boylan-Kolchin}(2023)}]{Boylan-Kolchin+2023}
{Boylan-Kolchin}, M. 2023, Nature Astronomy, 7, 731, \dodoi{10.1038/s41550-023-01937-7}

\bibitem[{{Cammelli} {et~al.}(2025){Cammelli}, {Tan}, {Young}, {Hayes}, {Singh}, {Ellis}, {Saxena}, {Laporte}, {Monaco}, \& {Keller}}]{Cammelli+2025}
{Cammelli}, V., {Tan}, J.~C., {Young}, A.~R., {et~al.} 2025, arXiv e-prints, arXiv:2501.17675, \dodoi{10.48550/arXiv.2501.17675}

\bibitem[{{Castellano} {et~al.}(2022){Castellano}, {Fontana}, {Treu}, {Santini}, {Merlin}, {Leethochawalit}, {Trenti}, {Vanzella}, {Mestric}, {Bonchi}, {Belfiori}, {Nonino}, {Paris}, {Polenta}, {Roberts-Borsani}, {Boyett}, {Brada{\v{c}}}, {Calabr{\`o}}, {Glazebrook}, {Grillo}, {Mascia}, {Mason}, {Mercurio}, {Morishita}, {Nanayakkara}, {Pentericci}, {Rosati}, {Vulcani}, {Wang}, \& {Yang}}]{Castellano+2022}
{Castellano}, M., {Fontana}, A., {Treu}, T., {et~al.} 2022, \apjl, 938, L15, \dodoi{10.3847/2041-8213/ac94d0}

\bibitem[{{Castellano} {et~al.}(2024){Castellano}, {Napolitano}, {Fontana}, {Roberts-Borsani}, {Treu}, {Vanzella}, {Zavala}, {Arrabal Haro}, {Calabr{\`o}}, {Llerena}, {Mascia}, {Merlin}, {Paris}, {Pentericci}, {Santini}, {Bakx}, {Bergamini}, {Cupani}, {Dickinson}, {Filippenko}, {Glazebrook}, {Grillo}, {Kelly}, {Malkan}, {Mason}, {Morishita}, {Nanayakkara}, {Rosati}, {Sani}, {Wang}, \& {Yoon}}]{Castellano+2024}
{Castellano}, M., {Napolitano}, L., {Fontana}, A., {et~al.} 2024, arXiv e-prints, arXiv:2403.10238, \dodoi{10.48550/arXiv.2403.10238}

\bibitem[{{Chiaki} {et~al.}(2023){Chiaki}, {Chon}, {Omukai}, {Trinca}, {Schneider}, \& {Valiante}}]{Chiaki+2023}
{Chiaki}, G., {Chon}, S., {Omukai}, K., {et~al.} 2023, \mnras, 521, 2845, \dodoi{10.1093/mnras/stad689}

\bibitem[{{Cohen} {et~al.}(2017){Cohen}, {Fialkov}, {Barkana}, \& {Lotem}}]{Cohen+2017}
{Cohen}, A., {Fialkov}, A., {Barkana}, R., \& {Lotem}, M. 2017, \mnras, 472, 1915, \dodoi{10.1093/mnras/stx2065}

\bibitem[{{Curtis-Lake} {et~al.}(2023){Curtis-Lake}, {Carniani}, {Cameron}, {Charlot}, {Jakobsen}, {Maiolino}, {Bunker}, {Witstok}, {Smit}, {Chevallard}, {Willott}, {Ferruit}, {Arribas}, {Bonaventura}, {Curti}, {D'Eugenio}, {Franx}, {Giardino}, {Looser}, {L{\"u}tzgendorf}, {Maseda}, {Rawle}, {Rix}, {Rodr{\'\i}guez del Pino}, {{\"U}bler}, {Sirianni}, {Dressler}, {Egami}, {Eisenstein}, {Endsley}, {Hainline}, {Hausen}, {Johnson}, {Rieke}, {Robertson}, {Shivaei}, {Stark}, {Tacchella}, {Williams}, {Willmer}, {Bhatawdekar}, {Bowler}, {Boyett}, {Chen}, {de Graaff}, {Helton}, {Hviding}, {Jones}, {Kumari}, {Lyu}, {Nelson}, {Perna}, {Sandles}, {Saxena}, {Suess}, {Sun}, {Topping}, {Wallace}, \& {Whitler}}]{Curtis-Lake+2023}
{Curtis-Lake}, E., {Carniani}, S., {Cameron}, A., {et~al.} 2023, Nature Astronomy, 7, 622, \dodoi{10.1038/s41550-023-01918-w}

\bibitem[{{Das} {et~al.}(2017){Das}, {Mesinger}, {Pallottini}, {Ferrara}, \& {Wise}}]{Das+2017}
{Das}, A., {Mesinger}, A., {Pallottini}, A., {Ferrara}, A., \& {Wise}, J.~H. 2017, \mnras, 469, 1166, \dodoi{10.1093/mnras/stx943}

\bibitem[{de~Lera~Acedo(2019)}]{Eloy+2019}
de~Lera~Acedo, E. 2019, in 2019 International Conference on Electromagnetics in Advanced Applications (ICEAA), 0626--0629, \dodoi{10.1109/ICEAA.2019.8879199}

\bibitem[{{Dekel} {et~al.}(2023){Dekel}, {Sarkar}, {Birnboim}, {Mandelker}, \& {Li}}]{Dekel+2023}
{Dekel}, A., {Sarkar}, K.~C., {Birnboim}, Y., {Mandelker}, N., \& {Li}, Z. 2023, \mnras, 523, 3201, \dodoi{10.1093/mnras/stad1557}

\bibitem[{{Dijkstra} {et~al.}(2014){Dijkstra}, {Ferrara}, \& {Mesinger}}]{Dijkstra+2014}
{Dijkstra}, M., {Ferrara}, A., \& {Mesinger}, A. 2014, \mnras, 442, 2036, \dodoi{10.1093/mnras/stu1007}

\bibitem[{{Dijkstra} {et~al.}(2008){Dijkstra}, {Haiman}, {Mesinger}, \& {Wyithe}}]{Dijkstra+2008}
{Dijkstra}, M., {Haiman}, Z., {Mesinger}, A., \& {Wyithe}, J. S.~B. 2008, \mnras, 391, 1961, \dodoi{10.1111/j.1365-2966.2008.14031.x}

\bibitem[{{Donnan} {et~al.}(2025){Donnan}, {Dunlop}, {McLure}, {McLeod}, \& {Cullen}}]{Donnan+2025}
{Donnan}, C.~T., {Dunlop}, J.~S., {McLure}, R.~J., {McLeod}, D.~J., \& {Cullen}, F. 2025, arXiv e-prints, arXiv:2501.03217, \dodoi{10.48550/arXiv.2501.03217}

\bibitem[{{Donnan} {et~al.}(2023{\natexlab{a}}){Donnan}, {McLeod}, {McLure}, {Dunlop}, {Carnall}, {Cullen}, \& {Magee}}]{Donnan+2023_candidates}
{Donnan}, C.~T., {McLeod}, D.~J., {McLure}, R.~J., {et~al.} 2023{\natexlab{a}}, \mnras, 520, 4554, \dodoi{10.1093/mnras/stad471}

\bibitem[{{Donnan} {et~al.}(2023{\natexlab{b}}){Donnan}, {McLeod}, {Dunlop}, {McLure}, {Carnall}, {Begley}, {Cullen}, {Hamadouche}, {Bowler}, {Magee}, {McCracken}, {Milvang-Jensen}, {Moneti}, \& {Targett}}]{Donnan+2023_UVLF}
{Donnan}, C.~T., {McLeod}, D.~J., {Dunlop}, J.~S., {et~al.} 2023{\natexlab{b}}, \mnras, 518, 6011, \dodoi{10.1093/mnras/stac3472}

\bibitem[{{Donnan} {et~al.}(2024){Donnan}, {McLure}, {Dunlop}, {McLeod}, {Magee}, {Arellano-C{\'o}rdova}, {Barrufet}, {Begley}, {Bowler}, {Carnall}, {Cullen}, {Ellis}, {Fontana}, {Illingworth}, {Grogin}, {Hamadouche}, {Koekemoer}, {Liu}, {Mason}, {Santini}, \& {Stanton}}]{Donnan+2024}
{Donnan}, C.~T., {McLure}, R.~J., {Dunlop}, J.~S., {et~al.} 2024, arXiv e-prints, arXiv:2403.03171, \dodoi{10.48550/arXiv.2403.03171}

\bibitem[{{Duras} {et~al.}(2020){Duras}, {Bongiorno}, {Ricci}, {Piconcelli}, {Shankar}, {Lusso}, {Bianchi}, {Fiore}, {Maiolino}, {Marconi}, {Onori}, {Sani}, {Schneider}, {Vignali}, \& {La Franca}}]{Duras+2020}
{Duras}, F., {Bongiorno}, A., {Ricci}, F., {et~al.} 2020, \aap, 636, A73, \dodoi{10.1051/0004-6361/201936817}

\bibitem[{{Ewall-Wice} {et~al.}(2018){Ewall-Wice}, {Chang}, {Lazio}, {Dor{\'e}}, {Seiffert}, \& {Monsalve}}]{Ewall-Wice+2018}
{Ewall-Wice}, A., {Chang}, T.~C., {Lazio}, J., {et~al.} 2018, \apj, 868, 63, \dodoi{10.3847/1538-4357/aae51d}

\bibitem[{{Ewall-Wice} {et~al.}(2020){Ewall-Wice}, {Chang}, \& {Lazio}}]{Ewall-Wice2020}
{Ewall-Wice}, A., {Chang}, T.-C., \& {Lazio}, T. J.~W. 2020, \mnras, 492, 6086, \dodoi{10.1093/mnras/stz3501}

\bibitem[{{Fan} {et~al.}(2023){Fan}, {Ba{\~n}ados}, \& {Simcoe}}]{Fan+2023}
{Fan}, X., {Ba{\~n}ados}, E., \& {Simcoe}, R.~A. 2023, \araa, 61, 373, \dodoi{10.1146/annurev-astro-052920-102455}

\bibitem[{{Feng} \& {Holder}(2018)}]{Feng+2018}
{Feng}, C., \& {Holder}, G. 2018, \apjl, 858, L17, \dodoi{10.3847/2041-8213/aac0fe}

\bibitem[{{Ferrara}(2024)}]{Ferrara+2024a}
{Ferrara}, A. 2024, \aap, 684, A207, \dodoi{10.1051/0004-6361/202348321}

\bibitem[{{Ferrara} {et~al.}(2024){Ferrara}, {Carniani}, {di Mascia}, {Bouwens}, {Oesch}, \& {Schouws}}]{Ferrara+2024b}
{Ferrara}, A., {Carniani}, S., {di Mascia}, F., {et~al.} 2024, arXiv e-prints, arXiv:2409.17223, \dodoi{10.48550/arXiv.2409.17223}

\bibitem[{{Fialkov} \& {Barkana}(2019)}]{Fialkov+2019}
{Fialkov}, A., \& {Barkana}, R. 2019, \mnras, 486, 1763, \dodoi{10.1093/mnras/stz873}

\bibitem[{{Finkelstein} {et~al.}(2022){Finkelstein}, {Bagley}, {Arrabal Haro}, {Dickinson}, {Ferguson}, {Kartaltepe}, {Papovich}, {Burgarella}, {Kocevski}, {Huertas-Company}, {Iyer}, {Koekemoer}, {Larson}, {P{\'e}rez-Gonz{\'a}lez}, {Rose}, {Tacchella}, {Wilkins}, {Chworowsky}, {Medrano}, {Morales}, {Somerville}, {Yung}, {Fontana}, {Giavalisco}, {Grazian}, {Grogin}, {Kewley}, {Kirkpatrick}, {Kurczynski}, {Lotz}, {Pentericci}, {Pirzkal}, {Ravindranath}, {Ryan}, {Trump}, {Yang}, {Almaini}, {Amor{\'\i}n}, {Annunziatella}, {Backhaus}, {Barro}, {Behroozi}, {Bell}, {Bhatawdekar}, {Bisigello}, {Bromm}, {Buat}, {Buitrago}, {Calabr{\`o}}, {Casey}, {Castellano}, {Ch{\'a}vez Ortiz}, {Ciesla}, {Cleri}, {Cohen}, {Cole}, {Cooke}, {Cooper}, {Cooray}, {Costantin}, {Cox}, {Croton}, {Daddi}, {Dav{\'e}}, {de La Vega}, {Dekel}, {Elbaz}, {Estrada-Carpenter}, {Faber}, {Fern{\'a}ndez}, {Finkelstein}, {Freundlich}, {Fujimoto}, {Garc{\'\i}a-Argum{\'a}nez}, {Gardner}, {Gawiser}, {G{\'o}mez-Guijarro}, {Guo}, {Hamblin}, {Hamilton},
  {Hathi}, {Holwerda}, {Hirschmann}, {Hutchison}, {Jaskot}, {Jha}, {Jogee}, {Juneau}, {Jung}, {Kassin}, {Le Bail}, {Leung}, {Lucas}, {Magnelli}, {Mantha}, {Matharu}, {McGrath}, {McIntosh}, {Merlin}, {Mobasher}, {Newman}, {Nicholls}, {Pandya}, {Rafelski}, {Ronayne}, {Santini}, {Seill{\'e}}, {Shah}, {Shen}, {Simons}, {Snyder}, {Stanway}, {Straughn}, {Teplitz}, {Vanderhoof}, {Vega-Ferrero}, {Wang}, {Weiner}, {Willmer}, {Wuyts}, {Zavala}, \& {Ceers Team}}]{Finkelstein+2022}
{Finkelstein}, S.~L., {Bagley}, M.~B., {Arrabal Haro}, P., {et~al.} 2022, \apjl, 940, L55, \dodoi{10.3847/2041-8213/ac966e}

\bibitem[{{Furlanetto} {et~al.}(2006){Furlanetto}, {Oh}, \& {Briggs}}]{Furlanetto+2006}
{Furlanetto}, S.~R., {Oh}, S.~P., \& {Briggs}, F.~H. 2006, \physrep, 433, 181, \dodoi{10.1016/j.physrep.2006.08.002}

\bibitem[{{Furtak} {et~al.}(2024){Furtak}, {Labb{\'e}}, {Zitrin}, {Greene}, {Dayal}, {Chemerynska}, {Kokorev}, {Miller}, {Goulding}, {de Graaff}, {Bezanson}, {Brammer}, {Cutler}, {Leja}, {Pan}, {Price}, {Wang}, {Weaver}, {Whitaker}, {Atek}, {Bogd{\'a}n}, {Charlot}, {Curtis-Lake}, {van Dokkum}, {Endsley}, {Feldmann}, {Fudamoto}, {Fujimoto}, {Glazebrook}, {Juneau}, {Marchesini}, {Maseda}, {Nelson}, {Oesch}, {Plat}, {Setton}, {Stark}, \& {Williams}}]{Furtak+2024}
{Furtak}, L.~J., {Labb{\'e}}, I., {Zitrin}, A., {et~al.} 2024, \nat, 628, 57, \dodoi{10.1038/s41586-024-07184-8}

\bibitem[{{Gilfanov} {et~al.}(2004){Gilfanov}, {Grimm}, \& {Sunyaev}}]{Gilfanov+2004}
{Gilfanov}, M., {Grimm}, H.~J., \& {Sunyaev}, R. 2004, \mnras, 347, L57, \dodoi{10.1111/j.1365-2966.2004.07450.x}

\bibitem[{{Glover}(2015{\natexlab{a}})}]{Glover+2015_chemical_model}
{Glover}, S. C.~O. 2015{\natexlab{a}}, \mnras, 451, 2082, \dodoi{10.1093/mnras/stv1059}

\bibitem[{{Glover}(2015{\natexlab{b}})}]{Glover+2015_rate_coefficient}
---. 2015{\natexlab{b}}, \mnras, 453, 2901, \dodoi{10.1093/mnras/stv1781}

\bibitem[{{Glover}(2016)}]{Glover+2016}
---. 2016, arXiv e-prints, arXiv:1610.05679, \dodoi{10.48550/arXiv.1610.05679}

\bibitem[{{Greene} {et~al.}(2024){Greene}, {Labbe}, {Goulding}, {Furtak}, {Chemerynska}, {Kokorev}, {Dayal}, {Volonteri}, {Williams}, {Wang}, {Setton}, {Burgasser}, {Bezanson}, {Atek}, {Brammer}, {Cutler}, {Feldmann}, {Fujimoto}, {Glazebrook}, {de Graaff}, {Khullar}, {Leja}, {Marchesini}, {Maseda}, {Matthee}, {Miller}, {Naidu}, {Nanayakkara}, {Oesch}, {Pan}, {Papovich}, {Price}, {van Dokkum}, {Weaver}, {Whitaker}, \& {Zitrin}}]{Greene+2024}
{Greene}, J.~E., {Labbe}, I., {Goulding}, A.~D., {et~al.} 2024, \apj, 964, 39, \dodoi{10.3847/1538-4357/ad1e5f}

\bibitem[{{Grimm} {et~al.}(2003){Grimm}, {Gilfanov}, \& {Sunyaev}}]{Grimm+2003}
{Grimm}, H.~J., {Gilfanov}, M., \& {Sunyaev}, R. 2003, \mnras, 339, 793, \dodoi{10.1046/j.1365-8711.2003.06224.x}

\bibitem[{{Habouzit} {et~al.}(2016){Habouzit}, {Volonteri}, {Latif}, {Dubois}, \& {Peirani}}]{Habouzit+2016}
{Habouzit}, M., {Volonteri}, M., {Latif}, M., {Dubois}, Y., \& {Peirani}, S. 2016, \mnras, 463, 529, \dodoi{10.1093/mnras/stw1924}

\bibitem[{{Hainline} {et~al.}(2024){Hainline}, {Johnson}, {Robertson}, {Tacchella}, {Helton}, {Sun}, {Eisenstein}, {Simmonds}, {Topping}, {Whitler}, {Willmer}, {Rieke}, {Suess}, {Hviding}, {Cameron}, {Alberts}, {Baker}, {Baum}, {Bhatawdekar}, {Bonaventura}, {Boyett}, {Bunker}, {Carniani}, {Charlot}, {Chevallard}, {Chen}, {Curti}, {Curtis-Lake}, {D'Eugenio}, {Egami}, {Endsley}, {Hausen}, {Ji}, {Looser}, {Lyu}, {Maiolino}, {Nelson}, {Pusk{\'a}s}, {Rawle}, {Sandles}, {Saxena}, {Smit}, {Stark}, {Williams}, {Willott}, \& {Witstok}}]{Hainline+2024}
{Hainline}, K.~N., {Johnson}, B.~D., {Robertson}, B., {et~al.} 2024, \apj, 964, 71, \dodoi{10.3847/1538-4357/ad1ee4}

\bibitem[{{Harikane} {et~al.}(2024){Harikane}, {Nakajima}, {Ouchi}, {Umeda}, {Isobe}, {Ono}, {Xu}, \& {Zhang}}]{Harikane+2024}
{Harikane}, Y., {Nakajima}, K., {Ouchi}, M., {et~al.} 2024, \apj, 960, 56, \dodoi{10.3847/1538-4357/ad0b7e}

\bibitem[{{Harikane} {et~al.}(2023{\natexlab{a}}){Harikane}, {Zhang}, {Nakajima}, {Ouchi}, {Isobe}, {Ono}, {Hatano}, {Xu}, \& {Umeda}}]{Harikane+2023_AGN}
{Harikane}, Y., {Zhang}, Y., {Nakajima}, K., {et~al.} 2023{\natexlab{a}}, \apj, 959, 39, \dodoi{10.3847/1538-4357/ad029e}

\bibitem[{{Harikane} {et~al.}(2023{\natexlab{b}}){Harikane}, {Ouchi}, {Oguri}, {Ono}, {Nakajima}, {Isobe}, {Umeda}, {Mawatari}, \& {Zhang}}]{Harikane+2023_LF}
{Harikane}, Y., {Ouchi}, M., {Oguri}, M., {et~al.} 2023{\natexlab{b}}, \apjs, 265, 5, \dodoi{10.3847/1538-4365/acaaa9}

\bibitem[{{Hayes} {et~al.}(2024){Hayes}, {Tan}, {Ellis}, {Young}, {Cammelli}, {Singh}, {Runnholm}, {Saxena}, {Lunnan}, {Keller}, {Monaco}, {Laporte}, \& {Melinder}}]{Hayes+2024}
{Hayes}, M.~J., {Tan}, J.~C., {Ellis}, R.~S., {et~al.} 2024, arXiv e-prints, arXiv:2403.16138, \dodoi{10.48550/arXiv.2403.16138}

\bibitem[{{Hicks} {et~al.}(2024){Hicks}, {Norman}, {Wells}, \& {Bordner}}]{Hicks2024}
{Hicks}, W.~M., {Norman}, M.~L., {Wells}, A.~I., \& {Bordner}, J.~O. 2024, arXiv e-prints, arXiv:2407.20429, \dodoi{10.48550/arXiv.2407.20429}

\bibitem[{{Hirano} {et~al.}(2021){Hirano}, {Machida}, \& {Basu}}]{Hirano+2021}
{Hirano}, S., {Machida}, M.~N., \& {Basu}, S. 2021, \apj, 917, 34, \dodoi{10.3847/1538-4357/ac0913}

\bibitem[{{Hirano} {et~al.}(2023){Hirano}, {Machida}, \& {Basu}}]{Hirano+2023}
---. 2023, \apj, 952, 56, \dodoi{10.3847/1538-4357/acda94}

\bibitem[{{Hirata}(2006)}]{Hirata+2006}
{Hirata}, C.~M. 2006, \mnras, 367, 259, \dodoi{10.1111/j.1365-2966.2005.09949.x}

\bibitem[{{Hui} \& {Gnedin}(1997)}]{Hui+1997}
{Hui}, L., \& {Gnedin}, N.~Y. 1997, \mnras, 292, 27, \dodoi{10.1093/mnras/292.1.27}

\bibitem[{{Inayoshi} \& {Omukai}(2012)}]{Inayoshi_2012MNRAS.422.2539I}
{Inayoshi}, K., \& {Omukai}, K. 2012, \mnras, 422, 2539, \dodoi{10.1111/j.1365-2966.2012.20812.x}

\bibitem[{{Inayoshi} \& {Tanaka}(2015)}]{Inayoshi+2015}
{Inayoshi}, K., \& {Tanaka}, T.~L. 2015, \mnras, 450, 4350, \dodoi{10.1093/mnras/stv871}

\bibitem[{{Inayoshi} {et~al.}(2020){Inayoshi}, {Visbal}, \& {Haiman}}]{Inayoshi+2020}
{Inayoshi}, K., {Visbal}, E., \& {Haiman}, Z. 2020, \araa, 58, 27, \dodoi{10.1146/annurev-astro-120419-014455}

\bibitem[{{Inayoshi} {et~al.}(2015){Inayoshi}, {Visbal}, \& {Kashiyama}}]{Inayoshi_2015MNRAS.453.1692I}
{Inayoshi}, K., {Visbal}, E., \& {Kashiyama}, K. 2015, \mnras, 453, 1692, \dodoi{10.1093/mnras/stv1654}

\bibitem[{{Ivezi{\'c}} {et~al.}(2002){Ivezi{\'c}}, {Menou}, {Knapp}, {Strauss}, {Lupton}, {Vanden Berk}, {Richards}, {Tremonti}, {Weinstein}, {Anderson}, {Bahcall}, {Becker}, {Bernardi}, {Blanton}, {Eisenstein}, {Fan}, {Finkbeiner}, {Finlator}, {Frieman}, {Gunn}, {Hall}, {Kim}, {Kinkhabwala}, {Narayanan}, {Rockosi}, {Schlegel}, {Schneider}, {Strateva}, {SubbaRao}, {Thakar}, {Voges}, {White}, {Yanny}, {Brinkmann}, {Doi}, {Fukugita}, {Hennessy}, {Munn}, {Nichol}, \& {York}}]{Ivezic+2002}
{Ivezi{\'c}}, {\v{Z}}., {Menou}, K., {Knapp}, G.~R., {et~al.} 2002, \aj, 124, 2364, \dodoi{10.1086/344069}

\bibitem[{{Jiang} {et~al.}(2016){Jiang}, {McGreer}, {Fan}, {Strauss}, {Ba{\~n}ados}, {Becker}, {Bian}, {Farnsworth}, {Shen}, {Wang}, {Wang}, {Wang}, {White}, {Wu}, {Wu}, {Yang}, \& {Yang}}]{Jiang+2016}
{Jiang}, L., {McGreer}, I.~D., {Fan}, X., {et~al.} 2016, \apj, 833, 222, \dodoi{10.3847/1538-4357/833/2/222}

\bibitem[{{Jiang} {et~al.}(2022){Jiang}, {Ning}, {Fan}, {Ho}, {Luo}, {Wang}, {Wu}, {Wu}, {Yang}, \& {Zheng}}]{Jiang+2022}
{Jiang}, L., {Ning}, Y., {Fan}, X., {et~al.} 2022, Nature Astronomy, 6, 850, \dodoi{10.1038/s41550-022-01708-w}

\bibitem[{{Johnson} \& {Dijkstra}(2017)}]{Johnson+2017}
{Johnson}, J.~L., \& {Dijkstra}, M. 2017, \aap, 601, A138, \dodoi{10.1051/0004-6361/201630010}

\bibitem[{{Kocevski} {et~al.}(2023){Kocevski}, {Onoue}, {Inayoshi}, {Trump}, {Arrabal Haro}, {Grazian}, {Dickinson}, {Finkelstein}, {Kartaltepe}, {Hirschmann}, {Aird}, {Holwerda}, {Fujimoto}, {Juneau}, {Amor{\'\i}n}, {Backhaus}, {Bagley}, {Barro}, {Bell}, {Bisigello}, {Calabr{\`o}}, {Cleri}, {Cooper}, {Ding}, {Grogin}, {Ho}, {Hutchison}, {Inoue}, {Jiang}, {Jones}, {Koekemoer}, {Li}, {Li}, {McGrath}, {Molina}, {Papovich}, {P{\'e}rez-Gonz{\'a}lez}, {Pirzkal}, {Wilkins}, {Yang}, \& {Yung}}]{Kocevski+2023}
{Kocevski}, D.~D., {Onoue}, M., {Inayoshi}, K., {et~al.} 2023, \apjl, 954, L4, \dodoi{10.3847/2041-8213/ace5a0}

\bibitem[{{Kocevski} {et~al.}(2024){Kocevski}, {Finkelstein}, {Barro}, {Taylor}, {Calabr{\`o}}, {Laloux}, {Buchner}, {Trump}, {Leung}, {Yang}, {Dickinson}, {P{\'e}rez-Gonz{\'a}lez}, {Pacucci}, {Inayoshi}, {Somerville}, {McGrath}, {Akins}, {Bagley}, {Bisigello}, {Bowler}, {Carnall}, {Casey}, {Cheng}, {Cleri}, {Costantin}, {Cullen}, {Davis}, {Donnan}, {Dunlop}, {Ellis}, {Ferguson}, {Fujimoto}, {Fontana}, {Giavalisco}, {Grazian}, {Grogin}, {Hathi}, {Hirschmann}, {Huertas-Company}, {Holwerda}, {Illingworth}, {Juneau}, {Kartaltepe}, {Koekemoer}, {Li}, {Lucas}, {Magee}, {Mason}, {McLeod}, {McLure}, {Napolitano}, {Papovich}, {Pirzkal}, {Rodighiero}, {Santini}, {Wilkins}, \& {Yung}}]{Kocevski+2024}
{Kocevski}, D.~D., {Finkelstein}, S.~L., {Barro}, G., {et~al.} 2024, arXiv e-prints, arXiv:2404.03576, \dodoi{10.48550/arXiv.2404.03576}

\bibitem[{{Kreckel} {et~al.}(2010){Kreckel}, {Bruhns}, {{\v{C}}{\'\i}{\v{z}}ek}, {Glover}, {Miller}, {Urbain}, \& {Savin}}]{Kreckel+2010}
{Kreckel}, H., {Bruhns}, H., {{\v{C}}{\'\i}{\v{z}}ek}, M., {et~al.} 2010, Science, 329, 69, \dodoi{10.1126/science.1187191}

\bibitem[{{Kulkarni} {et~al.}(2021){Kulkarni}, {Visbal}, \& {Bryan}}]{Kulkarni2021ApJ}
{Kulkarni}, M., {Visbal}, E., \& {Bryan}, G.~L. 2021, \apj, 917, 40, \dodoi{10.3847/1538-4357/ac08a3}

\bibitem[{{Latif} {et~al.}(2015){Latif}, {Bovino}, {Grassi}, {Schleicher}, \& {Spaans}}]{Latif+2015}
{Latif}, M.~A., {Bovino}, S., {Grassi}, T., {Schleicher}, D.~R.~G., \& {Spaans}, M. 2015, \mnras, 446, 3163, \dodoi{10.1093/mnras/stu2244}

\bibitem[{{Latif} {et~al.}(2014){Latif}, {Bovino}, {Van Borm}, {Grassi}, {Schleicher}, \& {Spaans}}]{Latif+2014}
{Latif}, M.~A., {Bovino}, S., {Van Borm}, C., {et~al.} 2014, \mnras, 443, 1979, \dodoi{10.1093/mnras/stu1230}

\bibitem[{{Latif} \& {Schleicher}(2023)}]{Latif2023ApJ...952L...9L}
{Latif}, M.~A., \& {Schleicher}, D. R.~G. 2023, \apjl, 952, L9, \dodoi{10.3847/2041-8213/ace34f}

\bibitem[{{Latif} {et~al.}(2023){Latif}, {Schleicher}, \& {Khochfar}}]{Latif2023ApJ...945..137L}
{Latif}, M.~A., {Schleicher}, D. R.~G., \& {Khochfar}, S. 2023, \apj, 945, 137, \dodoi{10.3847/1538-4357/acbcc2}

\bibitem[{{Latif} {et~al.}(2013){Latif}, {Schleicher}, {Schmidt}, \& {Niemeyer}}]{Latif2013MNRAS}
{Latif}, M.~A., {Schleicher}, D.~R.~G., {Schmidt}, W., \& {Niemeyer}, J. 2013, \mnras, 433, 1607, \dodoi{10.1093/mnras/stt834}

\bibitem[{{Latif} {et~al.}(2022){Latif}, {Whalen}, {Khochfar}, {Herrington}, \& {Woods}}]{Latif+2022}
{Latif}, M.~A., {Whalen}, D.~J., {Khochfar}, S., {Herrington}, N.~P., \& {Woods}, T.~E. 2022, \nat, 607, 48, \dodoi{10.1038/s41586-022-04813-y}

\bibitem[{{Leitherer} {et~al.}(2010){Leitherer}, {Ortiz Ot{\'a}lvaro}, {Bresolin}, {Kudritzki}, {Lo Faro}, {Pauldrach}, {Pettini}, \& {Rix}}]{Leitherer+2010}
{Leitherer}, C., {Ortiz Ot{\'a}lvaro}, P.~A., {Bresolin}, F., {et~al.} 2010, \apjs, 189, 309, \dodoi{10.1088/0067-0049/189/2/309}

\bibitem[{{Leitherer} {et~al.}(1999){Leitherer}, {Schaerer}, {Goldader}, {Delgado}, {Robert}, {Kune}, {de Mello}, {Devost}, \& {Heckman}}]{Leitherer+1999}
{Leitherer}, C., {Schaerer}, D., {Goldader}, J.~D., {et~al.} 1999, \apjs, 123, 3, \dodoi{10.1086/313233}

\bibitem[{{Lenzuni} {et~al.}(1991){Lenzuni}, {Chernoff}, \& {Salpeter}}]{Lenzuni+1991}
{Lenzuni}, P., {Chernoff}, D.~F., \& {Salpeter}, E.~E. 1991, \apjs, 76, 759, \dodoi{10.1086/191580}

\bibitem[{{Luo} {et~al.}(2020){Luo}, {Shlosman}, {Nagamine}, \& {Fang}}]{Luo+2020}
{Luo}, Y., {Shlosman}, I., {Nagamine}, K., \& {Fang}, T. 2020, \mnras, 492, 4917, \dodoi{10.1093/mnras/staa153}

\bibitem[{{Maiolino} {et~al.}(2023){Maiolino}, {Scholtz}, {Curtis-Lake}, {Carniani}, {Baker}, {de Graaff}, {Tacchella}, {{\"U}bler}, {D'Eugenio}, {Witstok}, {Curti}, {Arribas}, {Bunker}, {Charlot}, {Chevallard}, {Eisenstein}, {Egami}, {Ji}, {Jones}, {Lyu}, {Rawle}, {Robertson}, {Rujopakarn}, {Perna}, {Sun}, {Venturi}, {Williams}, \& {Willott}}]{Maiolino+2023}
{Maiolino}, R., {Scholtz}, J., {Curtis-Lake}, E., {et~al.} 2023, arXiv e-prints, arXiv:2308.01230, \dodoi{10.48550/arXiv.2308.01230}

\bibitem[{{Maselli} {et~al.}(2003){Maselli}, {Ferrara}, \& {Ciardi}}]{Maselli+2003}
{Maselli}, A., {Ferrara}, A., \& {Ciardi}, B. 2003, \mnras, 345, 379, \dodoi{10.1046/j.1365-8711.2003.06979.x}

\bibitem[{{Matsuoka} {et~al.}(2023){Matsuoka}, {Onoue}, {Iwasawa}, {Strauss}, {Kashikawa}, {Izumi}, {Nagao}, {Imanishi}, {Akiyama}, {Silverman}, {Asami}, {Bosch}, {Furusawa}, {Goto}, {Gunn}, {Harikane}, {Ikeda}, {Inayoshi}, {Ishimoto}, {Kawaguchi}, {Kikuta}, {Kohno}, {Komiyama}, {Lee}, {Lupton}, {Minezaki}, {Miyazaki}, {Murayama}, {Nishizawa}, {Oguri}, {Ono}, {Oogi}, {Ouchi}, {Price}, {Sameshima}, {Sugiyama}, {Tait}, {Takada}, {Takahashi}, {Takata}, {Tanaka}, {Toba}, {Wang}, \& {Yamashita}}]{Matsuoka+2023}
{Matsuoka}, Y., {Onoue}, M., {Iwasawa}, K., {et~al.} 2023, \apjl, 949, L42, \dodoi{10.3847/2041-8213/acd69f}

\bibitem[{{Matthee} {et~al.}(2024){Matthee}, {Naidu}, {Brammer}, {Chisholm}, {Eilers}, {Goulding}, {Greene}, {Kashino}, {Labbe}, {Lilly}, {Mackenzie}, {Oesch}, {Weibel}, {Wuyts}, {Xiao}, {Bordoloi}, {Bouwens}, {van Dokkum}, {Illingworth}, {Kramarenko}, {Maseda}, {Mason}, {Meyer}, {Nelson}, {Reddy}, {Shivaei}, {Simcoe}, \& {Yue}}]{Matthee+2024}
{Matthee}, J., {Naidu}, R.~P., {Brammer}, G., {et~al.} 2024, \apj, 963, 129, \dodoi{10.3847/1538-4357/ad2345}

\bibitem[{{Mayer} {et~al.}(2024){Mayer}, {Capelo}, {Zwick}, \& {Di Matteo}}]{Mayer_2024ApJ...961...76M}
{Mayer}, L., {Capelo}, P.~R., {Zwick}, L., \& {Di Matteo}, T. 2024, \apj, 961, 76, \dodoi{10.3847/1538-4357/ad11cf}

\bibitem[{{Mayer} {et~al.}(2015){Mayer}, {Fiacconi}, {Bonoli}, {Quinn}, {Ro{\v{s}}kar}, {Shen}, \& {Wadsley}}]{Mayer+2015}
{Mayer}, L., {Fiacconi}, D., {Bonoli}, S., {et~al.} 2015, \apj, 810, 51, \dodoi{10.1088/0004-637X/810/1/51}

\bibitem[{{Mayer} {et~al.}(2010){Mayer}, {Kazantzidis}, {Escala}, \& {Callegari}}]{Mayer+2010}
{Mayer}, L., {Kazantzidis}, S., {Escala}, A., \& {Callegari}, S. 2010, \nat, 466, 1082, \dodoi{10.1038/nature09294}

\bibitem[{{McBride} {et~al.}(2009){McBride}, {Fakhouri}, \& {Ma}}]{McBride2009MNRAS}
{McBride}, J., {Fakhouri}, O., \& {Ma}, C.-P. 2009, \mnras, 398, 1858, \dodoi{10.1111/j.1365-2966.2009.15329.x}

\bibitem[{{McLeod} {et~al.}(2024){McLeod}, {Donnan}, {McLure}, {Dunlop}, {Magee}, {Begley}, {Carnall}, {Cullen}, {Ellis}, {Hamadouche}, \& {Stanton}}]{McLeod+2024}
{McLeod}, D.~J., {Donnan}, C.~T., {McLure}, R.~J., {et~al.} 2024, \mnras, 527, 5004, \dodoi{10.1093/mnras/stad3471}

\bibitem[{{McQuinn}(2016)}]{McQuinn+2016}
{McQuinn}, M. 2016, \araa, 54, 313, \dodoi{10.1146/annurev-astro-082214-122355}

\bibitem[{{Mesinger}(2019)}]{Mesinger+2019}
{Mesinger}, A. 2019, {The Cosmic 21-cm Revolution; Charting the first billion years of our universe}, \dodoi{10.1088/2514-3433/ab4a73}

\bibitem[{{Mesinger} {et~al.}(2011){Mesinger}, {Furlanetto}, \& {Cen}}]{Mesinger+2011}
{Mesinger}, A., {Furlanetto}, S., \& {Cen}, R. 2011, \mnras, 411, 955, \dodoi{10.1111/j.1365-2966.2010.17731.x}

\bibitem[{{Mineo} {et~al.}(2012){Mineo}, {Gilfanov}, \& {Sunyaev}}]{Mineo+2012}
{Mineo}, S., {Gilfanov}, M., \& {Sunyaev}, R. 2012, \mnras, 419, 2095, \dodoi{10.1111/j.1365-2966.2011.19862.x}

\bibitem[{{Mirabel} \& {Rodr{\'\i}guez}(2022)}]{Mirabel+2022}
{Mirabel}, I.~F., \& {Rodr{\'\i}guez}, L.~F. 2022, \nar, 94, 101642, \dodoi{10.1016/j.newar.2022.101642}

\bibitem[{{Mirocha} \& {Furlanetto}(2019)}]{Mirocha+2019}
{Mirocha}, J., \& {Furlanetto}, S.~R. 2019, \mnras, 483, 1980, \dodoi{10.1093/mnras/sty3260}

\bibitem[{{Monsalve} {et~al.}(2017){Monsalve}, {Rogers}, {Bowman}, \& {Mozdzen}}]{Monsalve+2017}
{Monsalve}, R.~A., {Rogers}, A. E.~E., {Bowman}, J.~D., \& {Mozdzen}, T.~J. 2017, \apj, 847, 64, \dodoi{10.3847/1538-4357/aa88d1}

\bibitem[{{Morales} \& {Wyithe}(2010)}]{Morales+2010}
{Morales}, M.~F., \& {Wyithe}, J. S.~B. 2010, \araa, 48, 127, \dodoi{10.1146/annurev-astro-081309-130936}

\bibitem[{{Nabizadeh} {et~al.}(2024){Nabizadeh}, {Zackrisson}, {Pacucci}, {Peter Maksym}, {Li}, {Civano}, {Cohen}, {D'Silva}, {Koekemoer}, {Summers}, {Windhorst}, {Adams}, {Conselice}, {Coe}, {Driver}, {Frye}, {Grogin}, {Jansen}, {Marshall}, {Nonino}, {Pirzkal}, {Robotham}, {Rutkowski}, {Ryan}, {Tompkins}, {Willmer}, {Yan}, {Diego}, {Cheng}, {Finkelstein}, {Willner}, {Wang}, {Zitrin}, {Smith}, {Bhatawdekar}, \& {Gim}}]{Nabizadeh+2024}
{Nabizadeh}, A., {Zackrisson}, E., {Pacucci}, F., {et~al.} 2024, \aap, 683, A58, \dodoi{10.1051/0004-6361/202347724}

\bibitem[{{Naidu} {et~al.}(2022){Naidu}, {Oesch}, {van Dokkum}, {Nelson}, {Suess}, {Brammer}, {Whitaker}, {Illingworth}, {Bouwens}, {Tacchella}, {Matthee}, {Allen}, {Bezanson}, {Conroy}, {Labbe}, {Leja}, {Leonova}, {Magee}, {Price}, {Setton}, {Strait}, {Stefanon}, {Toft}, {Weaver}, \& {Weibel}}]{Naidu+2022}
{Naidu}, R.~P., {Oesch}, P.~A., {van Dokkum}, P., {et~al.} 2022, \apjl, 940, L14, \dodoi{10.3847/2041-8213/ac9b22}

\bibitem[{{Oesch} {et~al.}(2018){Oesch}, {Bouwens}, {Illingworth}, {Labb{\'e}}, \& {Stefanon}}]{Oesch+2018}
{Oesch}, P.~A., {Bouwens}, R.~J., {Illingworth}, G.~D., {Labb{\'e}}, I., \& {Stefanon}, M. 2018, \apj, 855, 105, \dodoi{10.3847/1538-4357/aab03f}

\bibitem[{{Oh} \& {Haiman}(2002)}]{Oh2002ApJ}
{Oh}, S.~P., \& {Haiman}, Z. 2002, \apj, 569, 558, \dodoi{10.1086/339393}

\bibitem[{{Omukai}(2000)}]{Omukai+2000}
{Omukai}, K. 2000, \apj, 534, 809, \dodoi{10.1086/308776}

\bibitem[{{Omukai}(2001)}]{Omukai+2001}
---. 2001, \apj, 546, 635, \dodoi{10.1086/318296}

\bibitem[{{Omukai} {et~al.}(2008){Omukai}, {Schneider}, \& {Haiman}}]{Omukai+2008}
{Omukai}, K., {Schneider}, R., \& {Haiman}, Z. 2008, \apj, 686, 801, \dodoi{10.1086/591636}

\bibitem[{{Onoue} {et~al.}(2023){Onoue}, {Inayoshi}, {Ding}, {Li}, {Li}, {Molina}, {Inoue}, {Jiang}, \& {Ho}}]{Onoue+2023}
{Onoue}, M., {Inayoshi}, K., {Ding}, X., {et~al.} 2023, \apjl, 942, L17, \dodoi{10.3847/2041-8213/aca9d3}

\bibitem[{{O'Shea} {et~al.}(2015){O'Shea}, {Wise}, {Xu}, \& {Norman}}]{O'Shea+2015}
{O'Shea}, B.~W., {Wise}, J.~H., {Xu}, H., \& {Norman}, M.~L. 2015, \apjl, 807, L12, \dodoi{10.1088/2041-8205/807/1/L12}

\bibitem[{{Pacucci} \& {Ferrara}(2015)}]{Pacucci+2015_simulation}
{Pacucci}, F., \& {Ferrara}, A. 2015, \mnras, 448, 104, \dodoi{10.1093/mnras/stv018}

\bibitem[{{Pacucci} {et~al.}(2015){Pacucci}, {Ferrara}, {Volonteri}, \& {Dubus}}]{Pacucci+2015_spectral}
{Pacucci}, F., {Ferrara}, A., {Volonteri}, M., \& {Dubus}, G. 2015, \mnras, 454, 3771, \dodoi{10.1093/mnras/stv2196}

\bibitem[{{Pacucci} {et~al.}(2014){Pacucci}, {Mesinger}, {Mineo}, \& {Ferrara}}]{Pacucci+2014}
{Pacucci}, F., {Mesinger}, A., {Mineo}, S., \& {Ferrara}, A. 2014, \mnras, 443, 678, \dodoi{10.1093/mnras/stu1240}

\bibitem[{{P{\'e}rez-Gonz{\'a}lez} {et~al.}(2023){P{\'e}rez-Gonz{\'a}lez}, {Costantin}, {Langeroodi}, {Rinaldi}, {Annunziatella}, {Ilbert}, {Colina}, {N{\o}rgaard-Nielsen}, {Greve}, {{\"O}stlin}, {Wright}, {Alonso-Herrero}, {{\'A}lvarez-M{\'a}rquez}, {Caputi}, {Eckart}, {Le F{\`e}vre}, {Labiano}, {Garc{\'\i}a-Mar{\'\i}n}, {Hjorth}, {Kendrew}, {Pye}, {Tikkanen}, {van der Werf}, {Walter}, {Ward}, {Bik}, {Boogaard}, {Bosman}, {G{\'o}mez}, {Gillman}, {Iani}, {Jermann}, {Melinder}, {Meyer}, {Moutard}, {van Dishoek}, {Henning}, {Lagage}, {Guedel}, {Peissker}, {Ray}, {Vandenbussche}, {Garc{\'\i}a-Argum{\'a}nez}, \& {Mar{\'\i}a M{\'e}rida}}]{Perez-Gonzalez2023ApJ}
{P{\'e}rez-Gonz{\'a}lez}, P.~G., {Costantin}, L., {Langeroodi}, D., {et~al.} 2023, \apjl, 951, L1, \dodoi{10.3847/2041-8213/acd9d0}

\bibitem[{{Planck Collaboration} {et~al.}(2016){Planck Collaboration}, {Ade}, {Aghanim}, {Arnaud}, {Ashdown}, {Aumont}, {Baccigalupi}, {Banday}, {Barreiro}, {Bartlett}, {Bartolo}, {Battaner}, {Battye}, {Benabed}, {Beno{\^\i}t}, {Benoit-L{\'e}vy}, {Bernard}, {Bersanelli}, {Bielewicz}, {Bock}, {Bonaldi}, {Bonavera}, {Bond}, {Borrill}, {Bouchet}, {Boulanger}, {Bucher}, {Burigana}, {Butler}, {Calabrese}, {Cardoso}, {Catalano}, {Challinor}, {Chamballu}, {Chary}, {Chiang}, {Chluba}, {Christensen}, {Church}, {Clements}, {Colombi}, {Colombo}, {Combet}, {Coulais}, {Crill}, {Curto}, {Cuttaia}, {Danese}, {Davies}, {Davis}, {de Bernardis}, {de Rosa}, {de Zotti}, {Delabrouille}, {D{\'e}sert}, {Di Valentino}, {Dickinson}, {Diego}, {Dolag}, {Dole}, {Donzelli}, {Dor{\'e}}, {Douspis}, {Ducout}, {Dunkley}, {Dupac}, {Efstathiou}, {Elsner}, {En{\ss}lin}, {Eriksen}, {Farhang}, {Fergusson}, {Finelli}, {Forni}, {Frailis}, {Fraisse}, {Franceschi}, {Frejsel}, {Galeotta}, {Galli}, {Ganga}, {Gauthier}, {Gerbino}, {Ghosh}, {Giard},
  {Giraud-H{\'e}raud}, {Giusarma}, {Gjerl{\o}w}, {Gonz{\'a}lez-Nuevo}, {G{\'o}rski}, {Gratton}, {Gregorio}, {Gruppuso}, {Gudmundsson}, {Hamann}, {Hansen}, {Hanson}, {Harrison}, {Helou}, {Henrot-Versill{\'e}}, {Hern{\'a}ndez-Monteagudo}, {Herranz}, {Hildebrandt}, {Hivon}, {Hobson}, {Holmes}, {Hornstrup}, {Hovest}, {Huang}, {Huffenberger}, {Hurier}, {Jaffe}, {Jaffe}, {Jones}, {Juvela}, {Keih{\"a}nen}, {Keskitalo}, {Kisner}, {Kneissl}, {Knoche}, {Knox}, {Kunz}, {Kurki-Suonio}, {Lagache}, {L{\"a}hteenm{\"a}ki}, {Lamarre}, {Lasenby}, {Lattanzi}, {Lawrence}, {Leahy}, {Leonardi}, {Lesgourgues}, {Levrier}, {Lewis}, {Liguori}, {Lilje}, {Linden-V{\o}rnle}, {L{\'o}pez-Caniego}, {Lubin}, {Mac{\'\i}as-P{\'e}rez}, {Maggio}, {Maino}, {Mandolesi}, {Mangilli}, {Marchini}, {Maris}, {Martin}, {Martinelli}, {Mart{\'\i}nez-Gonz{\'a}lez}, {Masi}, {Matarrese}, {McGehee}, {Meinhold}, {Melchiorri}, {Melin}, {Mendes}, {Mennella}, {Migliaccio}, {Millea}, {Mitra}, {Miville-Desch{\^e}nes}, {Moneti}, {Montier}, {Morgante}, {Mortlock},
  {Moss}, {Munshi}, {Murphy}, {Naselsky}, {Nati}, {Natoli}, {Netterfield}, {N{\o}rgaard-Nielsen}, {Noviello}, {Novikov}, {Novikov}, {Oxborrow}, {Paci}, {Pagano}, {Pajot}, {Paladini}, {Paoletti}, {Partridge}, {Pasian}, {Patanchon}, {Pearson}, {Perdereau}, {Perotto}, {Perrotta}, {Pettorino}, {Piacentini}, {Piat}, {Pierpaoli}, {Pietrobon}, {Plaszczynski}, {Pointecouteau}, {Polenta}, {Popa}, {Pratt}, {Pr{\'e}zeau}, {Prunet}, {Puget}, {Rachen}, {Reach}, {Rebolo}, {Reinecke}, {Remazeilles}, {Renault}, {Renzi}, {Ristorcelli}, {Rocha}, {Rosset}, {Rossetti}, {Roudier}, {Rouill{\'e} d'Orfeuil}, {Rowan-Robinson}, {Rubi{\~n}o-Mart{\'\i}n}, {Rusholme}, {Said}, {Salvatelli}, {Salvati}, {Sandri}, {Santos}, {Savelainen}, {Savini}, {Scott}, {Seiffert}, {Serra}, {Shellard}, {Spencer}, {Spinelli}, {Stolyarov}, {Stompor}, {Sudiwala}, {Sunyaev}, {Sutton}, {Suur-Uski}, {Sygnet}, {Tauber}, {Terenzi}, {Toffolatti}, {Tomasi}, {Tristram}, {Trombetti}, {Tucci}, {Tuovinen}, {T{\"u}rler}, {Umana}, {Valenziano}, {Valiviita}, {Van Tent},
  {Vielva}, {Villa}, {Wade}, {Wandelt}, {Wehus}, {White}, {White}, {Wilkinson}, {Yvon}, {Zacchei}, \& {Zonca}}]{Planck2016}
{Planck Collaboration}, {Ade}, P.~A.~R., {Aghanim}, N., {et~al.} 2016, \aap, 594, A13, \dodoi{10.1051/0004-6361/201525830}

\bibitem[{{Pritchard} \& {Loeb}(2012)}]{Pritchard+2012}
{Pritchard}, J.~R., \& {Loeb}, A. 2012, Reports on Progress in Physics, 75, 086901, \dodoi{10.1088/0034-4885/75/8/086901}

\bibitem[{{Ranalli} {et~al.}(2003){Ranalli}, {Comastri}, \& {Setti}}]{Ranalli+2003}
{Ranalli}, P., {Comastri}, A., \& {Setti}, G. 2003, \aap, 399, 39, \dodoi{10.1051/0004-6361:20021600}

\bibitem[{{Regan} {et~al.}(2014){Regan}, {Johansson}, \& {Wise}}]{Regan+2014}
{Regan}, J.~A., {Johansson}, P.~H., \& {Wise}, J.~H. 2014, \apj, 795, 137, \dodoi{10.1088/0004-637X/795/2/137}

\bibitem[{{Regan} {et~al.}(2016){Regan}, {Johansson}, \& {Wise}}]{Regan+2016}
---. 2016, \mnras, 461, 111, \dodoi{10.1093/mnras/stw1307}

\bibitem[{{Reis} {et~al.}(2021){Reis}, {Fialkov}, \& {Barkana}}]{Reis+2021}
{Reis}, I., {Fialkov}, A., \& {Barkana}, R. 2021, \mnras, 506, 5479, \dodoi{10.1093/mnras/stab2089}

\bibitem[{{Schaerer}(2003)}]{Schaerer+2003}
{Schaerer}, D. 2003, \aap, 397, 527, \dodoi{10.1051/0004-6361:20021525}

\bibitem[{{Schauer} {et~al.}(2017){Schauer}, {Regan}, {Glover}, \& {Klessen}}]{Schauer+2017}
{Schauer}, A. T.~P., {Regan}, J., {Glover}, S. C.~O., \& {Klessen}, R.~S. 2017, \mnras, 471, 4878, \dodoi{10.1093/mnras/stx1915}

\bibitem[{{Scholtz} {et~al.}(2023){Scholtz}, {Maiolino}, {D'Eugenio}, {Curtis-Lake}, {Carniani}, {Charlot}, {Curti}, {Silcock}, {Arribas}, {Baker}, {Bhatawdekar}, {Boyett}, {Bunker}, {Chevallard}, {Circosta}, {Eisenstein}, {Hainline}, {Hausen}, {Ji}, {Ji}, {Johnson}, {Kumari}, {Looser}, {Lyu}, {Maseda}, {Parlanti}, {Perna}, {Rieke}, {Robertson}, {Rodr{\'\i}guez Del Pino}, {Sun}, {Tacchella}, {{\"U}bler}, {Venturi}, {Williams}, {Willmer}, {Willott}, \& {Witstok}}]{Scholtz+2023}
{Scholtz}, J., {Maiolino}, R., {D'Eugenio}, F., {et~al.} 2023, arXiv e-prints, arXiv:2311.18731, \dodoi{10.48550/arXiv.2311.18731}

\bibitem[{{Shang} {et~al.}(2010){Shang}, {Bryan}, \& {Haiman}}]{Shang+2010}
{Shang}, C., {Bryan}, G.~L., \& {Haiman}, Z. 2010, \mnras, 402, 1249, \dodoi{10.1111/j.1365-2966.2009.15960.x}

\bibitem[{{Sheth} \& {Tormen}(2002)}]{Sheth+2002}
{Sheth}, R.~K., \& {Tormen}, G. 2002, \mnras, 329, 61, \dodoi{10.1046/j.1365-8711.2002.04950.x}

\bibitem[{{Singh} {et~al.}(2018){Singh}, {Subrahmanyan}, {Udaya Shankar}, {Sathyanarayana Rao}, {Fialkov}, {Cohen}, {Barkana}, {Girish}, {Raghunathan}, {Somashekar}, \& {Srivani}}]{Singh+2018}
{Singh}, S., {Subrahmanyan}, R., {Udaya Shankar}, N., {et~al.} 2018, \apj, 858, 54, \dodoi{10.3847/1538-4357/aabae1}

\bibitem[{{Singh} {et~al.}(2022){Singh}, {Jishnu}, {Subrahmanyan}, {Udaya Shankar}, {Girish}, {Raghunathan}, {Somashekar}, {Srivani}, \& {Sathyanarayana Rao}}]{Singh+2022}
{Singh}, S., {Jishnu}, N.~T., {Subrahmanyan}, R., {et~al.} 2022, Nature Astronomy, 6, 607, \dodoi{10.1038/s41550-022-01610-5}

\bibitem[{{Stenrup} {et~al.}(2009){Stenrup}, {Larson}, \& {Elander}}]{Stenrup+2009}
{Stenrup}, M., {Larson}, {\r{A}}., \& {Elander}, N. 2009, \pra, 79, 012713, \dodoi{10.1103/PhysRevA.79.012713}

\bibitem[{{Suazo} {et~al.}(2019){Suazo}, {Prieto}, {Escala}, \& {Schleicher}}]{Suazo+2019}
{Suazo}, M., {Prieto}, J., {Escala}, A., \& {Schleicher}, D. R.~G. 2019, \apj, 885, 127, \dodoi{10.3847/1538-4357/ab45eb}

\bibitem[{{Sugimura} {et~al.}(2014){Sugimura}, {Omukai}, \& {Inoue}}]{Sugimura+2014}
{Sugimura}, K., {Omukai}, K., \& {Inoue}, A.~K. 2014, \mnras, 445, 544, \dodoi{10.1093/mnras/stu1778}

\bibitem[{{Tripodi} {et~al.}(2024){Tripodi}, {Martis}, {Markov}, {Brada{\v{c}}}, {Di Mascia}, {Cammelli}, {D'Eugenio}, {Willott}, {Curti}, {Bhatt}, {Gallerani}, {Rihtar{\v{s}}i{\v{c}}}, {Singh}, {Gaspar}, {Harshan}, {Jude{\v{z}}}, {Merida}, {Desprez}, {Sawicki}, {Goovaerts}, {Muzzin}, {Noirot}, {Sarrouh}, {Abraham}, {Asada}, {Brammer}, {Estrada Carpenter}, {Felicioni}, {Fujimoto}, {Iyer}, {Mowla}, \& {Strait}}]{Tripodi+2024}
{Tripodi}, R., {Martis}, N., {Markov}, V., {et~al.} 2024, arXiv e-prints, arXiv:2412.04983, \dodoi{10.48550/arXiv.2412.04983}

\bibitem[{{Vald{\'e}s} \& {Ferrara}(2008)}]{Valdes+2008}
{Vald{\'e}s}, M., \& {Ferrara}, A. 2008, \mnras, 387, L8, \dodoi{10.1111/j.1745-3933.2008.00471.x}

\bibitem[{{V{\'a}zquez} \& {Leitherer}(2005)}]{Vazquez+2005}
{V{\'a}zquez}, G.~A., \& {Leitherer}, C. 2005, \apj, 621, 695, \dodoi{10.1086/427866}

\bibitem[{{Verner} {et~al.}(1996){Verner}, {Ferland}, {Korista}, \& {Yakovlev}}]{Verner+1996}
{Verner}, D.~A., {Ferland}, G.~J., {Korista}, K.~T., \& {Yakovlev}, D.~G. 1996, \apj, 465, 487, \dodoi{10.1086/177435}

\bibitem[{{Visbal} {et~al.}(2014{\natexlab{a}}){Visbal}, {Haiman}, \& {Bryan}}]{Visbal+2014_DCBH}
{Visbal}, E., {Haiman}, Z., \& {Bryan}, G.~L. 2014{\natexlab{a}}, \mnras, 445, 1056, \dodoi{10.1093/mnras/stu1794}

\bibitem[{{Visbal} {et~al.}(2014{\natexlab{b}}){Visbal}, {Haiman}, {Terrazas}, {Bryan}, \& {Barkana}}]{Visbal+2014_star_formation}
{Visbal}, E., {Haiman}, Z., {Terrazas}, B., {Bryan}, G.~L., \& {Barkana}, R. 2014{\natexlab{b}}, \mnras, 445, 107, \dodoi{10.1093/mnras/stu1710}

\bibitem[{{Vogelsberger} {et~al.}(2019){Vogelsberger}, {McKinnon}, {O'Neil}, {Marinacci}, {Torrey}, \& {Kannan}}]{Vogelsberger2019}
{Vogelsberger}, M., {McKinnon}, R., {O'Neil}, S., {et~al.} 2019, \mnras, 487, 4870, \dodoi{10.1093/mnras/stz1644}

\bibitem[{{Wang} {et~al.}(2023){Wang}, {Fujimoto}, {Labb{\'e}}, {Furtak}, {Miller}, {Setton}, {Zitrin}, {Atek}, {Bezanson}, {Brammer}, {Leja}, {Oesch}, {Price}, {Chemerynska}, {Cutler}, {Dayal}, {van Dokkum}, {Goulding}, {Greene}, {Fudamoto}, {Khullar}, {Kokorev}, {Marchesini}, {Pan}, {Weaver}, {Whitaker}, \& {Williams}}]{Wang+2023}
{Wang}, B., {Fujimoto}, S., {Labb{\'e}}, I., {et~al.} 2023, \apjl, 957, L34, \dodoi{10.3847/2041-8213/acfe07}

\bibitem[{{Wang} {et~al.}(2006){Wang}, {Wu}, \& {Kong}}]{Wang+2006}
{Wang}, R., {Wu}, X.-B., \& {Kong}, M.-Z. 2006, \apj, 645, 890, \dodoi{10.1086/504401}

\bibitem[{{Whalen} {et~al.}(2021){Whalen}, {Mezcua}, {Patrick}, {Meiksin}, \& {Latif}}]{Whalen2021ApJ}
{Whalen}, D.~J., {Mezcua}, M., {Patrick}, S.~J., {Meiksin}, A., \& {Latif}, M.~A. 2021, \apjl, 922, L39, \dodoi{10.3847/2041-8213/ac35e6}

\bibitem[{{Williams} {et~al.}(2018){Williams}, {Curtis-Lake}, {Hainline}, {Chevallard}, {Robertson}, {Charlot}, {Endsley}, {Stark}, {Willmer}, {Alberts}, {Amorin}, {Arribas}, {Baum}, {Bunker}, {Carniani}, {Crandall}, {Egami}, {Eisenstein}, {Ferruit}, {Husemann}, {Maseda}, {Maiolino}, {Rawle}, {Rieke}, {Smit}, {Tacchella}, \& {Willott}}]{Williams2018}
{Williams}, C.~C., {Curtis-Lake}, E., {Hainline}, K.~N., {et~al.} 2018, \apjs, 236, 33, \dodoi{10.3847/1538-4365/aabcbb}

\bibitem[{{Wise} {et~al.}(2019){Wise}, {Regan}, {O'Shea}, {Norman}, {Downes}, \& {Xu}}]{Wise+2019}
{Wise}, J.~H., {Regan}, J.~A., {O'Shea}, B.~W., {et~al.} 2019, \nat, 566, 85, \dodoi{10.1038/s41586-019-0873-4}

\bibitem[{{Wolcott-Green} {et~al.}(2017){Wolcott-Green}, {Haiman}, \& {Bryan}}]{Wolcott-Green+2017}
{Wolcott-Green}, J., {Haiman}, Z., \& {Bryan}, G.~L. 2017, \mnras, 469, 3329, \dodoi{10.1093/mnras/stx167}

\bibitem[{{Wolcott-Green} {et~al.}(2021){Wolcott-Green}, {Haiman}, \& {Bryan}}]{Wolcott-Green+2021}
---. 2021, \mnras, 500, 138, \dodoi{10.1093/mnras/staa3057}

\bibitem[{{Woods} {et~al.}(2019){Woods}, {Agarwal}, {Bromm}, {Bunker}, {Chen}, {Chon}, {Ferrara}, {Glover}, {Haemmerl{\'e}}, {Haiman}, {Hartwig}, {Heger}, {Hirano}, {Hosokawa}, {Inayoshi}, {Klessen}, {Kobayashi}, {Koliopanos}, {Latif}, {Li}, {Mayer}, {Mezcua}, {Natarajan}, {Pacucci}, {Rees}, {Regan}, {Sakurai}, {Salvadori}, {Schneider}, {Surace}, {Tanaka}, {Whalen}, \& {Yoshida}}]{Woods+2019}
{Woods}, T.~E., {Agarwal}, B., {Bromm}, V., {et~al.} 2019, \pasa, 36, e027, \dodoi{10.1017/pasa.2019.14}

\bibitem[{{Xu} {et~al.}(2018){Xu}, {Yue}, \& {Chen}}]{Xu2018ApJ}
{Xu}, Y., {Yue}, B., \& {Chen}, X. 2018, \apj, 869, 42, \dodoi{10.3847/1538-4357/aae97b}

\bibitem[{{Xu} {et~al.}(2021){Xu}, {Yue}, \& {Chen}}]{Xu2021ApJ}
---. 2021, \apj, 923, 98, \dodoi{10.3847/1538-4357/ac30da}

\bibitem[{{Yan} {et~al.}(1998){Yan}, {Sadeghpour}, \& {Dalgarno}}]{Yan+1998}
{Yan}, M., {Sadeghpour}, H.~R., \& {Dalgarno}, A. 1998, \apj, 496, 1044, \dodoi{10.1086/305420}

\bibitem[{{Yang} {et~al.}(2003){Yang}, {Mo}, \& {van den Bosch}}]{Yang+2003}
{Yang}, X., {Mo}, H.~J., \& {van den Bosch}, F.~C. 2003, \mnras, 339, 1057, \dodoi{10.1046/j.1365-8711.2003.06254.x}

\bibitem[{{Yue} \& {Ferrara}(2021)}]{Yue+2021}
{Yue}, B., \& {Ferrara}, A. 2021, \mnras, 506, 5606, \dodoi{10.1093/mnras/stab2121}

\bibitem[{{Yue} {et~al.}(2017){Yue}, {Ferrara}, {Pacucci}, \& {Omukai}}]{Yue+2017}
{Yue}, B., {Ferrara}, A., {Pacucci}, F., \& {Omukai}, K. 2017, \apj, 838, 111, \dodoi{10.3847/1538-4357/aa6627}

\bibitem[{{Yue} {et~al.}(2014){Yue}, {Ferrara}, {Salvaterra}, {Xu}, \& {Chen}}]{Yue+2014}
{Yue}, B., {Ferrara}, A., {Salvaterra}, R., {Xu}, Y., \& {Chen}, X. 2014, \mnras, 440, 1263, \dodoi{10.1093/mnras/stu351}

\bibitem[{{Zhang} {et~al.}(2023){Zhang}, {Yue}, {Xu}, {Ma}, {Chen}, \& {Liu}}]{Zhang+2023}
{Zhang}, Z., {Yue}, B., {Xu}, Y., {et~al.} 2023, \prd, 107, 083013, \dodoi{10.1103/PhysRevD.107.083013}

\bibitem[{{Ziparo} {et~al.}(2022){Ziparo}, {Gallerani}, {Ferrara}, \& {Vito}}]{Ziparo+2022}
{Ziparo}, F., {Gallerani}, S., {Ferrara}, A., \& {Vito}, F. 2022, \mnras, 517, 1086, \dodoi{10.1093/mnras/stac2705}

\bibitem[{{Zwick} {et~al.}(2023){Zwick}, {Mayer}, {Haemmerl{\'e}}, \& {Klessen}}]{Zwick_2023MNRAS.518.2076Z}
{Zwick}, L., {Mayer}, L., {Haemmerl{\'e}}, L., \& {Klessen}, R.~S. 2023, \mnras, 518, 2076, \dodoi{10.1093/mnras/stac3204}

\end{thebibliography}
\bibliographystyle{aasjournal}




\end{document}